\documentclass[aps,showpacs,twocolumn,preprintnumbers,eqsecnum,bibnotes]{revtex4-1}
\usepackage{ifpdf}
\usepackage{graphicx}
\usepackage{color}
\usepackage{dcolumn}
\usepackage{bm}
\usepackage{mathrsfs}
\usepackage{amsfonts}
\usepackage{amssymb}
\usepackage{amsmath}
\usepackage{dsfont}
\usepackage{longtable}
\usepackage{fixmath}
\usepackage{upgreek}
\usepackage{latexsym}
\usepackage{bbding}
\usepackage{bold-extra}
\usepackage[T1]{fontenc}% Has bold small caps - relevant to \textsc in sec. headings
\newcommand{\SC}[1]{{\usefont{T1}{cmr}{bx}{sc}#1}}%
\usepackage{ae,aecompl}
\usepackage{url}
\usepackage[dvipsnames*,svgnames]{xcolor}
\usepackage[pdfauthor={Behnam Farid},%
            pdftitle={Some rigorous results concerning the uniform metallic ground states of single-band Hamiltonians in arbitrary dimensions},%
            plainpages=false,pdfpagelabels,pagebackref,%
            breaklinks=true,%
            hyperfootnotes=true,%
            bookmarks=true]{hyperref}
\hypersetup{colorlinks=true,%
            citecolor=blue,%DarkBlue,
            filecolor=black,%
            linkcolor=blue,%DarkBlue,
            urlcolor=blue,%
            pdftex,%
            pdfstartview={FitH}}
\usepackage[all]{hypcap}
\makeatletter
\renewcommand{\@fnsymbol}[1]{\ensuremath{\ifcase#1\or *\or \S \else\@ctrerr\fi}}
\makeatother
\def\t#1{\tilde{#1}}
\def\wh#1{\widehat{#1}}
\def\h#1{\hat{#1}}
\def\b#1{\bar{#1}}
\def\ol#1{\overline{#1}}

\def\Sc#1{\textsc{#1}}

\DeclareMathOperator{\re}{Re}
\DeclareMathOperator{\im}{Im}
\DeclareMathOperator{\e}{e}
\DeclareMathOperator{\rd}{d\!}
\DeclareMathOperator{\sgn}{sgn}

\newcounter{dummy}
\begin{document}
\ifx\href\undefined\else\hypersetup{linktocpage=true}\fi
% You should use BibTeX and apsrev.bst for references
\bibliographystyle{apsrev}

%\preprint{2013-1}
\title{Some rigorous results concerning the uniform metallic ground states of single-band Hamiltonians in arbitrary dimensions}
\protect\thanks{Dedicated to the memory of John Tjon [John Alexander Tjon Joe Gin] (7 December 1937 - 20 September 2010).}
\author{Behnam Farid}
\protect\email{behnam.farid@btinternet.com}

%\date{December 8, 2009}
%\date{May 8, 2013}
\date{\today}

\begin{abstract}
We reproduce and review some of the main results of three of our earlier papers, utilizing in doing so a considerably more transparent formalism than originally utilized. The most fundamental result to which we pay especial attention in this paper, is that the exact Fermi surface of the $N$-particle \textsl{uniform} metallic ground state of any single-band Hamiltonian, describing fermions interacting through an isotropic two-body potential whose Fourier transform exists, is a subset of the Fermi surface within the framework of the \textsl{exact} Hartree-Fock theory, in general to be distinguished from the one corresponding to a single-Slater-determinant approximation of the ground-state wave function. We also review some of the physical implications of the latter result. Our considerations reveal that the interacting Fermi surface of a uniform metallic ground state (whether isotropic or anisotropic) \textsl{cannot} be calculated exactly to order $\nu$, with $\nu\ge 2$, in the coupling constant $\lambda$ of the interaction potential in terms of the self-energy calculated to order $\nu$ in a non-self-consistent fashion. We show this to be interlinked with the failure of the Luttinger-Ward identity, and thus of the Luttinger theorem, for a self-energy that is not appropriately (e.g., self-consistently) related to the single-particle Green function from which the Fermi surface is deduced. We further show that the same mechanism that embodies the Luttinger theorem within the framework of the exact theory, accounts for a non-trivial dependence of the exact self-energy on $\lambda$ that cannot be captured within a non-self-consistent framework. We thus establish that the extant calculations that purportedly prove deformation of the interacting Fermi surface of the $N$-particle metallic ground state of the single-band Hubbard Hamiltonian with respect to its Hartree-Fock counterpart at the second order in the on-site interaction energy $U$, are fundamentally deficient. In an appendix we show that the number-density distribution function, to be distinguished from the site-occupation distribution function, corresponding to the $N$-particle ground state of the Hubbard Hamiltonian is \textsl{not} non-interacting $v$-representable, a fact established earlier numerically. This property is of particular relevance in respect of the zero-temperature formalism of the many-body perturbation theory.
\end{abstract}


\maketitle

{\footnotesize{\tableofcontents}}
%\tableofcontents

% 1.
\section{Introduction}
\label{s1}
The main purpose of this paper is to present an overview of some of the salient findings of Refs.~\cite{BF03,BF04a,BF04b} that have thus far not received the attention that we believe they deserve. Here we provide, with the advantage of hindsight, simplified demonstrations of these findings. For this purpose, in this paper we explicitly deal with the single-band Hubbard Hamiltonian \cite{PWA59,ThWR62,JH63}, for which one has
\begin{equation}
\wh{\mathcal{H}} = \!\sum_{\bm{k},\sigma}\! \varepsilon_{\bm{k}}\,
\h{a}_{\bm{k};\sigma}^{\dag} \h{a}_{\bm{k};\sigma} + \frac{U}{2\hspace{0.5pt} N_{\textsc{s}}}\! \sum_{\sigma,\sigma'} \sum_{\bm{k}, \bm{p}, \bm{q}}\!
\h{a}_{\bm{k}+\bm{q};\sigma}^{\dag} \h{a}_{\bm{p}-\bm{q};\sigma'}^{\dag}
\h{a}_{\bm{p};\sigma'}^{\phantom{\dag}} \h{a}_{\bm{k};\sigma}^{\phantom{\dag}}.
\label{e11}
\end{equation}
Where appropriate, we shall indicate the way in which the results corresponding to this Hamiltonian are extended and made to correspond to a more general single-band Hamiltonian that accounts for an arbitrary isotropic interaction potential, Eq.~(\ref{ea1}) \cite{BF04a}. In Eq.~(\ref{e11}), $\varepsilon_{\bm{k}}$ is the non-interacting energy dispersion, which may or may not be of the strictly tight-binding form, $U$ the on-site interaction energy, and $N_{\textsc{s}}$ the number of the lattice sites $\{\bm{R}_j\}$ on which $\wh{\mathcal{H}}$ is defined. Since we are interested in the $N$-particle \textsl{metallic} ground state (GS) of $\wh{\mathcal{H}}$, in the following $N_{\textsc{s}}$ will be macroscopically large.

For definiteness, we assume that $\{ {\bm R}_j \}$ is a Bravais lattice embedded in $\mathds{R}^d$ so that where we do not state otherwise, the summations over wave vectors, such as those in Eq.~(\ref{e11}), are over the $d$-dimensional first Brillouin zone (1BZ) corresponding to $\{ {\bm R}_j \}$. The operators $\h{a}_{\bm{k};\sigma}^{\phantom{\dag}}$ and $\h{a}_{\bm{k};\sigma}^{\dag}$ are canonical annihilation and creation operators in the Schr\"odinger picture, corresponding to fermions with spin index $\sigma$. They are periodic over the complete wave-vector space, with the 1BZ the fundamental region of periodicity. This property is enforced by identifying, for instance, $\h{a}_{\bm{k}+\bm{q};\sigma}^{\dag}$ with $\h{a}_{\bm{k}+\bm{q}+\bm{K}_0;\sigma}^{\dag}$, where $\bm{K}_0$ is a reciprocal-lattice vector for which $\bm{k}+\bm{q}+\bm{K}_0 \in \mathrm{1BZ}$.

A byproduct of the considerations in this paper is some new (from the perspective of either Refs.~\cite{BF03,BF04a,BF04b} or other earlier relevant publications by others known to us) insights regarding a number of properties of the exact self-energy $\Sigma_{\sigma}(\bm{k};\varepsilon)$ and the failure of \textsl{non-self-consistent} many-body perturbation expansions, to arbitrary order $\nu$ in the coupling constant of interaction, to reproduce these correctly. Of particular interest is our explicit demonstration of the vital role that satisfaction of the Luttinger theorem \cite{LW60,JML60,ID03,BF07a,BF07-12} plays in correctly, albeit qualitatively, reproducing the dependence of the exact $\Sigma_{\sigma}(\bm{k};\varepsilon)$ on the coupling constant of interaction in approximate calculations.

% 1.a.
\subsection{Generalities}
\label{s1a}
The $N$-particle \textsl{uniform} GS of $\wh{\mathcal{H}}$ for spin-$\frac{1}{2}$ fermions is characterized by two site-occupation numbers $\{n_{\uparrow}, n_{\downarrow}\}$, where
\begin{equation}
n_{\sigma} = \frac{N_{\sigma}}{N_{\textsc{s}}},\;\; \sigma \in \{\uparrow,\downarrow\},
\label{e12}
\end{equation}
in which $N_{\sigma}$ is the total number of particles with spin index $\sigma$ in the GS. One has
\begin{equation}
N = \sum_{\sigma} N_{\sigma}, \;\; n = \sum_{\sigma} n_{\sigma} \equiv \frac{N}{N_{\textsc{s}}}.
\label{e13}
\end{equation}
With $N_{\textsc{s}}$ assumed to be macroscopically large, a non-vanishing $n$ corresponds to a macroscopically large $N$.

Let $\vert\Psi_{N_{\sigma},N_{\b\sigma};0}\rangle$ denote the $N$-particle GS of $\wh{\mathcal{H}}$ and $E_{N_{\sigma},N_{\b\sigma};0}$ the corresponding eigenenergy, where $\b\sigma$ is the spin index complementary to $\sigma$; for $\sigma = \uparrow (\downarrow)$, $\b\sigma = \downarrow (\uparrow)$. In this paper, the numbers $\{ N_{\sigma}, N_{\b\sigma}\}$, and thus $\{n_{\sigma}, n_{\b\sigma}\}$, are special in that $E_{N_{\sigma},N_{\b\sigma};0}$ is minimal with respect to variations of $N_{\b\sigma}$, $0\le N_{\b\sigma}\le N$, in $N_{\sigma} = N - N_{\b\sigma}$. In contrast, in dealing with such \textsl{eigenstates} of $\wh{\mathcal{H}}$ as $\vert\Psi_{N_{\sigma}\pm 1,N_{\b\sigma};0}\rangle$, we shall be merely considering the \textsl{lowest-lying} $(N\pm 1)$-particle \textsl{eigenstates} of $\wh{\mathcal{H}}$ corresponding to the $\{ N_{\sigma}, N_{\b\sigma}\}$ specific to $\vert\Psi_{N_{\sigma},N_{\b\sigma};0}\rangle$. In general, the exact $(N\pm 1)$-particle \textsl{GS}s of $\wh{\mathcal{H}}$ coincide with $\vert\Psi_{N_{\sigma}+m,N_{\b\sigma}+m';0}\rangle$, where $m+m' = \pm 1$ \cite[\S\hspace{0.0pt}B.1.1]{BF07a}. It should be evident, however, that the \textsl{eigenenergies} $E_{N_{\sigma}\pm 1,N_{\b\sigma};0}$ are variational \textsl{upper bounds} to the energies $E_{N\pm 1;0}$ of the $(N\pm 1)$-particle GSs of $\wh{\mathcal{H}}$. Below we shall use $E_{N;0}$ and $E_{N_{\sigma},N_{\b\sigma};0}$ interchangeably, and similarly as regards $\vert\Psi_{N;0}\rangle$ and $\vert\Psi_{N_{\sigma},N_{\b\sigma};0}\rangle$.

% 2.
\section{On the Fermi surface \texorpdfstring{$\mathcal{S}_{\SC{\textsc{f}};\sigma}$}{}}
\label{s2}
In this section we introduce two energy dispersions, $\varepsilon_{\bm{k};\sigma}^{<}$ and $\varepsilon_{\bm{k};\sigma}^{>}$, which we demonstrate to satisfy
\begin{equation}
\varepsilon_{\bm{k};\sigma}^{<} < \mu < \varepsilon_{\bm{k};\sigma}^{>},\;\;\forall \bm{k} \in \mathrm{1BZ},
\label{e21}
\end{equation}
where $\mu$ is the chemical potential. The locus of the $\bm{k}$ points of the $\mathrm{1BZ}$ for which $\varepsilon_{\bm{k};\sigma}^{<}$ and $\varepsilon_{\bm{k};\sigma}^{>}$ are up to a \textsl{microscopic} deviation of the order of $1/N$ equal, defines a $(d-1)$-dimensional subset of the $\mathrm{1BZ}$, which we denote by $\mathcal{S}_{\textsc{f};\sigma}'$ and which may in principle be empty. The inequalities in Eq.~(\ref{e21}) imply that at any $\bm{k} \in \mathcal{S}_{\textsc{f};\sigma}'$ the energies $\varepsilon_{\bm{k};\sigma}^{<}$ and $\varepsilon_{\bm{k};\sigma}^{>}$ must be up to errors of the order of $1/N$ equal to $\mu$. We demonstrate that not only $\mathcal{S}_{\textsc{f};\sigma}'$ is equal to the exact Fermi surface $\mathcal{S}_{\textsc{f};\sigma}$ of the $N$-particle metallic GS of $\wh{\mathcal{H}}$, but also it is a subset of $\mathcal{S}_{\textsc{f};\sigma}^{\textsc{hf}}$, the Fermi surface within the framework of the \textsl{exact} Hartree-Fock theory. For the cases where $\mathcal{S}_{\textsc{f};\sigma}'$ is a \textsl{proper} subset of $\mathcal{S}_{\textsc{f};\sigma}^{\textsc{hf}}$, the difference set $\mathcal{S}_{\textsc{f};\sigma}^{\textsc{hf}}\backslash\mathcal{S}_{\textsc{f};\sigma}'$ constitutes the \textsl{pseudogap} region \cite[\S10]{BF03} of the Fermi surface of the $N$-particle metallic GS under consideration \cite{BF03,BF04a}. It is interesting to note that the property $\mathcal{S}_{\textsc{f};\sigma} = \mathcal{S}_{\textsc{f};\sigma}^{\textsc{hf}}$ is known to be exact for the Hubbard model in $d=\infty$ \cite{EMH89b} \cite{MV89,EMH89a,GKKR96,PF03}, on account of the constancy of the self-energy in this limit with respect to variations of $\bm{k}$ \cite{EMH89a,EMH89b}.

% 2.a.
\subsection{Preliminaries}
\label{s2a}
Defining
\begin{eqnarray}
\mu_{N;\sigma}^- &\doteq& E_{N;0} - E_{N_{\sigma}-1,N_{\b\sigma};0},
\label{e22} \\
\mu_{N;\sigma}^+ &\doteq& E_{N_{\sigma}+1,N_{\b\sigma};0} - E_{N;0},
\label{e23}
\end{eqnarray}
on account of the Jensen inequality \cite{J06}
\begin{equation}
E_{N;0} < \frac{1}{2} (E_{N-1;0} + E_{N+1;0}),
\label{e24}
\end{equation}
one arrives at
\begin{equation}
E_{N;0} < \frac{1}{2} (E_{N_{\sigma}-1,N_{\b\sigma};0} + E_{N_{\sigma}+1,N_{\b\sigma};0})
\Leftrightarrow \mu_{N;\sigma}^- < \mu_{N;\sigma}^+.
\label{e25}
\end{equation}
The Jensen inequality, signifying the \textsl{strict convexity} of $E_{N;0}$ as a function of $N$ (for the consequence of $E_{N;0}$ not being \textsl{strictly convex}, but merely \textsl{convex}, see Ref.~\cite[\S\hspace{0.0pt}B.1]{BF07a}), applies for thermodynamically stable systems \cite{WT90}. Hence, the last inequality in Eq.~(\ref{e25}) amounts to an exact relationship for the system under consideration, which by assumption is thermodynamically stable \cite[\S\hspace{0.0pt}B.1]{BF07a}. One can more generally show that the chemical potential at zero temperature, $\mu$, specific to the grand canonical ensemble of the states spanning the Fock space of $\wh{\mathcal{H}}$, in which the mean value $\b{N}$ of the number of particles is equal to $N$, satisfies the following inequalities \cite[\S\hspace{0.0pt}B.1]{BF07a}:
\begin{equation}
\mu_{N;\sigma}^- < \mu < \mu_{N;\sigma}^+,\;\; \forall\sigma.
\label{e26}
\end{equation}
Since in this paper we are dealing with \textsl{metallic} $N$-particle GSs, up to microscopic corrections of the order of $1/N$ \cite[\S\hspace{0.0pt}B.1]{BF07a} one has
\begin{equation}
\mu = \varepsilon_{\textsc{f}} = \max_{\sigma} \mu_{N;\sigma}^- = \min_{\sigma} \mu_{N;\sigma}^+,
\label{e27}
\end{equation}
where $\varepsilon_{\textsc{f}}$ is the Fermi energy corresponding to the $N$-particle metallic GS of $\wh{\mathcal{H}}$. While deviating from the most general formulation \cite[\S\hspace{0.0pt}B.1]{BF07a}, in the following we shall for simplicity assume that $\mu_{N;\sigma}^{\mp} = \mu_{N;\b\sigma}^{\mp}$. \emph{In the following we shall therefore denote $\mu_{N;\sigma}^{\mp}$ more concisely by $\mu_{N}^{\mp}$}.

With
\begin{equation}
\mathsf{n}_{\sigma}(\bm{k}) \doteq \langle\Psi_{N;0}\vert \h{a}_{\bm{k};\sigma}^{\dag} \h{a}_{\bm{k};\sigma}^{\phantom{\dag}} \vert\Psi_{N;0}\rangle,
\label{e28}
\end{equation}
the GS momentum-distribution function corresponding to particles with spin index $\sigma$, for $\mathsf{n}_{\sigma}(\bm{k}) \not= 0$ and $\mathsf{n}_{\sigma}(\bm{k}) \not= 1$ we introduce the following \textsl{normalized} $(N\mp 1)$-particle states \cite{BF03}:
\begin{equation}
\vert\Phi_{N_{\sigma}-1,N_{\b\sigma};\bm{k}}\rangle \doteq \frac{1}{\sqrt{\mathsf{n}_{\sigma}(\bm{k})}}\, \h{a}_{\bm{k};\sigma}^{\phantom{\dag}} \vert\Psi_{N;0}\rangle,
\label{e29}
\end{equation}
\begin{equation}
\vert\Phi_{N_{\sigma}+1,N_{\b\sigma};\bm{k}}\rangle \doteq \frac{1}{\sqrt{1-\mathsf{n}_{\sigma}(\bm{k})}}\, \h{a}_{\bm{k};\sigma}^{\dag} \vert\Psi_{N;0}\rangle.
\label{e210}
\end{equation}
With reference to the latter expression, we note that following the anti-commutation relation $[\h{a}_{\bm{k};\sigma}^{\phantom{\dag}}, \h{a}_{\bm{k};\sigma}^{\dag}]_{+} = 1$, one has
\begin{equation}
\langle\Psi_{N;0}\vert \h{a}_{\bm{k};\sigma}^{\phantom{\dag}} \h{a}_{\bm{k};\sigma}^{\dag} \vert\Psi_{N;0}\rangle \equiv 1 - \mathsf{n}_{\sigma}(\bm{k}).
\label{e211}
\end{equation}

With $\h{\bm P}$ denoting the total-momentum operator \cite[Eq.~(7.50)]{FW03}, making use of the property $\h{\bm P}\, \vert\Psi_{N;0}\rangle =0$, one verifies that $\{\vert\Phi_{N_{\sigma} \pm 1,N_{\b\sigma};\bm{k}}\rangle\}$ are eigenstates of $\h{\bm P}$ corresponding to eigenvalues $\{\pm \hbar\bm{k}\}$.

Defining (recall that $\langle\Phi_{N_{\sigma}\pm 1,N_{\b\sigma};\bm{k}}\vert\Phi_{N_{\sigma}\pm 1,N_{\b\sigma};\bm{k}}\rangle = 1$ for all $\bm{k} \in \textrm{1BZ}$ \cite{Note1})
\begin{equation}
E_{N_{\sigma} \pm 1,N_{\b\sigma};{\bm k}}' \doteq \langle\Phi_{N_{\sigma}\pm 1,N_{\b\sigma};\bm{k}}\vert \wh{\mathcal{H}} \vert\Phi_{N_{\sigma}\pm 1,N_{\b\sigma};{\bm k}}\rangle,
\label{e212}
\end{equation}
by the variational principle one has
\begin{equation}
E_{N_{\sigma} \pm 1,N_{\b\sigma};0} \le E_{N_{\sigma} \pm 1,N_{\b\sigma};\bm{k}}', \; \forall \bm{k} \in \mathrm{1BZ}.
\label{e213}
\end{equation}
Hence, by writing
\begin{equation}
E_{N_{\sigma} \pm 1,N_{\b\sigma};\bm{k}}' \equiv E_{N_{\sigma} \pm 1,N_{\b\sigma};0} + \delta E_{\pm}(\bm{k}), \label{e214}
\end{equation}
in the light of the inequality in Eq.~(\ref{e213}) one has $\delta E_{\pm}(\bm{k}) \ge 0$, $\forall \bm{k} \in \mathrm{1BZ}$. Introducing the single-particle energy dispersions (cf. Eqs.~(\ref{e22}) and (\ref{e23}))
\begin{eqnarray}
\varepsilon_{\bm{k};\sigma}^{<} &\doteq& E_{N;0} - E_{N_{\sigma} -1,N_{\b\sigma};\bm{k}}',
\label{e215} \\
\varepsilon_{\bm{k};\sigma}^{>} &\doteq& E_{N_{\sigma}+1,N_{\b\sigma};\bm{k}}' - E_{N;0},
\label{e216}
\end{eqnarray}
on account of $\delta E_{\pm}(\bm{k}) \ge 0$, $\forall \bm{k} \in \mathrm{1BZ}$, the following inequalities apply (see Eqs.~(\ref{e21}) and (\ref{e26})):
\begin{equation}
\varepsilon_{\bm{k};\sigma}^{<} \le \mu_{N}^- < \mu < \mu_{N}^+ \le \varepsilon_{\bm{k};\sigma}^{>}, \;\; \forall \bm{k} \in \mathrm{1BZ}.
\label{e217}
\end{equation}

Making use of the expression for $\wh{\mathcal{H}}$ in Eq.~(\ref{e11}), and of the canonical anti-commutation relations for $\h{a}_{\bm{k};\sigma}$ and $\h{a}_{\bm{k}';\sigma'}^{\dag}$, one readily obtains that \cite{BF03,BF04a}
\begin{equation}
\varepsilon_{\bm{k};\sigma}^{<} = \varepsilon_{\bm{k}} + U\, \frac{\beta_{\bm{k};\sigma}^{<}}{\mathsf{n}_{\sigma}(\bm{k})},\;\;
\varepsilon_{\bm{k};\sigma}^{>} = \varepsilon_{\bm{k}} + U\, \frac{\beta_{\bm{k};\sigma}^{>}}{1-\mathsf{n}_{\sigma}(\bm{k})},
\label{e218}
\end{equation}
where
\begin{eqnarray}
\beta_{\bm{k};\sigma}^{<} &\doteq& \frac{1}{N_{\textsc{s}}} \sum_{\sigma'}
\sum_{\bm{p},\bm{q}}\, \langle\Psi_{N;0}\vert \h{a}_{\bm{k};\sigma}^{\dag} \h{a}_{\bm{p}-\bm{q};\sigma'}^{\dag} \h{a}_{\bm{p};\sigma'} \h{a}_{\bm{k}-\bm{q};\sigma} \vert\Psi_{N;0}\rangle,\nonumber\\
\label{e219} \\
\beta_{\bm{k};\sigma}^{>} &\doteq& \frac{1}{N_{\textsc{s}}} \sum_{\sigma'}
\sum_{\bm{p},\bm{q}} \langle\Psi_{N;0}\vert \h{a}_{\bm{p}-\bm{q};\sigma'}^{\dag} \h{a}_{\bm{p};\sigma'} \h{a}_{\bm{k}-\bm{q};\sigma} \h{a}_{\bm{k};\sigma}^{\dag}
\vert\Psi_{N;0}\rangle\nonumber\\
&\equiv& n_{\bar\sigma} - \beta_{\bm{k};\sigma}^{<}.
\label{e220}
\end{eqnarray}

% 2.b.
\subsection{The surface \texorpdfstring{$\mathcal{S}_{\SC{\textsc{f}};\sigma}'$}{} and its relation to the exact Hartree-Fock Fermi surface \texorpdfstring{$\mathcal{S}_{\SC{\textsc{f}};\sigma}^{\SC{\textsc{hf}}}$}{}}
\label{s2b}
Let
\begin{equation}
\mathcal{S}_{\textsc{f};\sigma}' \doteq \big\{ \bm{k} \,\|\, \varepsilon_{\bm{k};\sigma}^{<} = \varepsilon_{\bm{k};\sigma}^{>} \big\},
\label{e221}
\end{equation}
where, in the light of the inequalities in Eq.~(\ref{e217}), $\varepsilon_{\bm{k};\sigma}^{<} = \varepsilon_{\bm{k};\sigma}^{>}$ is meant to signify an equality up to a microscopic correction of the order of $1/N$. It cannot \emph{a priori} be ruled out that the set $\mathcal{S}_{\textsc{f};\sigma}'$ may be empty.

From the expressions in Eq.~(\ref{e218}) and the identity in Eq.~(\ref{e220}), one deduces that
\begin{equation}
\varepsilon_{\bm{k};\sigma}^{<} = \varepsilon_{\bm{k};\sigma}^{>} \iff \beta_{\bm{k};\sigma}^{<} = \mathsf{n}_{\sigma}(\bm{k})\hspace{0.7pt} n_{\b\sigma}.
\label{e222}
\end{equation}
Combining the equality on the right-hand side (RHS) of the $\Leftrightarrow$ in Eq.~(\ref{e222}) with the first expression in Eq.~(\ref{e218}), in view of the results in Eqs.~(\ref{e27}) and (\ref{e217}) one immediately obtains
\begin{equation}
\varepsilon_{\bm{k};\sigma}^{<} = \varepsilon_{\bm{k};\sigma}^{>} \Longrightarrow \varepsilon_{\bm{k}} + U n_{\b\sigma} = \varepsilon_{\textsc{f}}.
\label{e223}
\end{equation}
The $\Rightarrow$ in Eq.~(\ref{e223}) and the $\Leftrightarrow$ in Eq.~(\ref{e222})  should be noted. Thus, whereas $\beta_{\bm{k};\sigma}^{<} = \mathsf{n}_{\sigma}(\bm{k})\hspace{0.7pt} n_{\b\sigma}$ is a necessary \textsl{and} sufficient condition for $\varepsilon_{\bm{k};\sigma}^{<} = \varepsilon_{\bm{k};\sigma}^{>}$, $\varepsilon_{\bm{k}} + U n_{\b\sigma} = \varepsilon_{\textsc{f}}$ is \textsl{only} a necessary condition for $\varepsilon_{\bm{k};\sigma}^{<} = \varepsilon_{\bm{k};\sigma}^{>}$. With reference to the expression in Eq.~(\ref{e224}) below, this is underlined by the right-most exact inequalities in Eq.~(\ref{e253}) below.

We point out that for the exact Hartree-Fock self-energy $\Sigma_{\sigma}^{\textsc{hf}}(\bm{k})$ corresponding to the $N$-particle uniform GS of $\wh{\mathcal{H}}$, one has (see Eq.~(\ref{ea6}))
\begin{equation}
\Sigma_{\sigma}^{\textsc{hf}}(\bm{k}) = \frac{1}{\hbar} U n_{\b\sigma},
\label{e224}
\end{equation}
which is \textsl{independent} of $\bm{k}$. Thus, the $U n_{\b\sigma}$ on the RHS of Eq.~(\ref{e223}) may be replaced by $\hbar\Sigma_{\sigma}^{\textsc{hf}}(\bm{k})$. Interestingly, one can demonstrate that in dealing with the $N$-particle uniform GS of the Hamiltonian in Eq.~(\ref{ea1}) one similarly has \cite{BF04a}
\begin{equation}
\varepsilon_{\bm{k};\sigma}^{<} = \varepsilon_{\bm{k};\sigma}^{>} \Longrightarrow \varepsilon_{\bm{k}} + \hbar\hspace{0.5pt}\Sigma_{\sigma}^{\textsc{hf}}(\bm{k}) = \varepsilon_{\textsc{f}},
\label{e225}
\end{equation}
where $\Sigma_{\sigma}^{\textsc{hf}}(\bm{k})$ non-trivially depends on $\bm{k}$ for non-contact-type two-particle interaction functions, Eqs.~(\ref{ea3}) -- (\ref{ea5}). For the explicit expressions of the $\varepsilon_{\bm{k};\sigma}^{<}$ and $\varepsilon_{\bm{k};\sigma}^{>}$ specific to the $N$-particle uniform GS of the $\wh{H}$ in Eq.~(\ref{ea1}), the reader is referred to Eq.~(6) in Ref.~\cite{BF04a}. Although in the present paper we are explicitly dealing with the $N$-particle uniform GS of the Hamiltonian in Eq.~(\ref{e11}), below we shall often denote $U n_{\b\sigma}$ by $\hbar\hspace{0.5pt}\Sigma_{\sigma}^{\textsc{hf}}(\bm{k})$ as a reminder that many of the results to be presented in this paper are applicable to the $N$-particle uniform metallic GS of the more general Hamiltonian in Eq.~(\ref{ea1}) \cite{BF04a}.

Defining
\begin{equation}
\mathcal{S}_{\textsc{f};\sigma}^{\textsc{hf}} \doteq \big\{ \bm{k} \,\|\, \varepsilon_{\bm{k}} + \hbar\Sigma_{\sigma}^{\textsc{hf}}(\bm{k}) = \varepsilon_{\textsc{f}}\big\},
\label{e226}
\end{equation}
from the expressions in Eqs.~(\ref{e221}) and (\ref{e225}) one immediately observes that
\begin{equation}
\mathcal{S}_{\textsc{f};\sigma}' \subseteq \mathcal{S}_{\textsc{f};\sigma}^{\textsc{hf}}.
\label{e227}
\end{equation}
That $\mathcal{S}_{\textsc{f};\sigma}'$ is not necessarily identical to $\mathcal{S}_{\textsc{f};\sigma}^{\textsc{hf}}$, is a direct consequence of the $\Rightarrow$ in Eq.~(\ref{e225}) (or Eq.~(\ref{e223})).

We should emphasize that \textsl{at this stage} of the considerations, $\mathcal{S}_{{\Sc f};\sigma}^{\Sc h\Sc f}$ differs from the Hartree-Fock Fermi surface corresponding to particles with spin index $\sigma$, by the fact that the $\varepsilon_{\textsc{f}}$ in Eq.~(\ref{e226}) is the \textsl{exact} Fermi energy, Eq.~(\ref{e27}), to be in principle distinguished from the Fermi energy $\varepsilon_{\textsc{f}}^{\textsc{hf}}$ within the framework of the exact Hartree-Fock theory. The latter energy is obtained by solving the following equation:
\begin{equation}
\frac{1}{N_{\textsc{s}}} \sum_{\sigma} \sum_{\bm{k} \in \mathrm{1BZ}} \Theta(\varepsilon_{\textsc{f}}^{\textsc{hf}} - [\varepsilon_{\bm{k}} + \hbar \Sigma_{\sigma}^{\textsc{hf}}(\bm{k})]) = n.
\label{e228}
\end{equation}
Later, Sec.~\ref{s2f}, we shall demonstrate that
\begin{equation}
\varepsilon_{\textsc{f}}^{\textsc{hf}} = \varepsilon_{\textsc{f}}.
\label{e229}
\end{equation}
For now we only mention that the \textsl{exact} $\Sigma_{\sigma}^{\textsc{hf}}(\bm{k})$ is an explicit functional of the \textsl{exact} $\{ \mathsf{n}_{\sigma}(\bm{k}) \| \sigma\}$, Eqs.~(\ref{ea3}) -- (\ref{ea5}), to be distinguished from $\{ \mathsf{n}_{\sigma}^{\textsc{hf}}(\bm{k}) \| \sigma\}$, the GS momentum-distribution functions corresponding to the single-Slater-determinant approximation of the $N$-particle GS of $\wh{\mathcal{H}}$ (for which one has $\mathsf{n}_{\sigma}^{\textsc{hf}}(\bm{k}) \in \{0,1\}$, $\forall \bm{k}, \sigma$), so that violation of the equality in Eq.~(\ref{e229}) would amount to an internal inconsistency in the exact Hartree-Fock theory (a possibility that cannot \emph{a priori} be ruled out): in the event of the equality in Eq.~(\ref{e229}) failing, on replacing the $\varepsilon_{\textsc{f}}^{\textsc{hf}}$ on the left-hand side (LHS) of Eq.~(\ref{e228}) by $\varepsilon_{\textsc{f}}$, the equality in Eq.~(\ref{e228}) would fail to hold. This failure should be viewed in the light of the fact that, following the defining expression in Eq.~(\ref{e28}), one has
\begin{equation}
\sum_{\bm{k}\in \mathrm{1BZ}} \mathsf{n}_{\sigma}(\bm{k}) = N_{\sigma},\;\; \forall\sigma,
\label{e230}
\end{equation}
where $\{N_{\sigma}, N_{\b\sigma}\}$ are the \textsl{exact} partial particle numbers corresponding to $\vert\Psi_{N;0}\rangle$, Eq.~(\ref{e13}). We remark that in general $\varepsilon_{\textsc{f}}^{\textsc{hf}}$ is \textsl{not} equal to its counterpart within the framework in which the $N$-particle uniform GS of $\wh{\mathcal{H}}$ is approximated by a single Slater determinant, except when $\vert\Psi_{N;0}\rangle$ and its approximation are \textsl{paramagnetic}, for which one has $n_{\sigma} = n_{\b\sigma} = n/2$; following the equality in Eq.~(\ref{e224}), in this case the approximate and exact Hartree-Fock self-energies coincide.

We note in passing that the equality in Eq.~(\ref{e228}) amounts to the statement of the Luttinger theorem \cite{LW60,JML60,ID03,BF07a,BF07-12} within the framework where $\Sigma_{\sigma}^{\textsc{hf}}(\bm{k})$ is the \textsl{total} self-energy. With reference to the expression in Eq.~(\ref{e330}) specialized to the case of $\nu=1$, one observes that indeed in this framework the Luttinger-Ward identity \cite{LW60} is satisfied.

% 2.c.
\subsection{The exact Fermi surface \texorpdfstring{$\mathcal{S}_{\SC{\textsc{f}};\sigma}$}{} and its relation to \texorpdfstring{$\mathcal{S}_{\SC{\textsc{f}};\sigma}'$}{}}
\label{s2c}
Since $\mu-\varepsilon_{\bm{k};\sigma}^{<}$ and $\varepsilon_{\bm{k};\sigma}^{>}-\mu$ are \textsl{variational} single-particle excitation energies at point $\bm{k}$, in view of the inequalities in Eq.~(\ref{e217}) it trivially follows that
\begin{equation}
\mathcal{S}_{\textsc{f};\sigma}' \subseteq \mathcal{S}_{\textsc{f};\sigma},
\label{e231}
\end{equation}
where $\mathcal{S}_{\textsc{f};\sigma}$ denotes the exact Fermi surface specific to particles with spin index $\sigma$ of the metallic $N$-particle GS under consideration. For clarity, $\mathcal{S}_{\textsc{f};\sigma}$ is by definition the locus of the points of the $\mathrm{1BZ}$ at which the single-particle excitation energies, as measured from $\mu$, are microscopically small, of the order of $1/N$. With $\Sigma_{\sigma}(\bm{k};\varepsilon)$ denoting the energy-momentum representation of the exact proper self-energy operator pertaining to the GS under consideration, the exact Fermi surface $\mathcal{S}_{\textsc{f};\sigma}$ is mathematically defined according to (cf. Eq.~(\ref{e271}))
\begin{equation}
\mathcal{S}_{\textsc{f};\sigma} \doteq \big\{ \bm{k} \,\|\, \varepsilon_{\bm{k}} + \hbar \Sigma_{\sigma}(\bm{k};\varepsilon_{\textsc{f}}) = \varepsilon_{\textsc{f}}\big\}.
\label{e232}
\end{equation}
We note that $\Sigma_{\sigma}(\bm{k};\varepsilon_{\textsc{f}}) \in \mathds{R}$ for \textsl{all} $\bm{k} \in \mathrm{1BZ}$ \cite[\S2.1.2]{BF07a}.

By the Luttinger theorem \cite{LW60,JML60,ID03,BF07a,BF07-12}, for the $N$-particle uniform GS under investigation one has (cf. Eqs.~(\ref{e228}), (\ref{e275}) and (\ref{e276}))
\begin{equation}
\frac{1}{N_{\textsc{s}}} \sum_{\sigma} \sum_{\bm{k} \in \mathrm{1BZ}} \Theta(\varepsilon_{\textsc{f}} - [\varepsilon_{\bm{k}} + \hbar \Sigma_{\sigma}^{\textsc{hf}}(\bm{k}) +\hbar S_{\sigma}(\bm{k})]) = n, \label{e233}
\end{equation}
where we have used the decomposition \cite{BF04a}
\begin{equation}
\Sigma_{\sigma}(\bm{k};\varepsilon_{\textsc{f}}) \equiv \Sigma_{\sigma}^{\textsc{hf}}(\bm{k}) + \Sigma_{\sigma}'(\bm{k};\varepsilon_{\textsc{f}})\equiv \Sigma_{\sigma}^{\textsc{hf}}(\bm{k}) + S_{\sigma}(\bm{k}),
\label{e234}
\end{equation}
in which, by the Kramers-Kr\"onig relationship, \cite{BF04a}
\begin{equation}
S_{\sigma}(\bm{k}) = \frac{1}{\pi} \int_0^{\infty} \frac{\rd\varepsilon}{\varepsilon}\, \im[\Sigma_{\sigma}(\bm{k};\varepsilon_{\textsc{f}}+\varepsilon) + \Sigma_{\sigma}(\bm{k};\varepsilon_{\textsc{f}}-\varepsilon)].
\label{e235}
\end{equation}

Since $\im[\Sigma_{\sigma}(\bm{k};\varepsilon)] \ge 0$ for $\varepsilon < \varepsilon_{\textsc{f}}$ and $\im[\Sigma_{\sigma}(\bm{k};\varepsilon)] \le 0$ for $\varepsilon > \varepsilon_{\textsc{f}}$, $\forall \bm{k}$ \cite[\S2.1.2]{BF07a}, one observes that $S_{\sigma}(\bm{k})$ is comprised of two competing contributions, so that the possibility of $S_{\sigma}(\bm{k}) = 0$ for \textsl{some} $\bm{k}$ cannot \emph{a priori} be ruled out. In fact, in the light of the decomposition in Eq.~(\ref{e234}) and the expressions in Eqs.~(\ref{e226}) and (\ref{e232}), the fundamental relationship in Eq.~(\ref{e241}) below implies that
\begin{equation}
S_{\sigma}(\bm{k}) = 0,\;\; \forall \bm{k} \in \mathcal{S}_{\textsc{f};\sigma}.
\label{e236}
\end{equation}
Thus, following the expressions leading to the result in Eq.~(\ref{e236}), one can write
\begin{equation}
\mathcal{S}_{\textsc{f};\sigma} = \mathcal{S}_{\textsc{f};\sigma}^{\textsc{hf}} \cap \big\{ \bm{k} \,\|\, S_{\sigma}(\bm{k}) = 0 \big\},
\label{e237}
\end{equation}
and consequently
\begin{equation}
\mathcal{S}_{\textsc{f};\sigma}^{\textsc{hf}}\backslash \mathcal{S}_{\textsc{f};\sigma} =  \mathcal{S}_{\textsc{f};\sigma}^{\textsc{hf}} \cap \big\{ \bm{k} \,\|\, S_{\sigma}(\bm{k}) \not= 0 \big\}. \label{e238}
\end{equation}
With reference to the remarks in the opening paragraph of Sec.~\ref{s2}, the set in Eq.~(\ref{e238}), if non-empty, amounts to the pseudogap region \cite[\S10]{BF03} of the Fermi surface of the $N$-particle uniform GS of $\wh{\mathcal{H}}$ \cite{BF03,BF04a}. We note in passing that following the result in Eq.~(\ref{e236}), the vector $\bm{\nabla} S_{\sigma}(\bm{k})$, when it exists, stands normal to $\mathcal{S}_{\textsc{f};\sigma}$ for all $\bm{k} \in \mathcal{S}_{\textsc{f};\sigma}$ \cite{BF04a}.

The result in Eq.~(\ref{e236}) gains additional significance by comparing the expressions in Eqs.~(\ref{e228}) and (\ref{e233}), taking into account the equality in Eq.~(\ref{e229}). For instance, one observes that the combination of $S_{\sigma}(\bm{k}) \le 0$ for $\bm{k}$ inside \textsl{and} $S_{\sigma}(\bm{k}) > 0$ for $\bm{k}$ outside the Hartree-Fock Fermi sea, amounts to a \textsl{sufficient} condition for the validity of the Luttinger theorem \cite{LW60,JML60,ID03,BF07a,BF07-12} for the $N$-particle uniform metallic GS of $\wh{\mathcal{H}}$.

For illustration, by considering the data for $\im[\t{\Sigma}_{\sigma}(\bm{k};\varepsilon+i 0^+)]$ in Figs.~1(b) and 1(c) of Ref.~\cite{ZSS95}, taking into account the expressions in Eqs.~(2.5), (2.14), (B.55) and (B.59) of Ref.~\cite{BF07a}, and the fact that in these figures $\varepsilon_{\textsc{f}}$ is the origin of the energy axis, one can readily convince oneself that for at least the calculations reported in Ref.~\cite{ZSS95} the function $S_{\sigma}(\bm{k})$ behaves as described above. Explicitly, noting that the results displayed in Figs.~1(b) and 1(c) of Ref.~\cite{ZSS95} correspond to the non-interacting energy dispersion $\varepsilon_{\bm{k}} = -2 t (\cos(k_x) + \cos(k_y))$ with $t=1$, and the band-filling $n=0.97$, it is evident that for instance $\bm{k} \equiv (k_x, k_y) = (3\pi/4,3\pi/4)$ is located \textsl{outside} the underlying Fermi sea. Because of the long tail of the $\im[\t{\Sigma}_{\sigma}(\bm{k};\varepsilon+i 0^+)]$ corresponding to this $\bm{k}$ for \textsl{negative} values of $\varepsilon$, it is evident that the function $\im[\Sigma_{\sigma}(\bm{k};\varepsilon_{\textsc{f}}-\varepsilon)]$ dominates the value of the $S_{\sigma}(\bm{k})$ in Eq.~(\ref{e235}), resulting in $S_{\sigma}(\bm{k}) > 0$. In contrast, with $\bm{k} = (\pi/4,\pi/4)$ being located \textsl{inside} the underlying Fermi sea, by the same reasoning as above, from the long tail of the $\im[\t{\Sigma}_{\sigma}(\bm{k};\varepsilon+i 0^+)]$ corresponding to this $\bm{k}$ for \textsl{positive} values of $\varepsilon$, it follows that $S_{\sigma}(\bm{k}) < 0$. From the data corresponding to $\bm{k} = (0,0)$ one arrives at a similar conclusion, that $S_{\sigma}(\bm{k}) < 0$, and further that $-S_{\sigma}(\bm{k})$ is larger at $\bm{k} = (0,0)$ than at $\bm{k} = (\pi/4,\pi/4)$. For the $\bm{k}$ points close to the underlying Fermi surface, such as $\bm{k} = (\pi,0)$, one clearly observes that $\im[\t{\Sigma}_{\sigma}(\bm{k};\varepsilon+i 0^+)]$ is nearly symmetrical with respect to the origin (insofar as $\im[\Sigma_{\sigma}(\bm{k};\varepsilon)]$ is concerned, Eq.~(\ref{ea8}), nearly \textsl{anti-symmetrical} with respect to $\varepsilon = \varepsilon_{\textsc{f}}$), in conformity with the result in Eq.~(\ref{e236}).

We now proceed by demonstrating that
\begin{equation}
\mathcal{S}_{\textsc{f};\sigma} \subseteq \mathcal{S}_{\textsc{f};\sigma}',
\label{e239}
\end{equation}
which in conjunction with the relationship in Eq.~(\ref{e231}) results in
\begin{equation}
\mathcal{S}_{\textsc{f};\sigma}' = \mathcal{S}_{\textsc{f};\sigma}.
\label{e240}
\end{equation}
On the basis of this and the exact relationship in Eq.~(\ref{e227}), one arrives at the fundamental relationship \cite{BF03,BF04a} (cf. Eq.~(\ref{e237}))
\begin{equation}
\mathcal{S}_{\textsc{f};\sigma} \subseteq \mathcal{S}_{\textsc{f};\sigma}^{\textsc{hf}}.
\label{e241}
\end{equation}

For demonstrating the validity of the relationship in Eq.~(\ref{e239}), we consider the annihilation operator $\h{a}_{\bm{k};\sigma}$ in the Heisenberg picture, which we denote by $\h{a}_{\bm{k};\sigma}^{\textsc{h}}(t)$. From the Heisenberg equation of motion \cite[Eq.~(6.29)]{FW03},
\begin{equation}
i\hbar\, \frac{\partial}{\partial t} \h{a}_{\bm{k};\sigma}^{\textsc{h}}(t)= \big[\h{a}_{\bm{k};\sigma}^{\textsc{h}}(t),\wh{\mathcal{H}}\big]_{-},
\label{e242}
\end{equation}
one readily obtains that the function $\beta_{\bm{k};\sigma}^{<}$, defined in Eq.~(\ref{e219}), can be expressed as follows \cite{BF04a}:
\begin{equation}
\beta_{\bm{k};\sigma}^{<} = \frac{1}{U} \Big\{\left. \!\!\hbar \frac{\partial}{\partial t} \mathsf{G}_{\sigma}(\bm{k};t-t')\right|_{t'=t+0^+}\!\!\! - \varepsilon_{\bm{k}}\, \mathsf{n}_{\sigma}(\bm{k})\Big\},
\label{e243}
\end{equation}
where ${\sf G}_{\sigma}(\bm{k};t-t')$ is the single-particle Green function corresponding to particles with spin index $\sigma$, defined according to \cite[Eq.~(7.46)]{FW03}
\begin{equation}
\mathsf{G}_{\sigma}(\bm{k};t-t') \doteq -i \langle\Psi_{N;0}\vert\mathcal{T} \{ \h{a}_{\bm{k};\sigma}^{\textsc{h}}(t)\hspace{0.4pt}\h{a}_{\bm{k};\sigma}^{\textsc{h} \dag}(t')\}\vert\Psi_{N;0}\rangle, \label{e244}
\end{equation}
where $\mathcal{T}$ is the fermion time-ordering operator. One readily verifies that \cite{BF04a}
\begin{eqnarray}
\left. \hbar \frac{\partial}{\partial t} \mathsf{G}_{\sigma}(\bm{k};t-t')\right|_{t'=t+0^+}\!\!\!\! &=& \frac{1}{\hbar} \int_{-\infty}^{\infty}\! \frac{\rd\varepsilon}{2\pi i}\, \e^{i\varepsilon 0^+/\hbar}\, \varepsilon\, G_{\sigma}(\bm{k};\varepsilon) \nonumber\\
&\equiv& \frac{1}{\hbar} \int_{-\infty}^{\mu} \!\rd\varepsilon\, \varepsilon\hspace{0.5pt} A_{\sigma}(\bm{k};\varepsilon),
\label{e245}
\end{eqnarray}
where $G_{\sigma}(\bm{k};\varepsilon)$, to be encountered in Eq.~(\ref{e271}) below, is the time-Fourier transform of $\mathsf{G}_{\sigma}(\bm{k};t)$, and $A_{\sigma}(\bm{k};\varepsilon)$ the single-particle spectral function, defined according to
\begin{equation}
A_{\sigma}(\bm{k};\varepsilon) \doteq \mp \frac{1}{\pi} \im[\t{G}_{\sigma}({\bm k};\varepsilon \pm i 0^+)]. \label{e246}
\end{equation}
Here $\t{G}_{\sigma}(\bm{k};z)$, $z\in \mathds{C}$, is the single-particle Green function in terms of which the `physical' Green function $G_{\sigma}(\bm{k};\varepsilon)$, $\varepsilon \in \mathds{R}$, is defined according to the prescription in Eq.~(\ref{ea8}).

One has the following exact sum rules \cite{BF03,BF04a,BF07a}:
\begin{equation}
\frac{1}{\hbar} \int_{-\infty}^{\infty} \rd\varepsilon\, A_{\sigma}(\bm{k};\varepsilon) = 1,
\label{e247}
\end{equation}
\begin{equation}
\frac{1}{\hbar} \int_{-\infty}^{\infty} \rd\varepsilon\, \varepsilon\hspace{0.5pt} A_{\sigma}(\bm{k};\varepsilon) = \varepsilon_{\bm{k}} + \hbar\hspace{0.5pt} \Sigma_{\sigma}^{\textsc{hf}}(\bm{k}).
\label{e248}
\end{equation}
For the latter sum rule, see Eq.~(50) in Ref.~\cite{BF03}; noting that the expression on the LHS of Eq.~(\ref{e248}) is the function $G_{\sigma;\infty_2}(\bm{k})$ \cite[Eq.~(B.68)]{BF07a}, with reference to Eq.~(B.72) in Ref.~\cite{BF07a}, see Eq.~(73) in Ref.~\cite{BF02}; also compare the expressions in Eqs.~(162) and (173) of the latter reference.

Further, the GS momentum-distribution function, defined in Eq.~(\ref{e28}), can be expressed as follows (cf. Eq.~(\ref{ea7})):
\begin{equation}
\mathsf{n}_{\sigma}(\bm{k}) = \frac{1}{\hbar} \int_{-\infty}^{\mu} \rd\varepsilon\, A_{\sigma}(\bm{k};\varepsilon). \label{e249}
\end{equation}
The exact result in Eq.~(\ref{e230}), which follows from the defining expression in Eq.~(\ref{e28}), is seen to hold on employing the expression on the RHS of Eq.~(\ref{e249}), provided that the $\mu$ in this expression satisfies the inequalities in Eq.~(\ref{e26}). Combining the result in Eq.~(\ref{e249}) with that in Eq.~(\ref{e247}), one deduces that (cf. Eq.~(\ref{e211}))
\begin{equation}
\frac{1}{\hbar} \int_{\mu}^{\infty} \rd\varepsilon\, A_{\sigma}(\bm{k};\varepsilon) = 1 -\mathsf{n}_{\sigma}(\bm{k}).
\label{e250}
\end{equation}

On the basis of the above observations, the expressions in Eq.~(\ref{e218}) can be written as follows:
\begin{equation}
\varepsilon_{\bm{k};\sigma}^{<} = \frac{\frac{1}{\hbar} \int_{-\infty}^{\mu} \rd\varepsilon\, \varepsilon\hspace{0.5pt} A_{\sigma}(\bm{k};\varepsilon)}{\frac{1}{\hbar} \int_{-\infty}^{\mu} \rd\varepsilon\, A_{\sigma}(\bm{k};\varepsilon)},\;\; \varepsilon_{\bm{k};\sigma}^{>} = \frac{\frac{1}{\hbar} \int_{\mu}^{\infty} \rd\varepsilon\, \varepsilon\hspace{0.5pt} A_{\sigma}(\bm{k};\varepsilon)}{\frac{1}{\hbar} \int_{\mu}^{\infty} \rd\varepsilon\, A_{\sigma}(\bm{k};\varepsilon)}.
\label{e251}
\end{equation}
From these expressions and those in Eqs.~(\ref{e248}), (\ref{e249}) and (\ref{e250}), one deduces the following identity \cite{BF03,BF04a}:
\begin{equation}
\mathsf{n}_{\sigma}(\bm{k})\, \varepsilon_{\bm{k};\sigma}^{<} + \big(1-\mathsf{n}_{\sigma}(\bm{k})\big)\, \varepsilon_{\bm{k};\sigma}^{>} = \varepsilon_{\bm{k}} + \hbar\hspace{0.5pt} \Sigma_{\sigma}^{\textsc{hf}}(\bm{k}),\; \forall \bm{k} \in\mathrm{1BZ}.
\label{e252}
\end{equation}
On account of the exact property $\varepsilon_{\bm{k};\sigma}^{<} \le \varepsilon_{\bm{k};\sigma}^{>}$ (as regards the equality, up to a correction of the order of $1/N$), Eq.~(\ref{e217}), from the above identity one infers that
\begin{equation}
0 \le \mathsf{n}_{\sigma}(\bm{k}) \le 1 \iff \varepsilon_{\bm{k};\sigma}^{<} \le \varepsilon_{\bm{k}} + \hbar\hspace{0.5pt} \Sigma_{\sigma}^{\textsc{hf}}(\bm{k}) \le \varepsilon_{\bm{k};\sigma}^{>}.
\label{e253}
\end{equation}
The inequalities on the LHS of the $\Leftrightarrow$ being true for all ${\bm k} \in \mathrm{1BZ}$, it follows that the inequalities on the RHS of the $\Leftrightarrow$ must also be true for all $\bm{k} \in \mathrm{1BZ}$. In this light, it is to be noted that the inequalities on the RHS of the $\Leftrightarrow$ are in full conformity with the result in Eq.~(\ref{e227}) (see also Eq.~(\ref{e225})). Evidently, the right-most inequalities in Eq.~(\ref{e253}) do not rule out the possibility that $\mathcal{S}_{\textsc{f};\sigma}'$ may indeed be a \textsl{proper} subset of $\mathcal{S}_{\textsc{f};\sigma}^{\textsc{hf}}$.

Below we demonstrate the validity of the expression in Eq.~(\ref{e239}). In doing so, we consider Fermi-liquid and non-Fermi-liquid metallic GSs separately. In the following we consider an \textsl{arbitrary} $\bm{k} \in \mathcal{S}_{\textsc{f};\sigma}$, which we denote by $\bm{k}_{\textsc{f};\sigma}$. As elsewhere in this paper, below by $\bm{k}_{\textsc{f};\sigma}^-$ and $\bm{k}_{\textsc{f};\sigma}^+$ we signify radial vectors whose end points are displaced \textsl{infinitesimally} from the end point of $\bm{k}_{\textsc{f};\sigma}$, with the endpoint of  $\bm{k}_{\textsc{f};\sigma}^-$ located \textsl{inside} and that of  $\bm{k}_{\textsc{f};\sigma}^+$ \textsl{outside} the underlying Fermi sea.

% 2.c.1
\subsubsection{Fermi liquids}
\label{s2c1}
Here we assume that the $N$-particle uniform metallic GS under consideration is a Fermi liquid (not necessarily a conventional one \cite[\S11.2.2]{BF03} \cite{BF99}), for which one has \cite{BF04b}
\begin{equation}
A_{\sigma}(\bm{k}_{\textsc{f};\sigma}^-;\varepsilon) - A_{\sigma}(\bm{k}_{\textsc{f};\sigma}^+;\varepsilon) = \hbar Z_{\bm{k}_{\textsc{f};\sigma}} \big\{ \delta(\varepsilon-\mu_{N}^-) - \delta(\varepsilon-\mu_{N}^+)\big\}, \label{e254}
\end{equation}
where $Z_{\bm{k}_{\textsc{f};\sigma}} >0$ is the Landau quasi-particle weight $Z_{\bm{k}}$ at $\bm{k} = \bm{k}_{\textsc{f};\sigma}$. The equality in Eq.~(\ref{e254}) follows on account of $\| \bm{k}_{\textsc{f};\sigma}^- - \bm{k}_{\textsc{f};\sigma}^+\|$ being infinitesimally small, whereby the incoherent parts of $A_{\sigma}(\bm{k}_{\textsc{f};\sigma}^{\mp};\varepsilon)$ do not contribute to the difference on the LHS of Eq.~(\ref{e254}). Further, the equality of the quasi-particle weight at $\varepsilon= \mu_{N}^{-}$ with that at $\varepsilon= \mu_{N}^+$ is dictated by the exact sum rule in Eq.~(\ref{e247}), which applies for \textsl{all} $\bm{k} \in \mathrm{1BZ}$.

Following the inequalities in Eq.~(\ref{e217}), one immediately observes that for $Z_{\bm{k}_{\textsc{f};\sigma}} > 0$ the numerators and the denominators of the expressions in Eq.~(\ref{e251}) are finitely \textsl{discontinuous} at $\bm{k} = \bm{k}_{\textsc{f};\sigma}$. In this connection, with reference to the expression in Eq.~(\ref{e249}) and the inner inequalities in Eq.~(\ref{e217}), from the expression in Eq.~(\ref{e254}) one obtains that
\begin{equation}
\mathsf{n}_{\sigma}(\bm{k}_{\textsc{f};\sigma}^-) - \mathsf{n}_{\sigma}(\bm{k}_{\textsc{f};\sigma}^+) = Z_{\bm{k}_{\textsc{f};\sigma}},
\label{e255}
\end{equation}
which is the celebrated Migdal theorem \cite{ABM57}.

In view of the above observations, we consider the following function:
\begin{equation}
h(k) \doteq \frac{f(k)}{g(k)}.
\label{e256}
\end{equation}
We assume that while $f(k)$ and $g(k)$ are both finitely discontinuous at $k=k_0$, $h(k)$ is continuous at $k=k_0$. With $f(k_0^-) - f(k_0^+) = \Delta_f \not=0$ and $g(k_0^-) - g(k_0^+) = \Delta_g \not=0$, where $k_0^{\pm} \doteq k_0 \pm 0^+$, from $h(k_0^-) = h(k_0^+)$ one trivially obtains that \cite[\S3.3]{BF04b}
\begin{equation}
h(k_0) = \frac{\Delta_f}{\Delta_g}.
\label{e257}
\end{equation}
Following this result, by \textsl{assuming} that $\varepsilon_{\bm{k};\sigma}^{<}$ and $\varepsilon_{\bm{k};\sigma}^{>}$ are continuous at $\bm{k} = \bm{k}_{\textsc{f};\sigma} \in \mathcal{S}_{\textsc{f};\sigma}$, from the expressions in Eq.~(\ref{e251}) one deduces that
\begin{equation}
\varepsilon_{\bm{k}_{\textsc{f};\sigma};\sigma}^{<} = \frac{Z_{\bm{k}_{\textsc{f};\sigma}} \mu_{N}^-}{Z_{\bm{k}_{\textsc{f};\sigma}}} \equiv \mu_{N}^-,\;\; \varepsilon_{\bm{k}_{\textsc{f};\sigma};\sigma}^{>} = \frac{-Z_{\bm{k}_{\textsc{f};\sigma}} \mu_{N}^+}{-Z_{\bm{k}_{\textsc{f};\sigma}}} \equiv \mu_{N}^+.
\label{e258}
\end{equation}
Since for metallic GSs, $\mu_{N}^+ - \mu_{N}^-$ is infinitesimally small, of the order of $1/N$ \cite[\S\hspace{0.0pt}B.1.1]{BF07a}, Eq.~(\ref{e27}), the results in Eq.~(\ref{e258}) establish that $\bm{k}_{\textsc{f};\sigma} \in \mathcal{S}_{\textsc{f};\sigma}'$, Eq.~(\ref{e221}). \emph{Thus, under the assumption that the functions $\varepsilon_{\bm{k};\sigma}^{<}$ and $\varepsilon_{\bm{k};\sigma}^{>}$ are continuous for \textsl{all} $\bm{k} \in \mathcal{S}_{\textsc{f};\sigma}$, we have established the validity of the relationship in Eq.~(\ref{e239}), and thereby of those in Eqs.~(\ref{e240}) and (\ref{e241}), in the case of Fermi-liquid metallic states.} For the validity of the assumption of continuity of $\varepsilon_{\bm{k};\sigma}^{<}$ and $\varepsilon_{\bm{k};\sigma}^{>}$ at $\bm{k} = \bm{k}_{\textsc{f};\sigma} \in \mathcal{S}_{\textsc{f};\sigma}$, the reader is referred to Sec.~\ref{s2d}.

% 2.c.2.
\subsubsection{Non-Fermi liquids}
\label{s2c2}
Although we have deduced the results in Eq.~(\ref{e258}) by assuming that $Z_{\bm{k}_{\textsc{f};\sigma}} > 0$, these results are clearly independent of the actual value of $Z_{\bm{k}_{\textsc{f};\sigma}}$. We therefore posit that the results in Eq.~(\ref{e258}) are equally valid for non-Fermi liquid metallic GSs, for which $Z_{\bm{k}_{\textsc{f};\sigma}}=0$; the essence of the present postulate is rooted in the fact that for metallic GSs, $\mathsf{n}_{\sigma}(\bm{k})$ is \textsl{singular} on $\mathcal{S}_{\textsc{f};\sigma}$. This is exemplified by the $N$-particle metallic GS of the one-dimensional Tomonaga-Luttinger model \cite{ML65}, for which one has (partly in the prevailing notation of the present paper) $Z_{\pm k_{\textsc{f};\sigma}} =0$ (cf. Eq.~(\ref{e416})) \cite{ML65,JV94}, and further $\Sigma^{\textsc{hf}}(k) \equiv 0$ \cite{ML65,JV94}\cite{Note2}. For this GS, the set of interacting Fermi wave numbers $\{ -k_{\textsc{f};\sigma}, +k_{\textsc{f};\sigma}\} \equiv \mathcal{S}_{\textsc{f};\sigma}$ coincides with its non-interacting counterpart $\{ -k_{\textsc{f};\sigma}^{(0)}, +k_{\textsc{f};\sigma}^{(0)}\} \equiv \mathcal{S}_{\textsc{f};\sigma}^{(0)}$ \cite{JV94}, which, in view of the property $\Sigma_{\sigma}^{\textsc{hf}}(k) \equiv 0$, can be trivially expressed as $\{ -k_{\textsc{f};\sigma}^{\textsc{hf}}, +k_{\textsc{f};\sigma}^{\textsc{hf}}\} \equiv \mathcal{S}_{\textsc{f};\sigma}^{\textsc{hf}}$. The use here of the notation $\mathcal{S}_{\textsc{f};\sigma}^{\textsc{hf}}$ is justified, since for the GS under consideration the Luttinger theorem applies \cite{BB97}, directly leading to the result in Eq.~(\ref{e229}), Sec.~\ref{s2f}. In this connection, as can be observed from Eqs.~(6) and (7) in Ref.~\cite{BB97}, $\varepsilon_{\textsc{f}}^{\textsc{hf}} \equiv \varepsilon_{\textsc{f}}^{(0)} = 0 = \varepsilon_{\textsc{f}}$. With the equality $\mathcal{S}_{\textsc{f};\sigma} = \mathcal{S}_{\textsc{f};\sigma}^{\textsc{hf}}$, Eq.~(\ref{e241}), having been shown to apply in the case at hand, on the basis of the fact that the relationship in Eq.~(\ref{e231}) is trivially valid, the truth of the relationship in Eq.~(\ref{e239}) follows. Consequently, the expressions in Eq.~(\ref{e258}) apply for the $N$-particle metallic GS of the one-dimensional Tomonaga-Luttinger model, even though for this GS $Z_{\pm k_{\textsc{f};\sigma}} =0$. Thus, and importantly, for this GS the corresponding $\varepsilon_{\bm{k};\sigma}^{<}$ and $\varepsilon_{\bm{k};\sigma}^{>}$ \textsl{are} indeed continuous at $\bm{k} = \bm{k}_{\textsc{f};\sigma} \in \mathcal{S}_{\textsc{f};\sigma}$, Sec.~\ref{s2d}.

As we have shown in Refs.~\cite{BF03,BF04a} (see in particular Sec.~11 in Ref.~\cite{BF03}), the distinction between Fermi-liquid and non-Fermi-liquid metallic GSs is manifested in the specific way in which $\varepsilon_{\textsc{f}} -\varepsilon_{\bm{k};\sigma}^{<}$ and $\varepsilon_{\bm{k};\sigma}^{>} -\varepsilon_{\textsc{f}}$ approach zero for $\bm{k} \to \bm{k}_{\textsc{f};\sigma} \in \mathcal{S}_{\textsc{f};\sigma}$, Sec.~\ref{s4}. For instance, for these functions vanishing to leading order like $\| \bm{k}-\bm{k}_{\textsc{f};\sigma}\|^{\gamma}$, with $0<\gamma <1$, one explicitly demonstrates that $Z_{\bm{k}_{\textsc{f};\sigma}}=0$. Sec.~\ref{s4a}.

% 2.d.
\subsection{On the continuity of \texorpdfstring{$\varepsilon_{\bm{k};\sigma}^{<}$}{} and \texorpdfstring{$\varepsilon_{\bm{k};\sigma}^{>}$}{} on \texorpdfstring{$\mathcal{S}_{\SC{\textsc{f}};\sigma}$}{}}
\label{s2d}
We first note that the $N$-particle uniform GS under consideration being \textsl{metallic} by assumption, it is \textsl{compressible}. Consequently, there is no \emph{a priori} reason why over $\mathcal{S}_{\textsc{f};\sigma}$, the set of $\bm{k}$ points at which the deviations of the exact single-particle excitation energies from $\varepsilon_{\textsc{f}}$ are of the order of $1/N$, the \textsl{variational} single-particle excitation energies $\varepsilon_{\bm{k};\sigma}^{<}$ and $\varepsilon_{\bm{k};\sigma}^{>}$ should be discontinuous. In this connection, it is to be noted that the variational $(N\mp 1)$-particle states introduced in Eqs.~(\ref{e29}) and (\ref{e210}) are eigenstates of the total-momentum operator $\h{\bm{P}}$, corresponding to continuous eigenvalues for continuous variations of $\bm{k}$ (see the remark in the paragraph following Eq.~(\ref{e211}) above).

We proceed by first considering the non-interacting (NI) $N$-particle GS $\vert\Psi_{N;0}^{(0)}\rangle$, approximating the exact $N$-particle GS $\vert\Psi_{N;0}\rangle$ in the weak-coupling region. We denote the Fermi surface and the momentum-distribution function corresponding to $\vert\Psi_{N;0}^{(0)}\rangle$ by respectively $\mathcal{S}_{\textsc{f};\sigma}^{(0)}$ and $\mathsf{n}_{\sigma}^{(0)}(\bm{k})$. Since $\mathsf{n}_{\sigma}^{(0)}(\bm{k})$ is equal to $1$ for $\bm{k}$ inside the NI Fermi sea, and equal to $0$ for $\bm{k}$ outside, the state in Eq.~(\ref{e29}) is not defined in the latter region, and similarly the state in Eq.~(\ref{e210}) is not defined in the former region. Consequently, in the present approximation $\varepsilon_{\bm{k};\sigma}^{<}$ and $\varepsilon_{\bm{k};\sigma}^{>}$ are defined \textsl{only} inside and outside the NI Fermi sea, respectively. In fact, for
\begin{equation}
\varepsilon_{\bm{k};\sigma}^{(0)} \doteq \left\{ \begin{array}{ll} \varepsilon_{\bm{k};\sigma}^{<}, & \bm{k}\; \text{inside the NI Fermi sea},\\ \\ \varepsilon_{\bm{k};\sigma}^{>}, & \bm{k}\; \text{outside the NI Fermi sea}, \end{array} \right.
\label{e259}
\end{equation}
in the framework of the present approximation one has \cite{BF03,BF04a}
\begin{equation}
\varepsilon_{\bm{k};\sigma}^{(0)} \equiv \varepsilon_{\bm{k}} + \hbar\Sigma_{\sigma;0}^{\textsc{hf}}(\bm{k}), \label{e260}
\end{equation}
where the `non-interacting' Hartree-Fock self-energy $\Sigma_{\sigma;0}^{\textsc{hf}}(\bm{k})$ is to be distinguished from the exact $\Sigma_{\sigma}^{\textsc{hf}}(\bm{k})$; with reference to Eqs.~(\ref{ea3}) -- (\ref{ea5}), $\Sigma_{\sigma;0}^{\textsc{hf}}(\bm{k})$ is defined in terms of $\mathsf{n}_{\sigma}^{(0)}(\bm{k})$. In the case of the $N$-particle uniform GS of the Hubbard Hamiltonian, for which $\Sigma_{\sigma}^{\textsc{hf}}(\bm{k})$ is determined by $n_{\b\sigma}$, Eq.~(\ref{e224}), $\Sigma_{\sigma;0}^{\textsc{hf}}(\bm{k})$ coincides with $\Sigma_{\sigma}^{\textsc{hf}}(\bm{k})$ for $n_{\b\sigma}^{(0)}$ coinciding with $n_{\b\sigma}$, $\forall\b\sigma$. It should be recalled that since $\vert\Psi_{N;0}^{(0)}\rangle$ is an $N$-particle state, one has $\sum_{\sigma} n_{\sigma}^{(0)} = \sum_{\sigma} n_{\sigma} = n$, Eq.~(\ref{e13}).

Evidently, $\varepsilon_{\bm{k};\sigma}^{(0)}$ varies continuously for $\bm{k}$ passing continuously through $\mathcal{S}_{\textsc{f};\sigma}^{(0)}$. This is particularly obvious in the case of the Hubbard Hamiltonian, for which $\Sigma_{\sigma;0}^{\textsc{hf}}(\bm{k})$ is \textsl{independent} of $\bm{k}$. Thus, in the `non-interacting' limit one has $\varepsilon_{\bm{k}_{\textsc{f};\sigma}^-;\sigma}^{<} = \varepsilon_{\bm{k}_{\textsc{f};\sigma}^+;\sigma}^{>}$ for $\bm{k}_{\textsc{f};\sigma} \in \mathcal{S}_{\textsc{f};\sigma}^{(0)}$, even though $\mathsf{n}_{\sigma}^{(0)}(\bm{k})$ is discontinuous at $\bm{k} = \bm{k}_{\textsc{f};\sigma}$; the origin of the continuity $\varepsilon_{\bm{k}_{\textsc{f};\sigma}^-;\sigma}^{<} = \varepsilon_{\bm{k}_{\textsc{f};\sigma}^+;\sigma}^{>}$ is rooted in the right-most equality in Eq.~(\ref{e222}), which applies also in the `non-interacting' limit.

The continuity of the type $\varepsilon_{\bm{k}_{\textsc{f};\sigma}^-;\sigma}^{<} = \varepsilon_{\bm{k}_{\textsc{f};\sigma}^+;\sigma}^{>}$, for $\bm{k}_{\textsc{f};\sigma} \in \mathcal{S}_{\textsc{f};\sigma}$, which in the previous paragraph we showed to be valid in the most singular case of the `non-interacting' limit (in the light of $\mathsf{n}_{\sigma}^{(0)}(\bm{k})$ undergoing the largest possible amount of discontinuity at the points of $\mathcal{S}_{\textsc{f};\sigma}^{(0)}$), is sufficient for demonstrating that $\varepsilon_{\bm{k};\sigma}^{<}$ and $\varepsilon_{\bm{k};\sigma}^{>}$ are continuous functions of $\bm{k}$ at $\bm{k} = \bm{k}_{\textsc{f};\sigma} \in \mathcal{S}_{\textsc{f};\sigma}$. This follows from the fact that since $\varepsilon_{\bm{k};\sigma}^{<} \le \mu_{N}^{-}$ and $\varepsilon_{\bm{k};\sigma}^{>} \ge \mu_{N}^{+}$, Eq.~(\ref{e217}), the equality $\varepsilon_{\bm{k}_{\textsc{f};\sigma}^-;\sigma}^{<} = \varepsilon_{\bm{k}_{\textsc{f};\sigma}^+;\sigma}^{>}$ implies each of these quantities to be, up to a correction of the order of $1/N$, equal to $\varepsilon_{\textsc{f}}$, Eq.~(\ref{e27}). Thus, for $\varepsilon_{\bm{k};\sigma}^{<}$ and $\varepsilon_{\bm{k};\sigma}^{>}$ to be discontinuous at $\bm{k} = \bm{k}_{\textsc{f};\sigma}$, one must have $\varepsilon_{\bm{k}_{\textsc{f};\sigma}^+;\sigma}^{<} = \varepsilon_{\textsc{f}} - \Delta_{\sigma}^{<}$ and $\varepsilon_{\bm{k}_{\textsc{f};\sigma}^-;\sigma}^{>} = \varepsilon_{\textsc{f}} + \Delta_{\sigma}^{>}$, where $\Delta_{\sigma}^{<}, \Delta_{\sigma}^{>} > 0$. On account of the continuity of $\Sigma_{\sigma}^{\textsc{hf}}(\bm{k})$ for $\bm{k}$ in a neighbourhood of $\mathcal{S}_{\textsc{f};\sigma}$, and of the identity in Eq.~(\ref{e252}), one must thus have
\begin{equation}
\mathsf{n}_{\sigma}(\bm{k}_{\textsc{f};\sigma}^+)\hspace{1pt} \Delta_{\sigma}^{<} + \big(1-\mathsf{n}_{\sigma}(\bm{k}_{\textsc{f};\sigma}^-)\big) \Delta_{\sigma}^{>} = 0.
\label{e261}
\end{equation}
It follows that unless $\mathsf{n}_{\sigma}(\bm{k}_{\textsc{f};\sigma}^+) = 0$, one cannot have $\Delta_{\sigma}^{<} \not= 0$, and unless $\mathsf{n}_{\sigma}(\bm{k}_{\textsc{f};\sigma}^-) = 1$, one cannot have $\Delta_{\sigma}^{>} \not= 0$. Since from the outset we have excluded from consideration those $\bm{k} \in \mathrm{1BZ}$ for which $\mathsf{n}_{\sigma}(\bm{k}) = 0$ and $\mathsf{n}_{\sigma}(\bm{k}) = 1$ \cite{Note1}, it follows that within the framework of our formalism, $\varepsilon_{\bm{k}_{\textsc{f};\sigma}^-;\sigma}^{<} = \varepsilon_{\bm{k}_{\textsc{f};\sigma}^+;\sigma}^{>}$ implies continuity of $\varepsilon_{\bm{k};\sigma}^{<}$ and $\varepsilon_{\bm{k};\sigma}^{>}$ at $\bm{k} = \bm{k}_{\textsc{f};\sigma}$, whereby the expressions in Eq.~(\ref{e258}) are shown to be valid.

\begin{figure}[t!]
\includegraphics[angle=0, width=0.43\textwidth]{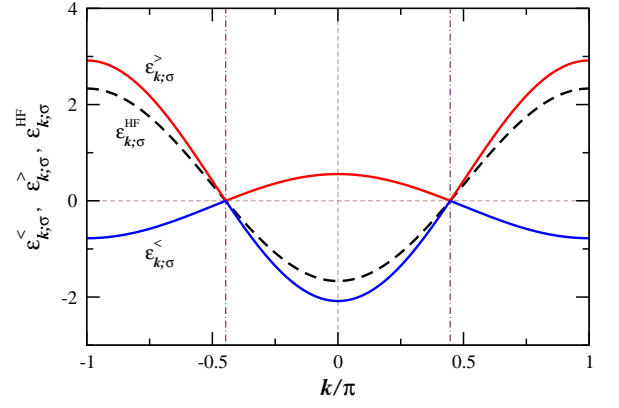}
\caption{(Colour) Schematic representations of the three energy dispersions $\varepsilon_{\bm{k};\sigma}^{<}$, $\varepsilon_{\bm{k};\sigma}^{>}$ and $\varepsilon_{\bm{k};\sigma}^{\textsc{hf}} \doteq \varepsilon_{\bm{k}} + \hbar\hspace{0.5pt} \Sigma_{\sigma}^{\textsc{hf}}(\bm{k})$ (in arbitrary units) specific to the metallic GS of a one-dimensional lattice model, with lattice constant $a=1$. Here $\varepsilon_{\textsc{f}}$ is chosen as the origin of energy, and further $\pm \bm{k}_{\textsc{f};\sigma}/\pi \approx \pm 0.447$ (the GS is less than half-filled). One notes \textsl{cusps} in $\varepsilon_{\bm{k};\sigma}^{<}$ and $\varepsilon_{\bm{k};\sigma}^{>}$ at $\bm{k} \in \{ -\bm{k}_{\textsc{f};\sigma}, +\bm{k}_{\textsc{f};\sigma}\} \equiv \mathcal{S}_{\textsc{f};\sigma}$. Similar cusps are seen in the single-particle energy dispersions calculated with the aid of a variational quantum Monte-Carlo method by Yunoki \emph{et al.} \protect\cite{YDS05} for a $t$-$J$ model on a square lattice (see Fig.~1 in Ref.~\protect\cite{YDS05}). Since by the Luttinger theorem $\|\bm{k}_{\textsc{f};\sigma}\|/\pi = n_{\sigma}$ (cf. Eq.~(\protect\ref{e233})), with reference to Eq.~(\protect\ref{e228}) from the depicted behaviour of $\varepsilon_{\bm{k};\sigma}^{\textsc{hf}}$ one observes that indeed the equality in Eq.~(\protect\ref{e229}) is satisfied. Note that the energy dispersions displayed here satisfy the inequalities in Eq.~(\protect\ref{e253}). Further, with reference to the expression in Eq.~(\protect\ref{e262}), the weights of the spectral peaks in $\mathcal{A}_{\sigma}(\bm{k};\varepsilon)$ at $\varepsilon = \varepsilon_{\bm{k};\sigma}^{<}$ and $\varepsilon = \varepsilon_{\bm{k};\sigma}^{>}$ are determined by respectively $\mathsf{n}_{\sigma}(\bm{k})$ and $1 - \mathsf{n}_{\sigma}(\bm{k})$. This clarifies the reason underlying $\varepsilon_{\bm{k};\sigma}^{<}$ and $\varepsilon_{\bm{k};\sigma}^{>}$ collapsing into the single-branch energy dispersion $\varepsilon_{\bm{k};\sigma}^{(0)}$, Eq.~(\protect\ref{e260}), in the non-interacting limit. The present diagram may be compared with the more schematic diagram in Fig.~3 of Ref.~\protect\cite{BF03}.}\label{f1}
\end{figure}

The question arises as to the consequences of the non-analyticity of $\mathsf{n}_{\sigma}(\bm{k})$ at $\bm{k} \in \mathcal{S}_{\textsc{f};\sigma}$ \cite{BF03} for $\varepsilon_{\bm{k};\sigma}^{<}$ and $\varepsilon_{\bm{k};\sigma}^{>}$. One can trivially demonstrate that non-analyticity of $\mathsf{n}_{\sigma}(\bm{k})$ (finite discontinuity in the case of Fermi liquids) leads to \textsl{cusps} in the latter energy dispersions at the points of $\mathcal{S}_{\textsc{f};\sigma}$ \cite{BF03}, Fig.~\ref{f1}. This follows from (i) the inequalities in Eq.~(\ref{e217}), (ii) the equality $\varepsilon_{\bm{k};\sigma}^{<} = \varepsilon_{\bm{k};\sigma}^{>}$, up to a deviation of the order of $1/N$, for $\bm{k} \in \mathcal{S}_{\textsc{f};\sigma} \subseteq \mathcal{S}_{\textsc{f};\sigma}^{\textsc{hf}}$, and (iii) the fact that, following the expressions in Eq.~(\ref{e218}), both $\varepsilon_{\bm{k};\sigma}^{<}$ and $\varepsilon_{\bm{k};\sigma}^{>}$ contain the contribution $\varepsilon_{\bm{k}}$, which, barring a possible subset of measure zero of the $\bm{k}$ space, is a monotonically increasing function of $\bm{k}$ for $\bm{k}$ transposed through $\mathcal{S}_{\textsc{f};\sigma}$ from the interior to the exterior of the underlying Fermi sea. In the light of these observations, one readily verifies that the \textsl{only} way for the inequalities in Eq.~(\ref{e217}) to remain satisfied for $\bm{k}$ in finite neighbourhoods of the points of $\mathcal{S}_{\textsc{f};\sigma}$, is that $\varepsilon_{\bm{k};\sigma}^{<}$ and $\varepsilon_{\bm{k};\sigma}^{>}$ be cusped at all $\bm{k}\in \mathcal{S}_{\textsc{f};\sigma}$. Detailed analysis \cite{BF03,BF04a} shows that the distinction between a variety of metallic GSs, whether of the Fermi-liquid type or otherwise, is reflected in the nature of the cusps in $\varepsilon_{\bm{k};\sigma}^{<}$ and $\varepsilon_{\bm{k};\sigma}^{>}$ at the points of $\mathcal{S}_{\textsc{f};\sigma}$. We shall briefly explore this aspect in Sec.~\ref{s4}. For now, we draw the attention of the reader to the expressions in Eqs.~(\ref{e43}), (\ref{e44}), (\ref{e421}), (\ref{e422}), (\ref{e423}), (\ref{e425}) and the inequalities in Eqs.~(\ref{e427}) and (\ref{e428}). In particular, the strict inequalities in Eq.~(\ref{e428}) unmistakably signify that the underlying $\varepsilon_{\bm{k};\sigma}^{<}$ and $\varepsilon_{\bm{k};\sigma}^{>}$ are indeed cusped at all $\bm{k} \in \mathcal{S}_{\textsc{f};\sigma}$.

% 2.e.
\subsection{Some salient properties of \texorpdfstring{$\varepsilon_{\bm{k};\sigma}^{<}$}{} and
\texorpdfstring{$\varepsilon_{\bm{k};\sigma}^{>}$}{}}
\label{s2e}
Consider the following single-particle spectral function:
\begin{equation}
\mathcal{A}_{\sigma}(\bm{k};\varepsilon) \doteq \hbar \big\{ \mathsf{n}_{\sigma}(\bm{k})\hspace{0.6pt} \delta(\varepsilon -\varepsilon_{\bm{k};\sigma}^{<}) + \big(1 - \mathsf{n}_{\sigma}(\bm{k})\big)\hspace{0.6pt} \delta(\varepsilon -\varepsilon_{\bm{k};\sigma}^{>}) \big\}.
\label{e262}
\end{equation}
Clearly, this function satisfies the exact sum rule in Eq.~(\ref{e247}). On account of the inequalities in Eq.~(\ref{e217}), one trivially verifies that in addition the equalities in Eqs.~(\ref{e249}) and (\ref{e250}) remain valid on substituting $\mathcal{A}_{\sigma}(\bm{k};\varepsilon)$ for the $A_{\sigma}(\bm{k};\varepsilon)$ in these equations. Similarly as regards the sum rule in Eq.~(\ref{e248}), this on account of the identity in Eq.~(\ref{e252}).

For the energy $E_{N;0}$ of the $N$-particle uniform GS under consideration, one has the following exact expression due to Galitskii and Migdal \cite{GM58} \cite[Eq.~(7.27)]{FW03}:
\begin{equation}
E_{N;0} = \frac{1}{2\hbar} \sum_{\bm{k},\sigma}\int_{-\infty}^{\mu} \rd\varepsilon\, (\varepsilon_{\bm{k}} + \varepsilon)\hspace{0.7pt} A_{\sigma}(\bm{k};\varepsilon).
\label{e263}
\end{equation}
Defining
\begin{equation}
\mathcal{E}_{N;0} \doteq \frac{1}{2\hbar} \sum_{\bm{k},\sigma}\int_{-\infty}^{\mu} \rd\varepsilon\, (\varepsilon_{\bm{k}} + \varepsilon)\hspace{0.7pt} \mathcal{A}_{\sigma}(\bm{k};\varepsilon),
\label{e264}
\end{equation}
following the inequalities in Eq.~(\ref{e217}) one obtains
\begin{equation}
\mathcal{E}_{N;0} = \frac{1}{2} \sum_{\bm{k},\sigma} \big\{\varepsilon_{\bm{k}} + \varepsilon_{\bm{k};\sigma}^{<} \big\}\hspace{0.7pt} \mathsf{n}_{\sigma}(\bm{k}).
\label{e265}
\end{equation}
In view of the expressions for $\varepsilon_{\bm{k};\sigma}^{<}$ and $\mathsf{n}_{\sigma}(\bm{k})$ in Eqs.~(\ref{e251}) and (\ref{e249}), the RHS of Eq.~(\ref{e265}) is seen to be identical to that of Eq.~(\ref{e263}). Thus \cite{BF03,BF04a}
\begin{equation}
\mathcal{E}_{N;0} = E_{N;0}.
\label{e266}
\end{equation}
This general result, whose validity is not restricted to the GS energy of the Hubbard Hamiltonian \cite{BF04a}, is in the case of the Hubbard Hamiltonian directly deduced through substituting the $\varepsilon_{\bm{k};\sigma}^{<}$ on the RHS of Eq.~(\ref{e265}) by the expression for $\varepsilon_{\bm{k};\sigma}^{<}$ in Eq.~(\ref{e218}); with reference to Eqs.~(\ref{e11}) and (\ref{e28}), and on account of $\langle\Psi_{N;0}\vert\Psi_{N;0}\rangle = 1$, one obtains
\begin{equation}
\mathcal{E}_{N;0} = \langle\Psi_{N;0}\vert \wh{\mathcal{H}}\vert\Psi_{N;0}\rangle \equiv E_{N;0}.
\label{e267}
\end{equation}

For the single-particle Green function $\t{G}_{\sigma}(\bm{k};z)$, $z\in \mathds{C}$, one has the exact spectral representation \cite[\S\hspace{0.0pt}B.2]{BF07a}
\begin{equation}
\t{G}_{\sigma}(\bm{k};z) = \int_{-\infty}^{\infty} \rd\varepsilon'\, \frac{A_{\sigma}(\bm{k};\varepsilon')}{z - \varepsilon'}
\label{e268}
\end{equation}
in terms of which the \textsl{physical} Green function $G_{\sigma}(\bm{k};\varepsilon)$, $\varepsilon \in \mathds{R}$, is determined according to the prescription in Eq.~(\ref{ea8}). In analogy, we define
\begin{equation}
\t{\mathcal{G}}_{\sigma}(\bm{k};z) \doteq \int_{-\infty}^{\infty} \rd\varepsilon'\, \frac{\mathcal{A}_{\sigma}(\bm{k};\varepsilon')}{z - \varepsilon'},
\label{e269}
\end{equation}
which is related to the `physical' Green function $\mathcal{G}_{\sigma}(\bm{k};\varepsilon)$, $\varepsilon\in \mathds{R}$, according to a prescription similar to that in Eq.~(\ref{ea8}). Making use of the expression in Eq.~(\ref{e262}), one obtains
\begin{equation}
\t{\mathcal{G}}_{\sigma}(\bm{k};z) = \hbar\Big\{ \frac{\mathsf{n}_{\sigma}(\bm{k})}{z -\varepsilon_{\bm{k};\sigma}^{<}} + \frac{1- \mathsf{n}_{\sigma}(\bm{k})}{z -\varepsilon_{\bm{k};\sigma}^{>}} \Big\}.
\label{e270}
\end{equation}
In Ref.~\cite{BF03} we have called this function a ``fictitious single-particle Green function'' (see Eqs.~(34) and (46) in Ref.~\cite{BF03}). Note that $\t{\mathcal{G}}_{\sigma}(\bm{k};z)$ shares the property  $\t{\mathcal{G}}_{\sigma}(\bm{k};z)\sim \hbar/z$ for $\vert z\vert \to \infty$, $\forall\bm{k}\in \textrm{1BZ}$, with the exact $\t{G}_{\sigma}(\bm{k};z)$ \cite[\S\hspace{0.0pt}B.3]{BF07a}, which is a direct consequence of $\mathcal{A}_{\sigma}(\bm{k};\varepsilon)$ satisfying the exact sum rule in Eq.~(\ref{e247}) for all $\bm{k} \in \textrm{1BZ}$.

By the Dyson equation \cite{FW03}, for the Fermi surface $\mathcal{S}_{\textsc{f};\sigma}$ as defined in Eq.~(\ref{e232}) one has \cite[Eq.~(23)]{BF03}
\begin{equation}
\mathcal{S}_{\textsc{f};\sigma} = \big\{ \bm{k} \,\|\, G_{\sigma}^{-1}(\bm{k};\varepsilon_{\textsc{f}}) = 0\big\}. \label{e271}
\end{equation}
By defining, in analogy,
\begin{equation}
\mathcal{S}_{\textsc{f};\sigma}'' \doteq \big\{ \bm{k} \,\|\, \mathcal{G}_{\sigma}^{-1}(\bm{k};\varepsilon_{\textsc{f}}) = 0\big\},
\label{e272}
\end{equation}
from the expression in Eq.~(\ref{e270}) one obtains
\begin{equation}
\mathcal{S}_{\textsc{f};\sigma}'' = \big\{ \bm{k} \,\|\, \varepsilon_{\bm{k};\sigma}^{<} = \varepsilon_{\textsc{f}}\big\} \cup \big\{ \bm{k} \,\|\, \varepsilon_{\bm{k};\sigma}^{>} = \varepsilon_{\textsc{f}}\big\}.
\label{e273}
\end{equation}
On replacing the $\cup$ on the RHS of this expression by $\cap$, one recovers the set $\mathcal{S}_{\textsc{f};\sigma}'$ as defined in Eq.~(\ref{e221}) (recall the equalities in Eq.~(\ref{e27}) and the inequalities in Eq.~(\ref{e217})). On account of the relationship $A \cap B \subseteq A \cup B$, it follows that $\mathcal{S}_{\textsc{f};\sigma}' \subseteq \mathcal{S}_{\textsc{f};\sigma}''$. Since by the same trivial reasoning as leading to the relationship in Eq.~(\ref{e231}) one has $\mathcal{S}_{\textsc{f};\sigma}'' \subseteq \mathcal{S}_{\textsc{f};\sigma}$, in the light of the result in Eq.~(\ref{e240}) one arrives at the conclusion that
\begin{equation}
\mathcal{S}_{\textsc{f};\sigma}'' = \mathcal{S}_{\textsc{f};\sigma}'.
\label{e274}
\end{equation}

For both metallic and insulating uniform GSs, one defines the Luttinger number $N_{\textsc{l};\sigma}$ according to \cite[Eq.~(2.21)]{BF07a} \cite{LW60,JML60,ID03,BF07-12}
\begin{equation}
N_{\textsc{l};\sigma} \doteq \sum_{\bm{k}} \Theta\big(G_{\sigma}(\bm{k};\mu)\big) \equiv \sum_{\bm{k}} \Theta\big(G_{\sigma}^{-1}(\bm{k};\mu)\big),
\label{e275}
\end{equation}
where $\Theta(x)$ is the unit-step function and $\mu$ the zero-temperature limit of the chemical potential $\mu_{\beta}$, where $\beta \doteq 1/(k_{\textsc{b}} T)$, satisfying the equation of state corresponding to the grand-canonical ensemble of the system under consideration whose mean value of particles $\b{N}$ is equal to $N$ \cite[\S2.3]{BF07a}. The Luttinger theorem states that (cf. Eq.~(\ref{e233})) \cite[Eq.~(2.22)]{BF07a} \cite{LW60,JML60,ID03,BF07-12}
\begin{equation}
N_{\textsc{l};\sigma} = N_{\sigma},\;\,\forall \sigma.
\label{e276}
\end{equation}

In analogy with the expression in Eq.~(\ref{e275}), we define \cite[Eq.~(35)]{BF03}
\begin{equation}
\mathcal{N}_{\textsc{l};\sigma} \doteq \sum_{\bm{k}} \Theta\big(\mathcal{G}_{\sigma}(\bm{k};\mu)\big). \label{e277}
\end{equation}
From the identities in Eq.~(45) of Ref.~\cite{BF03} it follows that
\begin{equation}
\mathcal{N}_{\textsc{l};\sigma} = N_{\sigma},\;\,\forall \sigma.
\label{e278}
\end{equation}
For clarity, the identities in Eq.~(45) of Ref.~\cite{BF03} follow from the fact that the set of $\bm{k}$ points for which $G_{\sigma}(\bm{k};\mu) > 0$, i.e. the set comprising the Fermi \textsl{sea} of the underlying uniform interacting metallic GS, is exactly the same set for which $\mathcal{G}_{\sigma}(\bm{k};\mu) > 0$. This observation follows from the exact Lehmann-type representation of $\t{\mathcal{G}}_{\sigma}(\bm{k};z)$ in Eq.~(33) of Ref.~\cite{BF03} with which the expression in Eq.~(\ref{e270}) is exactly equivalent, this on account of the expressions in Eqs.~(34) and (46) of Ref.~\cite{BF03}.

From the expression in Eq.~(\ref{e270}) one readily obtains that
\begin{eqnarray}
\mathcal{G}_{\sigma}(\bm{k};\mu) \ge 0 &\iff& \mathsf{n}_{\sigma}(\bm{k})\hspace{0.7pt} \varepsilon_{\bm{k};\sigma}^{>} + \big(1-\mathsf{n}_{\sigma}(\bm{k})\big)\hspace{0.7pt} \varepsilon_{\bm{k};\sigma}^{<} \ge \mu \nonumber\\
&\iff& \mathsf{n}_{\sigma}(\bm{k}) \ge \frac{\mu - \varepsilon_{\bm{k};\sigma}^{<}}{\varepsilon_{\bm{k};\sigma}^{>} - \varepsilon_{\bm{k};\sigma}^{<}}. \label{e279}
\end{eqnarray}
With $\mathrm{FS}_{\sigma}$ (Fermi or Luttinger sea \cite[\S2]{BF07a}) denoting the interior of $\mathcal{S}_{\textsc{f};\sigma}$, and $\ol{\mathrm{FS}}_{\sigma}$ its complementary part with respect to the underlying $\mathrm{1BZ}$ \cite[Eqs.~(18) and (22)]{BF03}, from the remark following Eq.~(\ref{e278}) it follows that
\begin{equation}
\mathsf{n}_{\sigma}(\bm{k}) \gtrless \frac{\mu - \varepsilon_{\bm{k};\sigma}^{<}}{\varepsilon_{\bm{k};\sigma}^{>} - \varepsilon_{\bm{k};\sigma}^{<}}\;\; \text{for}\;\; \bm{k} \in \left\{\begin{array}{l}  \mathrm{FS}_{\sigma}, \\ \\ \ol{\mathrm{FS}}_{\sigma}. \end{array} \right.
\label{e280}
\end{equation}
These inequalities amount to strict bounds on the range of variation of $\mathsf{n}_{\sigma}(\bm{k})$ for $\bm{k}$ over the entire $\mathrm{1BZ}$. For illustration, consider a not too strongly-correlated GS. Since for this GS $\textsf{n}_{\sigma}(\bm{k})$ is close to $1$ for $\bm{k}$ deep inside the $\mathrm{FS}_{\sigma}$, the upper inequality in Eq.~(\ref{e280}) implies that for this $\bm{k}$ the energy $\varepsilon_{\bm{k};\sigma}^{>}$ must be far closer to $\mu$ than $\varepsilon_{\bm{k};\sigma}^{<}$ is to $\mu$. Similarly, since for the same GS $\mathsf{n}_{\sigma}(\bm{k})$ is relatively small for $\bm{k}$ deep inside the $\ol{\mathrm{FS}}_{\sigma}$, the lower inequality in Eq.~(\ref{e280}) implies that for this $\bm{k}$ the energy $\varepsilon_{\bm{k};\sigma}^{<}$ must be far closer to $\mu$ than $\varepsilon_{\bm{k};\sigma}^{>}$ is to $\mu$. It is interesting to note that the right-most inequalities in Eq.~(\ref{e253}) provide strict upper and lowers bounds for respectively $\varepsilon_{\bm{k};\sigma}^{<}$ inside the $\mathrm{FS}_{\sigma}$ and $\varepsilon_{\bm{k};\sigma}^{>}$ inside the $\ol{\mathrm{FS}}_{\sigma}$. See Fig.~\ref{f1}.

Although the considerations in this paper are strictly confined to uniform \textsl{metallic} GSs, one can readily verify that of the results in this paper that have no bearing on $\mathcal{S}_{\textsc{f};\sigma}$ and $\varepsilon_{\textsc{f}}$, none is undermined by assuming the underlying uniform GSs to be insulating; it is on account of this consideration that in our above discussions concerning the Luttinger theorem, we have avoided use of $\varepsilon_{\textsc{f}}$ and used $\mu$ instead (for $\mu$, see the remark following Eq.~(\ref{e275}) above). Thus, by defining the \textsl{Luttinger surface} $\mathcal{S}_{\textsc{l};\sigma}$ as (for a more encompassing definition, see Ref.~\cite[\S2.4]{BF07a})
\begin{equation}
\mathcal{S}_{\textsc{l};\sigma} \doteq \big\{ \bm{k} \,\|\, G_{\sigma}(\bm{k};\mu) = 0\big\},
\label{e281}
\end{equation}
\textsl{it remains to be established} whether (cf. Eq.~(\ref{e240}))
\begin{equation}
\mathcal{S}_{\textsc{l};\sigma}' = \mathcal{S}_{\textsc{l};\sigma},
\label{e282}
\end{equation}
where
\begin{equation}
\mathcal{S}_{\textsc{l};\sigma}' \doteq
\big\{ \bm{k} \,\|\, \mathcal{G}_{\sigma}(\bm{k};\mu) = 0\big\}.
\label{e283}
\end{equation}
With reference to the second inequality in Eq.~(\ref{e279}), one has
\begin{equation}
\mathcal{S}_{\textsc{l};\sigma}' =
\big\{ \bm{k} \,\|\, \mathsf{n}_{\sigma}(\bm{k})\, \varepsilon_{\bm{k};\sigma}^{>} + \big(1-\mathsf{n}_{\sigma}(\bm{k})\big)\, \varepsilon_{\bm{k};\sigma}^{<} = \mu  \big\}.
\label{e284}
\end{equation}
Note the distinction between the expression to the left of the $\mu$ in Eq.~(\ref{e284}) and that on the LHS of the identity in Eq.~(\ref{e252}). Since for insulating GSs $\mathsf{n}_{\sigma}(\bm{k})$ is continuous at $\bm{k} \in \mathcal{S}_{\textsc{l};\sigma}$ (more generally, it is for any finite number of times differentiable with respect to $\bm{k}$ at this $\bm{k}$), in the event of the equality in Eq.~(\ref{e282}) being valid, one has the following well-defined expression (cf. Eq.~(\ref{e279})):
\begin{equation}
\mathsf{n}_{\sigma}(\bm{k}) = \frac{\mu - \varepsilon_{\bm{k};\sigma}^{<}}{\varepsilon_{\bm{k};\sigma}^{>} - \varepsilon_{\bm{k};\sigma}^{<}} \;\; \text{for}\;\; \bm{k} \in \mathcal{S}_{\textsc{l};\sigma}.
\label{e285}
\end{equation}
This result is by construction valid for all $\bm{k} \in \mathcal{S}_{\textsc{l};\sigma}'$.

% 2.f.
\subsection{Concerning \texorpdfstring{$\varepsilon_{\SC{\textsc{f}}}^{\SC{\textsc{hf}}} = \varepsilon_{\SC{\textsc{f}}}$}{}}
\label{s2f}
The result in Eq.~(\ref{e229}) is a consequence of two facts. First, the validity of the relationship in Eq.~(\ref{e241}) and, second, the validity of the Luttinger theorem \cite{LW60,JML60,ID03,BF07a,BF07-12}, whereby the number of the points comprising the underlying Fermi sea for \textsl{interacting} spin-$\sigma$ particles is equal to $N_{\sigma}$, Eq.~(\ref{e276}) and Fig.~\ref{f1}. We point out that neither the relationship in Eq.~(\ref{e241}) (unless $\mathcal{S}_{\textsc{f};\sigma} = \mathcal{S}_{\textsc{f};\sigma}^{\textsc{hf}}$), nor the inequalities on the RHS the $\Leftrightarrow$ in Eq.~(\ref{e253}) (in conjunction with the inequalities in Eq.~(\ref{e217})) preclude the possibility of $\varepsilon_{\bm{k};\sigma}^{\textsc{hf}} \doteq \varepsilon_{\bm{k}} + \hbar\Sigma_{\sigma}^{\textsc{hf}}(\bm{k})$ exceeding (falling short of) $\mu$ for $\bm{k}$ inside (outside) the Fermi sea of the interacting spin-$\sigma$ particles, however this possibility is \emph{a priori} ruled out in the case of the $N$-particle uniform GS of the Hubbard Hamiltonian, for which $\Sigma_{\sigma}^{\textsc{hf}}(\bm{k})$ is a constant, independent of $\bm{k}$, Eq.~(\ref{e224}). The functional form of the most general $\Sigma_{\sigma}^{\textsc{hf}}(\bm{k})$, Eqs.~(\ref{ea3}) -- (\ref{ea5}), practically rules out the above-mentioned possibility also for the $N$-particle uniform GS of the more general Hamiltonian in Eq.~(\ref{ea1}). Hence, the Fermi \textsl{sea} of the spin-$\sigma$ particles within the framework of the exact Hartree-Fock theory coincides with its exact counterpart. This coincides with the conclusion arrived at in Ref.~\cite{BF03}, to which we have referred in the remarks following Eq.~(\ref{e278}) above.

We note that the above-mentioned properties, leading to the result in Eq.~(\ref{e229}), as well as to that in Eq.~(\ref{e241}), are fully accounted for by the one-to-one mappings $\bm{\Phi}_{\sigma}^{<}(\bm{k})$ and $\bm{\Phi}_{\sigma}^{>}(\bm{k})$ introduced and employed in Ref.~\cite{BF03}. One noteworthy aspect that we have not emphasized in Ref.~\cite{BF03}, is that the latter mappings can be defined only in the limit where the underlying discrete set of $\bm{k}$ points is replaced by a continuum set. This follows from the fact that \textsl{bijective} mappings cannot be defined between two countable sets with different cardinal numbers \cite[p.~14]{HS91}. Even though in this paper we expressly do not employ the mappings $\bm{\Phi}_{\sigma}^{<}(\bm{k})$ and $\bm{\Phi}_{\sigma}^{>}(\bm{k})$ of Ref.~\cite{BF03}, the requirement for effecting the continuum limit prior to defining these mappings has its match in the considerations of the present paper, where we systematically neglect deviations of the order of $1/N$, in for instance considering the equality $\varepsilon_{\bm{k};\sigma}^{<} = \varepsilon_{\bm{k};\sigma}^{>}$. In view of the exact inequalities in Eq.~(\ref{e217}), without neglecting such deviations the latter equality \textsl{cannot} be satisfied for any $\bm{k}$.

% 3.
\section{On some extant calculations purportedly proving \texorpdfstring{$\mathcal{S}_{\SC{\textsc{f}};\sigma} \not\subseteq \mathcal{S}_{\SC{\textsc{f}};\sigma}^{\SC{\textsc{hf}}}$}{}}
\label{s3}
Results of several (numerical) calculations \cite{SG88,HM97,YY99,ZEG96} would suggest that for at least the paramagnetic $N$-particle uniform metallic GS of the single-band Hubbard Hamiltonian on two-dimensional lattices, one had $\mathcal{S}_{\textsc{f};\sigma} \not\subseteq \mathcal{S}_{\textsc{f};\sigma}^{(0)}$, where $\mathcal{S}_{\textsc{f};\sigma}^{(0)}$ is the Fermi surface of the underlying non-interacting $N$-particle GS, contradicting the result in Eq.~(\ref{e241}). In this connection, owing to the $\bm{k}$ and $\sigma$ independence of $\Sigma_{\sigma}^{\textsc{hf}}(\bm{k})$ in the case of the paramagnetic $N$-particle uniform GS of $\wh{\mathcal{H}}$ (for which $n_{\sigma} = n_{\b\sigma} = n/2$), Eq.~(\ref{e224}), one has $\mathcal{S}_{\textsc{f};\sigma}^{\textsc{hf}} =\mathcal{S}_{\textsc{f};\sigma}^{(0)}$, $\forall\sigma$. See Eq.~(\ref{e229}) as well as Sec.~\ref{s2f}. We remark however that since in this section we do not presuppose the equality in Eq.~(\ref{e229}), in discussing the observations in Refs.~\cite{SG88,HM97,YY99,ZEG96}, we assume $\mathcal{S}_{\textsc{f};\sigma}^{\textsc{hf}}$ to be defined in terms of $\varepsilon_{\textsc{f}}^{\textsc{hf}}$, and \textsl{not} $\varepsilon_{\textsc{f}}$, Eq.~(\ref{e226}) (cf. Eq.~(\ref{e32}) below).

In Ref.~\cite{BF03} we have argued that the calculations in Refs.~\cite{SG88,HM97,YY99,ZEG96} having been based on \textsl{non-self-consistent} many-body perturbation theory, the mere observation of Fermi-surface deformation (from $\mathcal{S}_{\textsc{f};\sigma}^{(0)}= \mathcal{S}_{\textsc{f};\sigma}^{\textsc{hf}}$ at $U=0$, into one violating $\mathcal{S}_{\textsc{f};\sigma} \subseteq \mathcal{S}_{\textsc{f};\sigma}^{\textsc{hf}}$ at $U>0$) rendered these calculations invalid. The failure of the conventional, that is \textsl{non-self-consistent}, many-body perturbation theory in \textsl{anisotropic} metallic GSs, as arising from the deformation of the underlying zeroth-order Fermi surface in consequence of interaction, has long since been recognized. For a discussion of this problem, and a possible way out of it, the reader is referred to Sec.~5.7 in Ref.~\cite{PN64}, as well as to the closing part of Sec.~\ref{s3c}, p.~\pageref{InClosing}.

\refstepcounter{dummy}
\label{Convex}
In this section we go into some details of the above-mentioned calculations (explicitly, those reported in Refs.~\cite{SG88,HM97,YY99}) and thus make our earlier qualitative argument in Ref.~\cite{BF03} quantitative. Since the calculations in Refs.~\cite{SG88,HM97,YY99} are restricted to $d=2$, in the following we mostly focus on $d=2$. Where we explicitly deal with the geometry of the Fermi surface, for transparency we exclusively consider the metallic GSs of which the underlying Fermi seas, corresponding to $\mathcal{S}_{\textsc{f};\sigma}^{\textsc{hf}}$ and $\mathcal{S}_{\textsc{f};\sigma}$, are \textsl{convex}. To avoid unnecessary notational complications, we further assume that both $\mathcal{S}_{\textsc{f};\sigma}^{\textsc{hf}}$ and $\mathcal{S}_{\textsc{f};\sigma}$ consist of closed curves. Naturally, in the following we do \textsl{not} presuppose the relationship $\mathcal{S}_{\textsc{f};\sigma} \subseteq \mathcal{S}_{\textsc{f};\sigma}^{\textsc{hf}}$, Eq.~(\ref{e241}).

% 3.a.
\subsection{Technical details}
\label{s3a}
In this section we first deduce two basic expressions that underly calculation of the possible deviation of $\mathcal{S}_{\textsc{f};\sigma}$ from $\mathcal{S}_{\textsc{f};\sigma}^{\textsc{hf}}$. Subsequently, we consider some details that are of relevance to the evaluation of these expressions in the framework of the many-body perturbation theory. The considerations reveal some interesting facts regarding the shortcomings of the self-energy calculated by means of a \textsl{non-self-consistent} many-body perturbation theory. To our knowledge, these shortcomings have to this date not been discussed elsewhere.

% 3.a.1.
\subsubsection{Two basic expressions}
\label{s3a1}
Following the above specifications regarding $\mathcal{S}_{\textsc{f};\sigma}$ and $\mathcal{S}_{\textsc{f};\sigma}^{\textsc{hf}}$, we introduce the outward planar unit vector $\h{\bm{n}}(\varphi)$ centred at the origin of the 1BZ under consideration and standing at angle $\varphi$ with respect to the positive $k_{x}$-axis. With $\bm{k}_{\textsc{f};\sigma}^{\textsc{hf}}(\varphi)$ and $\bm{k}_{\textsc{f};\sigma}(\varphi)$ denoting the wave vectors along $\h{\bm{n}}(\varphi)$ on respectively $\mathcal{S}_{\textsc{f};\sigma}^{\textsc{hf}}$ and $\mathcal{S}_{\textsc{f};\sigma}$, one can write
\begin{equation}
\bm{k}_{\textsc{f};\sigma}^{\textsc{hf}}(\varphi) = k_{\textsc{f};\sigma}^{\textsc{hf}}(\varphi)\hspace{0.7pt} \h{\bm{n}}(\varphi),\;\; \bm{k}_{\textsc{f};\sigma}(\varphi) = k_{\textsc{f};\sigma}(\varphi)\hspace{0.7pt} \h{\bm{n}}(\varphi).
\label{e31}
\end{equation}
Note that, by the assumed convexity of the relevant Fermi seas, $\bm{k}_{\textsc{f};\sigma}^{\textsc{hf}}$ and $\bm{k}_{\textsc{f};\sigma}$ are indeed uniquely specified by $\varphi$, which, by the further assumption that $\mathcal{S}_{\textsc{f};\sigma}^{\textsc{hf}}$ and $\mathcal{S}_{\textsc{f};\sigma}$ consist of closed curves, varies over the interval $[0,2\pi)$, and by periodicity over the entire $\mathds{R}$.

Following the expressions in Eqs.~(\ref{e224}) and (\ref{e226}), on replacing the $\varepsilon_{\textsc{f}}$ in the latter by $\varepsilon_{\textsc{f}}^{\textsc{hf}}$, one has
\begin{equation}
\varepsilon_{\bm{k}_{\textsc{f};\sigma}^{\textsc{hf}}(\varphi)} + U n_{\b\sigma} = \varepsilon_{\textsc{f}}^{\textsc{hf}},
\label{e32}
\end{equation}
and following the expressions in Eqs.~(\ref{e224}), (\ref{e232}) and (\ref{e234}),
\begin{equation}
\varepsilon_{\bm{k}_{\textsc{f};\sigma}(\varphi)} + U n_{\b\sigma} + \hbar\Sigma_{\sigma}'(\bm{k}_{\textsc{f};\sigma}(\varphi);\varepsilon_{\textsc{f}}) = \varepsilon_{\textsc{f}}.
\label{e33}
\end{equation}
With
\begin{equation}
\delta k_{\textsc{f};\sigma}^{\textsc{hf}}(\varphi) \doteq k_{\textsc{f};\sigma}(\varphi) - k_{\textsc{f};\sigma}^{\textsc{hf}}(\varphi),
\label{e34}
\end{equation}
\begin{equation}
\delta\varepsilon_{\textsc{f}}^{\textsc{hf}} \doteq \varepsilon_{\textsc{f}} - \varepsilon_{\textsc{f}}^{\textsc{hf}},
\label{e35}
\end{equation}
\begin{equation}
a_{\sigma}^{\textsc{hf}}(\varphi) \doteq \big(\left.\!\!\bm{\nabla}\varepsilon_{\bm{k}}\right|_{\bm{k} = \bm{k}_{\textsc{f};\sigma}^{\textsc{hf}}(\varphi)}\!\big) \cdot \h{\bm{n}}(\varphi),
\label{e36}
\end{equation}
and
\begin{equation}
\delta\t{\varepsilon}_{\sigma}^{\,\textsc{hf}}(\varphi) \doteq \varepsilon_{\bm{k}_{\textsc{f};\sigma}(\varphi)} - \varepsilon_{\bm{k}_{\textsc{f};\sigma}^{\textsc{hf}}(\varphi)} - a_{\sigma}^{\textsc{hf}}(\varphi)
\,\delta k_{\textsc{f};\sigma}^{\textsc{hf}}(\varphi),
\label{e37}
\end{equation}
on subtracting the equality in Eq.~(\ref{e32}) from that in Eq.~(\ref{e33}), one obtains the following \textsl{identity} for $\varphi$ over $[0,2\pi)$:
\begin{equation}
\delta k_{\textsc{f};\sigma}^{\textsc{hf}}(\varphi) \equiv \frac{\delta\varepsilon_{\textsc{f}}^{\textsc{hf}} - \delta\t{\varepsilon}_{\sigma}^{\,\textsc{hf}}(\varphi) -\hbar \Sigma_{\sigma}'(\bm{k}_{\textsc{f};\sigma}(\varphi);\varepsilon_{\textsc{f}})}
{a_{\sigma}^{\textsc{hf}}(\varphi)}.
\label{e38}
\end{equation}
In the light of the relationship in Eq.~(\ref{e241}), this is evidently the identity $0=0$, however we shall disregard this relationship in this section. We note that on taking into account the results in Eqs.~(\ref{e229}) and (\ref{e236}) (cf. Eq.~(\ref{e234})), the identity in Eq.~(\ref{e38}) would reduce to
\begin{equation}
\delta k_{\textsc{f};\sigma}^{\textsc{hf}}(\varphi) \equiv - \frac{\delta \t{\varepsilon}_{\sigma}^{\,\textsc{hf}}(\varphi)}{a_{\sigma}^{\textsc{hf}}(\varphi)},
\label{e39}
\end{equation}
through which the defining expression in Eq.~(\ref{e37}) would yield
\begin{equation}
\varepsilon_{\bm{k}_{\textsc{f};\sigma}(\varphi)} \equiv \varepsilon_{\bm{k}_{\textsc{f};\sigma}^{\textsc{hf}}(\varphi)},\;\; \forall\varphi,
\label{e310}
\end{equation}
which is in conformity with the identity $\delta k_{\textsc{f};\sigma}^{\textsc{hf}}(\varphi)\equiv 0$, Eqs.~(\ref{e34}) and (\ref{e241}).

The constant $\delta\varepsilon_{\textsc{f}}^{\textsc{hf}}$ (independent of $\varphi$) in Eq.~(\ref{e38}) can be eliminated, by invoking the Luttinger theorem \cite{LW60,JML60,ID03,BF07a,BF07-12} as follows. With the above-mentioned assumptions with regard to the exact and the Hartree-Fock Fermi seas, for the areas of these seas, $\mathscr{A}_{\sigma}$ and $\mathscr{A}_{\sigma}^{\textsc{hf}}$, one has
\begin{equation}
\mathscr{A}_{\sigma} = \int_0^{2\pi} \!\rd\varphi \int_0^{k_{\textsc{f};\sigma}(\varphi)} \!\rd k\, k \equiv \frac{1}{2}\int_0^{2\pi} \!\rd\varphi\, k_{\textsc{f};\sigma}^2(\varphi),
\label{e311}
\end{equation}
\begin{equation}
\mathscr{A}_{\sigma}^{\textsc{hf}} = \frac{1}{2}\int_0^{2\pi} \!\rd\varphi\, k_{\textsc{f};\sigma}^{\textsc{hf}\, 2}(\varphi).
\label{e312}
\end{equation}
With $\mathscr{A}_{\sigma} = \mathscr{A}_{\sigma}^{\textsc{hf}}$, as required by the Luttinger theorem \cite{LW60,JML60,ID03,BF07a,BF07-12}, from the expressions in Eqs.~(\ref{e311}) and (\ref{e312}) one deduces that
\begin{equation}
\int_0^{2\pi} \rd\varphi\, \big(k_{\textsc{f};\sigma}(\varphi) + k_{\textsc{f};\sigma}^{\textsc{hf}}(\varphi)\big)  \delta k_{\textsc{f};\sigma}^{\textsc{hf}}(\varphi) = 0.
\label{e313}
\end{equation}
Since $k_{\textsc{f};\sigma}(\varphi) + k_{\textsc{f};\sigma}^{\textsc{hf}}(\varphi) > 0$, $\forall\varphi$, for $n_{\sigma} >0$, the equality in Eq.~(\ref{e313}) implies that either $\delta k_{\textsc{f};\sigma}^{\textsc{hf}}(\varphi) \equiv 0$, or $\delta k_{\textsc{f};\sigma}^{\textsc{hf}}(\varphi)$ takes both positive and negative values over the interval $[0,2\pi)$, in which case by continuity $\delta k_{\textsc{f};\sigma}^{\textsc{hf}}(\varphi)$ must have a \textsl{finite} number of zeros over $[0,2\pi)$. In the light of the considerations of this paper, it is relevant to enquire about the specific characteristics of the points $\bm{k}$ at which $\delta k_{\textsc{f};\sigma}^{\textsc{hf}}(\varphi) = 0$. From the data displayed in for instance Fig.~3 of Ref.~\cite{HM97}, corresponding to different band fillings, one immediately infers that point-group symmetry \cite{JFC84} can clearly \textsl{not} be one such characteristic.

Multiplying both sides of the identity in Eq.~(\ref{e38}) by $\big(k_{\textsc{f};\sigma}(\varphi) + k_{\textsc{f};\sigma}^{\textsc{hf}}(\varphi)\big)$ and integrating the resulting expressions with respect to $\varphi$ over $[0,2\pi)$, in view of the equality in Eq.~(\ref{e313}) one obtains the following exact expression for $\delta\varepsilon_{\textsc{f}}^{\textsc{hf}}$:
\begin{widetext}
\begin{equation}
\delta\varepsilon_{\textsc{f}}^{\textsc{hf}} = \frac{\int_0^{2\pi} \rd\varphi\, \big(k_{\textsc{f};\sigma}(\varphi) + k_{\textsc{f};\sigma}^{\textsc{hf}}(\varphi)\big) \big(\delta\t{\varepsilon}_{\sigma}^{\,\textsc{hf}}(\varphi) + \hbar\Sigma_{\sigma}'(\bm{k}_{\textsc{f};\sigma}(\varphi);
\varepsilon_{\textsc{f}})\big)/a_{\sigma}^{\textsc{hf}}(\varphi)}{\int_0^{2\pi} \rd\varphi\, \big(k_{\textsc{f};\sigma}(\varphi) + k_{\textsc{f};\sigma}^{\textsc{hf}}(\varphi)\big)/a_{\sigma}^{\textsc{hf}}(\varphi)}.
\label{e314}
\end{equation}
\end{widetext}
Note that any constant shift, independent of $\varphi$, in either $\delta\t{\varepsilon}_{\sigma}^{\,\textsc{hf}}(\varphi)$ or $\Sigma_{\sigma}'(\bm{k}_{\textsc{f};\sigma}(\varphi);\varepsilon_{\textsc{f}})$, gives rise to an identical shift in $\delta\varepsilon_{\textsc{f}}^{\textsc{hf}}$. Consequently, on account of the identity in Eq.~(\ref{e38}), constant shifts in $\delta\t{\varepsilon}_{\sigma}^{\,\textsc{hf}}(\varphi)$ and $\Sigma_{\sigma}'(\bm{k}_{\textsc{f};\sigma}(\varphi);\varepsilon_{\textsc{f}})$ correctly do \textsl{not} affect the value of $\delta k_{\textsc{f};\sigma}^{\textsc{hf}}(\varphi)$, $\forall\varphi$.

% 3.a.2.
\subsubsection{Some theoretical considerations}
\label{s3a2}
The expressions relevant to the calculation of the deviation of $\mathcal{S}_{\textsc{f};\sigma}$ from $\mathcal{S}_{\textsc{f};\sigma}^{\textsc{hf}}$, presented above, only partly equip us with the means necessary for examining the calculations reported in Refs.~\cite{SG88,HM97,YY99}. For this, it is essential also to extend the notation for the self-energy that we have employed thus far in this paper. Formally, the extended notation has its origin in the perturbation expansion for the self-energy, which can in principle be based on the single-particle Green functions $\{G_{\sigma}^{\textsc{mf}} \| \sigma\}$ pertaining to a mean-field (MF) Hamiltonian (with the exception of the considerations in the closing part of Sec.~\ref{s3c}, p.~\pageref{InClosing}, in this paper $G_{\sigma}^{\textsc{mf}}$ is either $G_{\sigma}^{(0)}$, corresponding to the fully non-interacting Hamiltonian, or $G_{\sigma}^{\textsc{hf}}$, corresponding to the Hartree-Fock Hamiltonian), or the exact single-particle Green functions $\{G_{\sigma} \| \sigma\}$. In the former case all connected \textsl{proper} self-energy diagrams (those that do not become disconnected on removing a Green-function representing line), and in the latter only a subset of these, known as \textsl{skeleton} self-energy diagrams (those proper self-energy diagrams with no self-energy insertions) \cite{LW60,AGD75}, are to be taken into account. To make these aspects explicit, where necessary we denote the function that we have thus far denoted by $\Sigma_{\sigma}(\bm{k};\varepsilon)$, by $\Sigma_{\sigma}(\bm{k};\varepsilon;[\{G_{\sigma'}\}])$ ($\Upsigma_{\sigma}(\bm{k};\varepsilon;[\{G_{\sigma'}^{\textsc{mf}}\}])$) when formally it has been evaluated in terms of skeleton (proper) self-energy diagrams and $\{G_{\sigma}\| \sigma\}$ ($\{G_{\sigma}^{\textsc{mf}}\| \sigma\}$). Since in contrast to $\{G_{\sigma}^{\textsc{mf}}\| \sigma\}$, $\{G_{\sigma} \| \sigma\}$ is uniquely specified for the given Hamiltonian $\wh{\mathcal{H}}$, in many instances it will not be necessary to use the extended notation $\Sigma_{\sigma}(\bm{k};\varepsilon;[\{G_{\sigma'}\}])$ for $\Sigma_{\sigma}(\bm{k};\varepsilon)$. Therefore, \emph{in the following where no confusion can arise, we shall employ the shorter notation $\Sigma_{\sigma}(\bm{k};\varepsilon)$}. Similarly for $\Sigma_{\sigma}'(\bm{k};\varepsilon)$, Eqs.~(\ref{e234}) and (\ref{e316}).

We note in passing that the algebraic expressions corresponding to skeleton self-energy diagrams are free from the mathematical problems that plague non-skeleton proper self-energy diagrams, Refs.~\cite{JML61} and \cite[\S5.3.1]{BF07a}. Further, the notation introduced in the previous paragraph allows for viewing both $\Sigma_{\sigma}(\bm{k};\varepsilon;[\{G_{\sigma'}^{\textsc{mf}}\}])$ and $\Upsigma_{\sigma}(\bm{k};\varepsilon;[\{G_{\sigma'}\}])$ as meaningful functions.

Although one formally has
\begin{equation}
\Sigma_{\sigma}(\bm{k};\varepsilon;[\{G_{\sigma'}\}]) \equiv \Upsigma_{\sigma}(\bm{k};\varepsilon;[\{G_{\sigma'}^{\textsc{mf}}\}])
\label{e315}
\end{equation}
(since, as generally reasoned \cite{LW60}, by formally expanding $G_{\sigma}$ in `powers' of $G_{\sigma}^{\textsc{hf}}$, one recovers the set of all proper self-energy diagrams from the set of skeleton self-energy diagrams), this equivalence demonstrably fails for arbitrary mean-filed functions $\{G_{\sigma}^{\textsc{mf}}\| \sigma\}$. Notably, for an anisotropic $N$-particle uniform metallic GS it fails when the Fermi surfaces $\{\mathcal{S}_{\textsc{f};\sigma}^{\textsc{mf}} \| \sigma\}$ associated with $\{G_{\sigma}^{\textsc{mf}}\| \sigma\}$ are deformed with respect to their exact counterparts $\{\mathcal{S}_{\textsc{f};\sigma}\| \sigma\}$ \cite[\S5.7]{PN64}. Below we show that the equivalence in Eq.~(\ref{e315}) also fails for $\varepsilon_{\textsc{f}}^{\textsc{mf}} \not= \varepsilon_{\textsc{f}}$. More generally, \emph{even when $\varepsilon_{\textsc{f}}^{\textsc{mf}} = \varepsilon_{\textsc{f}}$, the equivalence in Eq.~(\ref{e315}) cannot hold exactly for $\bm{k}$ and $\varepsilon$ in a finite neighbourhood of respectively $\mathcal{S}_{\textsc{f};\sigma}$ and $\varepsilon_{\textsc{f}}$.} We arrive at this significant conclusion by demonstrating that for $\bm{k}$ and $\varepsilon$ in the latter neighbourhoods, $\Sigma_{\sigma}'(\bm{k};\varepsilon)/U^2$ is divergent for $U/t \to 0$, where (cf. Eq.~(\ref{e234}))
\begin{equation}
\t{\Sigma}_{\sigma}'(\bm{k};z;[\{G_{\sigma'}\}]) \doteq \t{\Sigma}_{\sigma}(\bm{k};z;[\{G_{\sigma'}\}]) -  \Sigma_{\sigma}^{\textsc{hf}}(\bm{k};[\{G_{\sigma'}\}]).
\label{e316}
\end{equation}
Failure of the perturbation theory (as described above) in terms of a fixed $\{G_{\sigma}^{\textsc{mf}}\| \sigma\}$, \emph{as opposed to the self-consistent perturbation theory}, has its root in the systematic failure of the non-self-consistent perturbation theory to comply with the Luttinger theorem \cite{LW60,JML60,ID03,BF07a,BF07-12}, or, what is the same, the Luttinger-Ward identity \cite{LW60}, Sec.~\ref{s3a3}.
\refstepcounter{dummy}
\label{ToBeExplicit}

To be explicit, for metallic GSs the Luttinger theorem embodies a very strict correspondence between all points of the Fermi surface $\mathcal{S}_{\textsc{f};\sigma}$, $\forall\sigma$, and the Fermi energy $\varepsilon_{\textsc{f}}$. This correspondence brings two singular aspects of metallic GSs into direct contact, one in the momentum space (as reflected in the singularity -- \textsl{not} necessarily a discontinuity -- of $\mathsf{n}_{\sigma}(\bm{k})$ at all $\bm{k} \in \mathcal{S}_{\textsc{f};\sigma}$, $\forall\sigma$), and one in the energy space (as reflected in the fact of $\t{\Sigma}_{\sigma}(\bm{k};z)$ \textsl{not} being arbitrary many times differentiable with respect to $z$ at $z=\varepsilon_{\textsc{f}}$ \cite{BF99} -- as we point out in appendix \ref{sab}, $z=\varepsilon_{\textsc{f}}$ is a branch-point singularity \cite[\S5.7]{WW62} of $\t{\Sigma}_{\sigma}(\bm{k};z)$ in the thermodynamic limit). The above-mentioned divergence of $\Sigma_{\sigma}'(\bm{k};\varepsilon)/U^2$ for in particular $\bm{k} \in \mathcal{S}_{\textsc{f};\sigma}$ and $\varepsilon = \varepsilon_{\textsc{f}}$ as $U/t \to 0$, is due to a coherence effect that is lost when in the calculation of the self-energy the above-mentioned strict correspondence, embodied by the Luttinger theorem, is systematically violated, Sec.~\ref{s3a3}.

There is one skeleton self-energy diagram to be considered for the evaluation of the second-order contribution to $\Sigma_{\sigma}(\bm{k};\varepsilon;[\{G_{\sigma'}\}])$, which we denote by $\Sigma_{\sigma}^{(2)}(\bm{k};\varepsilon;[\{G_{\sigma'}\}])$, and two connected proper self-energy diagrams for the evaluation of the second-order contribution to $\Upsigma_{\sigma}(\bm{k};\varepsilon;[\{G_{\sigma'}^{\textsc{mf}}\}])$, which we denote by $\Upsigma_{\sigma}^{(2)}(\bm{k};\varepsilon;[\{G_{\sigma'}^{\textsc{mf}}\}])$. For an arbitrary Hamiltonian, including the Hubbard Hamiltonian, the contribution of the second-order non-skeleton self-energy diagram to $\Upsigma_{\sigma}^{(2)}(\bm{k};\varepsilon;[\{G_{\sigma'}^{\textsc{mf}}\}])$ (an \textsl{anomalous} diagram \cite{KL60,LW60,NO98}, displayed in Fig.~2(b) of Ref.~\cite{HM97}) amounts to a real constant, independent of $\bm{k}$ and $\varepsilon$, which is non-vanishing only in the thermodynamic limit. With reference to the remarks following Eq.~(\ref{e314}) above, for the considerations of this section regarding $\delta k_{\textsc{f};\sigma}^{\textsc{hf}}(\varphi)$ one may therefore formally disregard the second-order non-skeleton self-energy diagram altogether and view the second-order self-energy $\Upsigma_{\sigma}^{(2)}(\bm{k};\varepsilon;[\{G_{\sigma'}^{\textsc{mf}}\}])$ as corresponding to the second-order \textsl{skeleton} self-energy diagram, to which $\Sigma_{\sigma}^{(2)}(\bm{k};\varepsilon;[\{G_{\sigma'}\}])$ corresponds. The equality in Eq.~(\ref{e314}) being deduced by the application of the Luttinger theorem, $N_{\textsc{l};\sigma} = N_{\sigma}$, Eq.~(\ref{e276}), one should however realize that it can only be used in the frameworks in which the Luttinger theorem applies (specifically for the considerations of Refs.~\cite{SG88,HM97,YY99,ZEG96}, to \textsl{at least} order $U^2$).

Without entering into details here, with reference to the statement in the abstract of the present paper, we remark that $\Upsigma_{\sigma}^{(2)}(\bm{k};\varepsilon;[\{G_{\sigma'}^{\textsc{mf}}\}])$ \textsl{not} being evaluated in terms of the single-particle Green functions $\{G_{\sigma}'\| \sigma\}$, Eq.~(\ref{e332}), associated with $\Upsigma_{\sigma}^{(2)}(\bm{k};\varepsilon;[\{G_{\sigma'}^{\textsc{mf}}\}])$ itself, the number $N_{\textsc{l};\sigma}$ of $\bm{k}$ points enclosed by the Fermi surface deduced on the basis of the latter self-energy \textsl{cannot} be equal to $N_{\sigma}$ (Eq.~(\ref{e325}), (\ref{e326}), (\ref{e327}), (\ref{e329}), (\ref{e332}), (\ref{e335})), which amounts to violation of the Luttinger theorem. The considerations in the following sections establish that theoretically the deviation $(N_{\sigma} - N_{\textsc{l};\sigma})/N_{\textsc{s}} \equiv n_{\sigma} -N_{\textsc{l};\sigma}/N_{\textsc{s}}$ can at best scale like $U^2$, a possibility that we believe to be unattainable in principle. Specifically for the $N$-particle uniform GS of the Hubbard Hamiltonian in $d \le 2$, in principle this deviation scales like $U^{\alpha} \ln^{\gamma}(\vert t\vert/U)$, where $1 \le \alpha \le 2$ and $\gamma \ge 0$, ruling out the possibilities of $\alpha = 1$, $\gamma > 0$, and $\alpha = 2$, $\gamma = 0$. See appendices \ref{sab} and \ref{sac}. \emph{For the reason that we shall present later in this section, p.~\pageref{SinceInD}, the restriction $d \le 2$, instead of $d <\infty$, is almost certainly superfluous.}

Both of the second-order self-energies referred to above, evaluated in terms of $\{G_{\sigma}^{(0)}\| \sigma\}$ and $\{G_{\sigma}^{\textsc{hf}}\| \sigma\}$, are directly proportional to $U^2$. While the function $\Upsigma_{\sigma}^{(2)}(\bm{k};\varepsilon;[\{G_{\sigma'}^{(0)}\}])/U^2$ is fully independent of $U$, this is not the case with $\Upsigma_{\sigma}^{(2)}(\bm{k};\varepsilon;[\{G_{\sigma'}^{\textsc{hf}}\}])/U^2$, as well as $\Sigma_{\sigma}^{(2)}(\bm{k};\varepsilon;[\{G_{\sigma'}\}])/U^2$. From the explicit expression for $\Upsigma_{\sigma}^{(2)}(\bm{k};\varepsilon;[\{G_{\sigma'}^{(0)}\}])$ (see, e.g., Eq.~(5) in Ref.~\cite{SG88} and note that the \textsl{minus} sign separating the products of the Fermi functions must be \textsl{plus}) one readily infers the following exact identity, specific to the case where within the framework of the Hartree-Fock approximation the $N$-particle uniform GS of the Hubbard Hamiltonian is paramagnetic:
\begin{equation}
\Upsigma_{\sigma}^{(2)}(\bm{k};\varepsilon;[\{G_{\sigma'}^{\textsc{hf}}\}]) \equiv \Upsigma_{\sigma}^{(2)}(\bm{k};\varepsilon - U n/2;[\{G_{\sigma'}^{(0)}\}]).
\label{e317}
\end{equation}
On identifying $\varepsilon$ with $\varepsilon_{\textsc{f}}^{\textsc{hf}} \equiv \varepsilon_{\textsc{f}}^{(0)} + U n/2$, Eq.~(\ref{e224}), the above identity reduces to the following less general but nonetheless important identity:
\begin{equation}
\Upsigma_{\sigma}^{(2)}(\bm{k};\varepsilon_{\textsc{f}}^{\textsc{hf}};[\{G_{\sigma'}^{\textsc{hf}}\}]) \equiv \Upsigma_{\sigma}^{(2)}(\bm{k};\varepsilon_{\textsc{f}}^{(0)};[\{G_{\sigma'}^{(0)}\}]).
\label{e318}
\end{equation}

The identity in Eq.~(\ref{e317}) reveals a fundamental shortcoming of the perturbation expansion of the self-energy in terms of the Green functions $\{G_{\sigma}^{\textsc{mf}}\| \sigma\}$ corresponding to a mean-field $N$-particle metallic GS whose relevant Fermi energy $\varepsilon_{\textsc{f}}^{\textsc{mf}}$ does not coincide with the exact Fermi energy $\varepsilon_{\textsc{f}}$ (more about this later). The problem is similar to that arising from the Fermi surface $\mathcal{S}_{\textsc{f};\sigma}^{\textsc{mf}}$ associated with $G_{\sigma}^{\textsc{mf}}$ being deformed with respect to the exact Fermi surface $\mathcal{S}_{\textsc{f};\sigma}$, $\forall\sigma$ \cite[\S5.7]{PN64}, to which we have referred earlier in this section.

The problem thus uncovered is not entirely unexpected, given the way in which contributions of self-energy diagrams are determined \cite[pp. 100-105]{FW03}: with $\mu^{\textsc{mf}}$ denoting the chemical potential associated with the underlying $N$-particle mean-field GS (for metallic GSs, $\mu^{\textsc{mf}}$ coincides with $\varepsilon_{\textsc{f}}^{\textsc{mf}}$ up to a correction of the order of $1/N$), the integrals with respect to the internal energy variables of self-energy diagrams are evaluated by employing the following spectral representation for $G_{\sigma}^{\textsc{mf}}(\bm{k};\varepsilon)$ \cite[Eq.~(7.45)]{FW03}:
\begin{equation}
G_{\sigma}^{\textsc{mf}}(\bm{k};\varepsilon) = \hbar\Big\{\frac{\Theta(\mu^{\textsc{mf}} - \varepsilon_{\bm{k}}^{\textsc{mf}})}{\varepsilon -\varepsilon_{\bm{k}}^{\textsc{mf}} - i 0^+} + \frac{\Theta(\varepsilon_{\bm{k}}^{\textsc{mf}} -\mu^{\textsc{mf}})}{\varepsilon - \varepsilon_{\bm{k}}^{\textsc{mf}} + i 0^+}\Big\}.
\label{e319}
\end{equation}
Clearly, unless $\varepsilon_{\textsc{f}}^{\textsc{mf}} = \varepsilon_{\textsc{f}}$, in particular the analytic properties of $\t{\Sigma}_{\sigma}(\bm{k};z;[\{G_{\sigma'}\}])$ in the neighbourhood of $z = \varepsilon_{\textsc{f}}$ \textsl{cannot} be correctly reproduced by $\t{\Upsigma}_{\sigma}(\bm{k};z;[\{G_{\sigma'}^{\textsc{mf}}\}])$. Rather, as the identity in Eq.~(\ref{e317}) also suggests, at best the analytic properties of $\t{\Upsigma}_{\sigma}(\bm{k};z;[\{G_{\sigma'}^{\textsc{mf}}\}])$ in a neighbourhood of $z = \varepsilon_{\textsc{f}}^{\textsc{mf}}$ are similar, but not necessarily identical, to those of $\t{\Sigma}_{\sigma}(\bm{k};z;[\{G_{\sigma'}\}])$ in a neighbourhood of $z = \varepsilon_{\textsc{f}}$.

With reference to the last remark in the previous paragraph, we note that one can readily demonstrate that
\begin{equation}
\im[\Upsigma_{\sigma}^{(2)}(\bm{k};\varepsilon_{\textsc{f}}^{\textsc{mf}};[\{G_{\sigma'}^{\textsc{mf}}\}])] \equiv 0,
\label{e320}
\end{equation}
which is to be contrasted with the exact property \cite{JML61}
\begin{equation}
\im[\Sigma_{\sigma}^{(\nu)}(\bm{k}; \varepsilon_{\textsc{f}};[\{G_{\sigma'}\}])] \equiv 0,\;\; \forall \nu \in \mathds{N}.
\label{e321}
\end{equation}
Given that for metallic GSs $\im[\Upsigma_{\sigma}^{(2)}(\bm{k};\varepsilon;[\{G_{\sigma'}^{\textsc{mf}}\}])] \not\equiv 0$ for $\bm{k}$ in a finite neighbourhood of $\mathcal{S}_{\textsc{f};\sigma}^{\textsc{mf}}$ and $\varepsilon$ in a finite neighbourhood of $\varepsilon =  \varepsilon_{\textsc{f}}^{\textsc{mf}}$ (for $d >1$, this function to leading order scales like $(\varepsilon - \varepsilon_{\textsc{f}}^{\textsc{mf}})^2$, which is a characteristic of Fermi-liquid metallic states) \cite{BF99}, unless $\varepsilon_{\textsc{f}} = \varepsilon_{\textsc{f}}^{\textsc{mf}}$, for any $U >0$ one must have
\begin{equation}
\im[\Upsigma_{\sigma}(\bm{k};\varepsilon_{\textsc{f}};[\{G_{\sigma'}^{\textsc{mf}}\}])] \not\equiv 0,
\label{e322}
\end{equation}
contradicting the exact identity \cite[Eq.~(2.9)]{BF07a}
\begin{equation}
\im[\Sigma_{\sigma}(\bm{k};\varepsilon_{\textsc{f}};[\{G_{\sigma'}\}])] \equiv 0.
\label{e323}
\end{equation}
For clarity, the result in Eq.~(\ref{e322}) follows from the fact that $\im[\Upsigma_{\sigma}^{(\nu)}(\bm{k};\varepsilon;[\{G_{\sigma'}^{\textsc{mf}}\}])] \gtrless 0$ for $\varepsilon \lessgtr \varepsilon_{\textsc{f}}^{\textsc{mf}}$, $\forall \nu \ge 2$ \cite[\S5.3.5]{BF07a}, whereby, unless $\varepsilon_{\textsc{f}} = \varepsilon_{\textsc{f}}^{\textsc{mf}}$, the above-mentioned non-vanishing value of $\im[\Upsigma_{\sigma}^{(2)}(\bm{k};\varepsilon_{\textsc{f}};[\{G_{\sigma'}^{\textsc{mf}}\}])]$ cannot be cancelled by higher-order terms in the perturbation expansion of the self-energy.

We should emphasize that since
\begin{equation}
\Sigma_{\sigma}[\{G_{\sigma'}\}] \equiv (G_{\sigma}^{(0)})^{-1} - G_{\sigma}^{-1}\;\,
\text{(Dyson's `equation')},
\label{e324}
\end{equation}
existence of $\Sigma_{\sigma}[\{G_{\sigma'}\}]$, for a given $G_{\sigma}$, is not in dispute \cite{Note3}. This can however \textsl{not} be said about $\Upsigma_{\sigma}[\{G_{\sigma'}^{\textsc{mf}}\}]$. The result in Eq.~(\ref{e322}) is thus formal, in that for an arbitrary mean-field GS the sum $\sum_{\nu = 1}^{\infty} \Upsigma_{\sigma}^{(\nu)}[\{G_{\sigma'}^{\textsc{mf}}\}]$ may not exist.

Neglecting the above-mentioned problem of non-existence, \emph{we have rigorously demonstrated that unless $\varepsilon_{\textsc{f}}^{\textsc{mf}} = \varepsilon_{\textsc{f}}$, $\Upsigma_{\sigma}(\bm{k};\varepsilon;[\{G_{\sigma'}^{\textsc{mf}}\}])$ cannot identically coincide with $\Sigma_{\sigma}(\bm{k};\varepsilon;[\{G_{\sigma'}\}])$, contradicting the equivalence relationship in Eq.~(\ref{e315}).} Stated differently, a \textsl{necessary} condition for the validity of the perturbation expansion for the self-energy in terms of $\{G_{\sigma}^{\textsc{mf}}\| \sigma\}$ is the equality $\varepsilon_{\textsc{f}}^{\textsc{mf}} = \varepsilon_{\textsc{f}}$. With reference to the result in Eq.~(\ref{e229}), the latter observation sheds additional light on the significance of the (exact) Hartree-Fock theory to the many-body perturbation theory as applied to metallic GSs.

% 3.a.3.
\subsubsection{On the Luttinger theorem}
\label{s3a3}
In view of the fact that in arriving at the expression in Eq.~(\ref{e314}) we have made use of the Luttinger theorem \cite{LW60,JML60,ID03,BF07a,BF07-12}, it is important to realize that this theorem does \textsl{not} apply within the framework in which $\Sigma_{\sigma}'[\{G_{\sigma'}\}]$ is approximated by $\Upsigma_{\sigma}^{(2)}[\{G_{\sigma'}^{\textsc{mf}}\}]$, this as a consequence of the failure of the Luttinger-Ward identity \cite{LW60} within this framework. To be explicit, for the mean number of particles with spin index $\sigma$ in the grand canonical ensemble of the Fock space of $\wh{\mathcal{H}}$ corresponding to the chemical potential $\mu$, that is $\b{N}_{\sigma}$, one has \cite[Eq.~(4.1)]{BF07a}
\begin{equation}
\b{N}_{\sigma} = \sum_{\bm{k}} \big( \b{\upnu}_{\sigma}'(\bm{k}) + \b{\upnu}_{\sigma}''(\bm{k})\big),
\label{e325}
\end{equation}
where, with $\beta \doteq 1/(k_{\textsc{b}}T)$, \cite[Eq.~(4.29)]{BF07a}
\begin{equation}
\lim_{\beta \to\infty} \sum_{\bm{k}} \b{\upnu}_{\sigma}'(\bm{k}) = N_{\textsc{l};\sigma},
\label{e326}
\end{equation}
the Luttinger number, Eq.~(\ref{e275}), and \cite[Eq.~(4.10)]{BF07a}
\begin{equation}
\lim_{\beta\to\infty} \sum_{\bm{k}} \b{\upnu}_{\sigma}''(\bm{k}) = \sum_{\bm{k}} \int_{\mathscr{C}(\mu)} \frac{\rd z}{2\pi i}\, \t{G}_{\sigma}(\bm{k};z) \frac{\partial}{\partial z} \t{\Sigma}_{\sigma}(\bm{k};z),
\label{e327}
\end{equation}
in which $\mathscr{C}(\mu)$ is a closed contour in the complex $z$ plane, crossing the real axis at $z = \mu$ and parameterizable as follows:
\begin{equation}
\mathscr{C}(\mu) = \big\{ z \| z = \mu + i y, y\uparrow_{-\infty}^{+\infty}\big\}.
\label{e328}
\end{equation}
The equality
\begin{equation}
\lim_{\beta\to\infty} \sum_{\bm{k}} \b{\upnu}_{\sigma}''(\bm{k}) = 0
\label{e329}
\end{equation}
is the above-mentioned Luttinger-Ward identity \cite{LW60}\cite[Eq.~(4.11)]{BF07a}. We note in passing that for reasons indicated in Refs.~\cite{BF07a} and \cite[\emph{a}]{BF07-12}, for \textsl{insulating} GSs the Luttinger-Ward identity may fail when $\mu$ is identified with a value in the single-particle excitation gap $(\mu_{N}^-,\mu_{N}^+)$ different from the zero-temperature limit of $\mu_{\beta}$, the chemical potential corresponding to $\b{N} = N$, where $\b{N} = \sum_{\sigma} \b{N}_{\sigma}$.

The Luttinger-Ward identity applies order-by-order as follows \cite[Eq.~(5.23)]{BF07a}:
\begin{equation}
\sum_{\bm{k}} \int_{\mathscr{C}(\mu)} \frac{\rd z}{2\pi i}\, \t{G}_{\sigma}(\bm{k};z) \frac{\partial}{\partial z} \t{\Sigma}_{\sigma}^{(\nu)}(\bm{k};z;[\{G_{\sigma'}\}]) = 0,\; \forall \nu.
\label{e330}
\end{equation}
The function $\t{\Sigma}_{\sigma}^{(\nu)}(\bm{k};z;[\{G_{\sigma'}\}])$, the total contribution of all $\nu$\hspace{0.6pt}th-order \textsl{skeleton} self-energy diagrams, being dependent on $z$ only for $\nu \ge 2$, the equality in Eq.~(\ref{e330}) trivially applies for $\nu = 1$. For $\nu \ge 2$, the validity of the equality in Eq.~(\ref{e330}) vitally depends on $\t{\Sigma}_{\sigma}^{(\nu)}$  being evaluated in terms of the exact single-particle Green functions $\{G_{\sigma}\| \sigma\}$. More generally, and importantly from the perspective of approximate but self-consistent calculations, the equality in Eq.~(\ref{e330}) remains on replacing the self-energy by one evaluated in terms of an in principle arbitrary set of single-particle Green functions $\{G_{\sigma}'\| \sigma\}$, provided that the \textsl{explicit} Green function on the LHS of Eq.~(\ref{e330}) be replaced by $G_{\sigma}'$ (cf. Eq.~(\ref{e336})). Thus, whereas \cite[Eqs.~(5.29), (5.30), (B.103)]{BF07a}
\begin{equation}
\sum_{\bm{k}} \int_{\mathscr{C}(\mu^{\textsc{mf}})} \frac{\rd z}{2\pi i}\, \t{G}_{\sigma}^{\textsc{mf}}(\bm{k};z) \frac{\partial}{\partial z} \t{\Sigma}_{\sigma}^{(\nu)}(\bm{k};z;[\{G_{\sigma'}^{\textsc{mf}}\}]) = 0,\; \forall \nu,
\label{e331}
\end{equation}
the equality fails on replacing the \textsl{explicit} Green function $\t{G}_{\sigma}^{\textsc{mf}}(\bm{k};z)$ on the LHS by a different Green function. Restricting oneself to the case of $\nu =2$, formally (see later) this different Green function may be one of the following two important single-particle Green functions, both of which we denote by $\t{G}_{\sigma}'$ for the economy of notation:
\begin{equation}
\t{G}_{\sigma}' \doteq \big(1 - \t{G}_{\sigma}^{\textsc{hf}} \t{\Upsigma}_{\sigma}^{(2)}[\{G_{\sigma'}^{\textsc{hf}}\}]\big)^{-1} \t{G}_{\sigma}^{\textsc{hf}},
\label{e332}
\end{equation}
\begin{equation}
\t{G}_{\sigma}' \doteq \big(1 - \t{G}_{\sigma}^{\textsc{hf}\prime} \t{\Sigma}_{\sigma}^{(2)}[\{G_{\sigma'}'\}]\big)^{-1} \t{G}_{\sigma}^{\textsc{hf}\prime},
\label{e333}
\end{equation}
where $\t{G}_{\sigma}^{\textsc{hf}}$ self-consistently corresponds to the mean-field energy dispersion $\varepsilon_{\bm{k}} + \hbar \Upsigma_{\sigma}^{\textsc{hf}}(\bm{k};[\{G_{\sigma'}^{\textsc{hf}}\}])$, Eq.~(\ref{e339}), and $\t{G}_{\sigma}^{\textsc{hf}\prime}$ self-consistently corresponds to the mean-field energy dispersion $\varepsilon_{\bm{k}} + \hbar \Sigma_{\sigma}^{\textsc{hf}}(\bm{k};[\{G_{\sigma'}'\}])$. For completeness, since $G_{\sigma}^{\textsc{hf}}$ ($G_{\sigma}^{\textsc{hf}\prime}$) takes account of the Hartree-Fock self-energy $\Upsigma_{\sigma}^{\textsc{hf}}[\{G_{\sigma'}^{\textsc{hf}}\}]$ ($\Sigma_{\sigma}^{\textsc{hf}}[\{G_{\sigma'}'\}]$), indeed not the full self-energy up to second-order, but only $\t{\Upsigma}_{\sigma}^{(2)}[\{G_{\sigma'}^{\textsc{hf}}\}]$ ($\t{\Sigma}_{\sigma}^{(2)}[\{G_{\sigma'}'\}]$) is to be encountered in the Dyson equations from which the expressions for $G_{\sigma}'$ in Eqs.~(\ref{e332}) and (\ref{e333}) are deduced.

Since the first-order proper self-energy diagram is also skeleton, the functionals $\Upsigma_{\sigma}^{\textsc{hf}}[\{X_{\sigma'}\}]$ and $\Sigma_{\sigma}^{\textsc{hf}}[\{X_{\sigma'}\}]$ identically coincide. However, since $G_{\sigma}^{\textsc{hf}}$ and $G_{\sigma}'$ are distinct, it follows that the chemical potentials $\mu^{\textsc{hf}}$ and $\mu^{\textsc{hf}\prime}$ associated with the mean-field $N$-particle uniform GSs to which respectively $\t{G}_{\sigma}^{\textsc{hf}}$ and $\t{G}_{\sigma}^{\textsc{hf}\prime}$ correspond, \textsl{cannot} be equal. Because of this fact, replacing the explicit $\t{G}_{\sigma}^{\textsc{hf}}$ on the LHS of Eq.~(\ref{e331}) by the $G_{\sigma}'$ introduced in Eq.~(\ref{e333}) is mathematically problematical (note the $\mu^{\textsc{hf}}$ as the argument of the contour $\mathscr{C}$ on the LHS of Eq.~(\ref{e331}) and consider the relationship in Eq.~(\ref{e320}), where $\varepsilon_{\textsc{f}}^{\textsc{hf}}$ is up to a deviation of the order of $1/N$ equal to $\mu^{\textsc{hf}}$). This problem disappears however by simultaneously changing the $[\{G_{\sigma'}^{\textsc{hf}}\}]$ and $\mu^{\textsc{hf}}$ by respectively $[\{G_{\sigma'}'\}]$ and $\mu^{\textsc{hf}\prime}$. \emph{Unless we indicate otherwise, below the function $G_{\sigma}'$ refers to that defined in Eq.~(\ref{e332}).}

It is interesting to note that for the exact Green function (cf. Eq.~(\ref{e333}))
\begin{equation}
\t{G}_{\sigma} = \big(1 - \t{G}_{\sigma}^{\textsc{hf}} \t{\Sigma}_{\sigma}'[\{G_{\sigma'}\}]\big)^{-1} \t{G}_{\sigma}^{\textsc{hf}},
\label{e334}
\end{equation}
where $\t{\Sigma}_{\sigma}'[\{G_{\sigma'}\}]$ is defined in Eq.~(\ref{e316}), and $\t{G}_{\sigma}^{\textsc{hf}}$ self-consistently corresponds to the mean-field single-particle energy dispersion $\varepsilon_{\bm{k}} + \hbar \Sigma_{\sigma}^{\textsc{hf}}(\bm{k};[\{G_{\sigma'}\}])$, the above-indicated problem arising from the deviation of two chemical potentials (one, i.e. $\mu$, pertaining to the exact $N$-particle GS to which $G_{\sigma}$ corresponds, and one, i.e. $\mu^{\textsc{hf}}$, pertaining to the Hartree-Fock theory) does \textsl{not} arise, this on account of the exact equality in Eq.~(\ref{e229}) (see Eq.~(\ref{e27})).

For later reference, one has
\begin{equation}
\sum_{\bm{k}} \int_{\mathscr{C}(\mu^{\textsc{hf}})} \frac{\rd z}{2\pi i}\, \t{G}_{\sigma}'(\bm{k};z) \frac{\partial}{\partial z} \t{\Sigma}_{\sigma}^{(2)}(\bm{k};z;[\{G_{\sigma'}^{\textsc{hf}}\}]) \not= 0,
\label{e335}
\end{equation}
while (cf. Eqs.~(\ref{e330}) and (\ref{e331}))
\begin{equation}
\sum_{\bm{k}} \int_{\mathscr{C}(\mu^{\textsc{hf}})} \frac{\rd z}{2\pi i}\, \t{G}_{\sigma}'(\bm{k};z) \frac{\partial}{\partial z} \t{\Sigma}_{\sigma}^{(2)}(\bm{k};z;[\{G_{\sigma'}'\}]) = 0.
\label{e336}
\end{equation}
The explicit and implicit function $G_{\sigma}'$ in Eq.~(\ref{e336}) is the one defined in Eq.~(\ref{e332}). The equality in Eq.~(\ref{e336}) remains on identifying the explicit and implicit function $G_{\sigma}'$ herein by that defined in Eq.~(\ref{e333}), provided that the $\mu^{\textsc{hf}}$ in the $\mathscr{C}(\mu^{\textsc{hf}})$ on the LHS be simultaneously replaced by $\mu^{\textsc{hf}\prime}$ (see the remark in the paragraph following that containing Eq.~(\ref{e333})).

The difference $\t{\Sigma}_{\sigma}^{(2)}(\bm{k};z;[\{G_{\sigma'}^{\textsc{hf}}\}]) -\t{\Upsigma}_{\sigma}^{(2)}(\bm{k};z;[\{G_{\sigma'}^{\textsc{hf}}\}])$ being independent of $z$ (as well as of $\bm{k}$), the expression in Eq.~(\ref{e335}) can be equivalently written as
\begin{equation}
\sum_{\bm{k}} \int_{\mathscr{C}(\mu^{\textsc{hf}})} \frac{\rd z}{2\pi i}\, \t{G}_{\sigma}'(\bm{k};z) \frac{\partial}{\partial z} \t{\Upsigma}_{\sigma}^{(2)}(\bm{k};z;[\{G_{\sigma'}^{\textsc{hf}}\}]) \not= 0,
\label{e337}
\end{equation}
to be contrasted with (cf. Eq.~(\ref{e331}))
\begin{equation}
\sum_{\bm{k}} \int_{\mathscr{C}(\mu^{\textsc{hf}})} \frac{\rd z}{2\pi i}\, \t{G}_{\sigma}^{\textsc{hf}}(\bm{k};z) \frac{\partial}{\partial z} \t{\Upsigma}_{\sigma}^{(2)}(\bm{k};z;[\{G_{\sigma'}^{\textsc{hf}}\}]) = 0.
\label{e338}
\end{equation}

Since for the $N$-particle uniform metallic GS of $\wh{\mathcal{H}}$ the function (cf. Eqs.~(\ref{e319}) and (\ref{ea8}))
\begin{equation}
\t{G}_{\sigma}^{\textsc{hf}}(\bm{k};z) = \frac{\hbar}{z - [\varepsilon_{\bm{k}} + \hbar\Upsigma_{\sigma}^{\textsc{hf}}(\bm{k};[\{G_{\sigma'}^{\textsc{hf}}\}])]},
\label{e339}
\end{equation}
is divergent at $z = \mu^{\textsc{hf}}$ and $\bm{k} \in \mathcal{S}_{\textsc{f};\sigma}^{\textsc{hf}}$ (see Eq.~(\ref{e226}) and recall that, as we have pointed out earlier, in the present section $\mathcal{S}_{\textsc{f};\sigma}^{\textsc{hf}}$ is defined in terms of $\varepsilon_{\textsc{f}}^{\textsc{hf}}$), by continuity for $z$ sufficiently close to $\mu^{\textsc{hf}}$ and $\bm{k}$ sufficiently close to $\mathcal{S}_{\textsc{f};\sigma}^{\textsc{hf}}$ the function $G_{\sigma}'(\bm{k};z)$ \textsl{cannot} to leading order in $U$ be approximated by $G_{\sigma}^{\textsc{hf}}(\bm{k};z)$, as a formal geometric series expansion of the expression on the RHS of Eq.~(\ref{e332}) would suggest. One trivially verifies that for $\Upsigma_{\sigma}^{(2)}(\bm{k};\mu^{\textsc{hf}};[\{G_{\sigma'}^{\textsc{hf}}\}]) \not=0$ and $\bm{k}$ sufficiently close to $\mathcal{S}_{\textsc{f};\sigma}^{\textsc{hf}}$,
\begin{equation}
\t{G}_{\sigma}'(\bm{k};z) \sim -1/\t{\Upsigma}_{\sigma}^{(2)}(\bm{k};z;[\{G_{\sigma'}^{\textsc{hf}}\}])\;\, \text{as}\;\, z \to \mu^{\textsc{hf}}.
\label{e340}
\end{equation}
The function on the RHS of this expression is the leading-order term in the geometric series expansion of $\t{G}_{\sigma}'$ in powers of $(\t{G}_{\sigma}^{\textsc{hf}} \t{\Upsigma}_{\sigma}^{(2)}[\{G_{\sigma'}^{\textsc{hf}}\}])^{-1}$, which is explicitly proportional to $1/U^{2}$. One observes that for $\bm{k}$ and $z$ sufficiently close to respectively $\mathcal{S}_{\textsc{f};\sigma}^{\textsc{hf}}$ and $\mu^{\textsc{hf}}$, the integrand of the integral on the LHS of the expression in Eq.~(\ref{e337}) is to leading order \textsl{independent} of $U$; this integrand is proportional to the logarithmic derivative of $\t{\Upsigma}_{\sigma}^{(2)}(\bm{k};z;[\{G_{\sigma'}^{\textsc{hf}}\}])$ with respect to $z$. Were it not for this fact, on account of the exact equality in Eq.~(\ref{e338}) the deviation of the LHS of the expression in Eq.~(\ref{e337}) from zero would be of the order of $U^4$ for $U/t \to 0$, appendix \ref{sab}. With reference to Eq.~(\ref{e236}), we note that the above assumption $\Upsigma_{\sigma}^{(2)}(\bm{k};\mu^{\textsc{hf}};[\{G_{\sigma'}^{\textsc{hf}}\}]) \not=0$ is in conformity with the observations in Refs.~\cite{SG88,HM97,YY99,ZEG96}, in that in particular on the exact Fermi surface it can fail only at a finite number of $\bm{k}$ points (see the remarks following Eq.~(\ref{e313}) above).

It is significant here to realize that $z=\mu^{\textsc{hf}}$ being a branch point \cite[\S5.7]{WW62} of $\t{\Upsigma}_{\sigma}^{(2)}(\bm{k};z;[\{G_{\sigma'}^{\textsc{hf}}\}])$ that separates two branch cuts of this function on the real axis of the $z$ plane (for details see appendix \ref{sab}), the point at which $\mathscr{C}(\mu^{\textsc{hf}})$ passes through this axis, that is  $z = \mu^{\textsc{hf}}$, is immovable. Considering the exact case, for the $N$-particle uniform GS of $\wh{\mathcal{H}}$ the contour $\mathscr{C}(\mu)$, which is to cross the real axis of the $z$ plane at $z = \mu$, where $\mu \in (\mu_{N;\sigma}^-,\mu_{N;\sigma}^+)$, is `pinched' \cite[\S6.3.1]{IZ80} when the GS is metallic, Eqs.~(\ref{e26}) and (\ref{e27}). For $N$-particle GSs, the immovability of the crossing point of $\mathscr{C}(\mu^{\textsc{hf}})$ (or of $\mathscr{C}(\mu)$ in the exact case) with the real axis of the $z$ plane non-trivially affects the functional form of the function on the LHS of the expression in Eq.~(\ref{e335}), or equivalently Eq.~(\ref{e337}), in particular in the asymptotic region \cite{WW62,ETC65,HAL74} $U/t \to 0$. This aspect is directly related to the fact that for $z$ in a neighbourhood of $\mu^{\textsc{hf}}$, the 1BZ can be subdivided into two non-overlapping regions: a region in the neighbourhood of $\mathcal{S}_{\textsc{f};\sigma}^{\textsc{hf}}$ where $\t{G}_{\sigma}'$ can be expanded in powers of $(\t{G}_{\sigma}^{\textsc{hf}} \t{\Upsigma}_{\sigma}^{(2)}[\{G_{\sigma'}^{\textsc{hf}}\}])^{-1}$, and a region where $\t{G}_{\sigma}'$ can be expanded in powers of $\t{G}_{\sigma}^{\textsc{hf}} \t{\Upsigma}_{\sigma}^{(2)}[\{G_{\sigma'}^{\textsc{hf}}\}]$.

The above considerations make explicit that the expressions on the LHSs of Eqs.~(\ref{e335}) and (\ref{e337}) do \textsl{not} to leading order scale like $U^4$ for $U/t \to 0$, the scaling of the form $U^4$ following from the direct proportionality of respectively $\t{\Sigma}_{\sigma}^{(2)}[\{G_{\sigma'}^{\textsc{hf}}\}]$ and $\t{\Upsigma}_{\sigma}^{(2)}[\{G_{\sigma'}^{\textsc{hf}}\}]$ with $U^2$, combined with the erroneous assumption that for $U/t \to 0$ the leading-order asymptotic contribution to $G_{\sigma}' - \t{G}_{\sigma}^{\textsc{hf}}$ were uniformly (i.e. independently of $\bm{k}$ and $z$) of the form $\t{G}_{\sigma}^{\textsc{hf}} \t{\Upsigma}_{\sigma}^{(2)}[\{G_{\sigma'}^{\textsc{hf}}\}] \t{G}_{\sigma}^{\textsc{hf}}$.

The considerations in appendix \ref{sab} reveal that the expressions on the LHSs of Eqs.~(\ref{e335}) and (\ref{e337}) diminish at the fastest like $U^2$ for $U/t \to 0$. Below we rigorously demonstrate that for at least $d \le 2$, these expressions are in the asymptotic region $U/t \to 0$ more dominant than $U^2$, scaling in principle like $U^{\alpha} \ln^{\gamma}(\vert t\vert/U)$, where $1 \le \alpha \le 2$ and $\gamma \ge 0$, with the possibilities $\alpha = 1$, $\gamma >0$, and $\alpha = 2$, $\gamma = 0$ ruled out, appendix \ref{sac}. \emph{For the reason that we indicate in Sec.~\ref{s3a4}, p.~\pageref{SinceInD}, these observations are almost certainly applicable for arbitrary values of $d$.}

% 3.a.4.
\subsubsection{The functional dependence of \texorpdfstring{$\Sigma_{\sigma}'(\bm{k};\varepsilon)$}{} on \texorpdfstring{$U$}{} for \texorpdfstring{$\bm{k} \to \bm{k}_{\SC{\textsc{f}};\sigma} \in \mathcal{S}_{\SC{\textsc{f}};\sigma}$}{} and \texorpdfstring{$\varepsilon \to \varepsilon_{\SC{\textsc{f}}}$}{}}
\label{s3a4}
Using the standard expression for the Landau quasi-particle weight $Z_{\sigma}(\bm{k})$ \cite{NO98,BF99,BF03} and the fact that the first-order self-energy $\Sigma_{\sigma}^{(1)}(\bm{k};\varepsilon) \equiv \Sigma_{\sigma}^{\textsc{hf}}(\bm{k})$ is independent of $\varepsilon$, one has (see Eq.~(\ref{e316}))
\begin{equation}
\left.\frac{\partial \Sigma_{\sigma}'(\bm{k};\varepsilon)}{\partial\varepsilon}\right|_{\varepsilon=\varepsilon_{\textsc{f}}} = \frac{1}{\hbar}\Big(1 - \frac{1}{Z_{\sigma}(\bm{k})}\Big).
\label{e341}
\end{equation}
With $Z_{\bm{k}_{\textsc{f};\sigma}} \equiv Z_{\sigma}(\bm{k}_{\textsc{f};\sigma})$, making use of the Migdal theorem, Eq.~(\ref{e255}), from the equality in Eq.~(\ref{e341}) for the specific case of $\bm{k} = \bm{k}_{\textsc{f};\sigma} \in \mathcal{S}_{\textsc{f};\sigma}$ one deduces the following general leading-order asymptotic expression:
\begin{eqnarray}
&&\hspace{-0.2cm} \left. \frac{\partial \Sigma_{\sigma}'(\bm{k};\varepsilon)}{\partial\varepsilon}\right|_{\varepsilon=\varepsilon_{\textsc{f}}} \hspace{-0.0cm} \sim -\frac{1}{\hbar} \big(\upalpha_{\sigma}^-(\bm{k}_{\textsc{f};\sigma}) + \upalpha_{\sigma}^+(\bm{k}_{\textsc{f};\sigma})\big) \Big(\frac{U}{\vert t\vert}\Big)^{\alpha}\nonumber\\
&&\hspace{3.1cm} \times \ln^{\gamma}\hspace{-0.1cm}\Big(\frac{\vert t\vert}{U}\Big)\;\;\text{for}\;\; \frac{U}{t} \to 0,
\label{e342}
\end{eqnarray}
where the \textsl{positive} constants $\upalpha_{\sigma}^{-}(\bm{k}_{\textsc{f};\sigma})$ and $\upalpha_{\sigma}^{+}(\bm{k}_{\textsc{f};\sigma})$ are the coefficients in the following asymptotic expressions corresponding to $U/t \to 0$:
\begin{eqnarray}
&&\mathsf{n}_{\sigma}(\bm{k}_{\textsc{f};\sigma}^-) \sim 1 - \upalpha_{\sigma}^-(\bm{k}_{\textsc{f};\sigma})  \Big(\frac{U}{\vert t\vert}\Big)^{\alpha} \ln^{\gamma}\hspace{-0.1cm}\Big(\frac{\vert t\vert}{U}\Big),
\nonumber\\
&&\mathsf{n}_{\sigma}(\bm{k}_{\textsc{f};\sigma}^+) \sim \upalpha_{\sigma}^+(\bm{k}_{\textsc{f};\sigma}) \Big(\frac{U}{\vert t\vert}\Big)^{\alpha} \ln^{\gamma}\hspace{-0.1cm}\Big(\frac{\vert t\vert}{U}\Big).
\label{e343}
\end{eqnarray}
In general, $\alpha$ and $\gamma$ are subject to one of the following three conditions: (i) $\alpha = 1$, $\gamma = 0$, (ii) $1 < \alpha < 2$, $\gamma \ge 0$, and (iii) $\alpha = 2$, $\gamma > 0$. See appendix \ref{sac}.

In the most general case, and away from half-filling, the constants $\alpha$ and $\gamma$ corresponding to $\bm{k} = \bm{k}_{\textsc{f};\sigma}^-$ and $\bm{k} = \bm{k}_{\textsc{f};\sigma}^+$ may be different, to be thus appropriately denoted by respectively $\alpha^-$, $\gamma^-$ and $\alpha^+$, $\gamma^+$. We have sacrificed this generality for the conciseness of notation. At half-filling however, $\mathsf{n}_{\sigma}(\bm{k}_{\textsc{f};\sigma}^-)$ and $\mathsf{n}_{\sigma}(\bm{k}_{\textsc{f};\sigma}^+)$ take values that are symmetric with respect to $1/2$ \cite{GST94}, whereby $\upalpha_{\sigma}^-(\bm{k}_{\textsc{f};\sigma}) = \upalpha_{\sigma}^+(\bm{k}_{\textsc{f};\sigma})$, $\alpha^- = \alpha^+$ and $\gamma^- = \gamma^+$.

To leading order in $U$ the function $\Sigma_{\sigma}'(\bm{k};\varepsilon)$ on the LHS of Eq.~(\ref{e342}) can be replaced by $\Sigma_{\sigma}^{(2)}(\bm{k};\varepsilon)$. Substituting the latter function by $\Upsigma_{\sigma}^{(2)}(\bm{k};\varepsilon;[\{G_{\sigma'}^{\textsc{mf}}\}])$, it trivially follows that for the exponents $\alpha$ and $\gamma$ one has $\alpha = 2$ and $\gamma = 0$. These values clearly equally apply to $\Upsigma_{\sigma}'(\bm{k};\varepsilon;[\{G_{\sigma'}^{\textsc{mf}}\}])$. The question arises as to whether these values of $\alpha$ and $\gamma$ also apply for the functions $\Sigma_{\sigma}^{(2)}(\bm{k};\varepsilon;[\{G_{\sigma'}\}])$ and $\Sigma_{\sigma}'(\bm{k};\varepsilon;[\{G_{\sigma'}\}])$. Below we show that \emph{the answer to this question is in the negative.} For now, we point out that the failure of $\Upsigma_{\sigma}^{(2)}(\bm{k};\varepsilon;[\{G_{\sigma'}^{\textsc{mf}}\}])$, as opposed to $\Sigma_{\sigma}^{(2)}(\bm{k};\varepsilon;[\{G_{\sigma'}\}])$, to reproduce the correct values for the exponents $\alpha$ and $\gamma$, is directly related to the violation of the Luttinger theorem \cite{LW60,JML60,ID03,BF07a,BF07-12} by the single-particle Green function corresponding to the former self-energy (see the remarks following Eq.~(\ref{e316}) and in the subsequent paragraph, p.~\pageref{ToBeExplicit}). To appreciate this fact more clearly, one should realize that the single-particle Green function $G_{\sigma}^{(0)}$ underlying the perturbational calculation of the function $\mathsf{n}_{\sigma}^{(2)}(\bm{k})$, to be introduced below, is consistent from the perspective of the Luttinger theorem.

The leading-order perturbational correction to the non-interacting momentum-distribution function $\mathsf{n}_{\sigma}^{(0)}(\bm{k})$, which is the characteristic function of the non-interacting Fermi sea, is of the order of $U^2$, to be thus denoted by $\mathsf{n}_{\sigma}^{(2)}(\bm{k})$, however this correction is logarithmically divergent in $d \le 2$ for $\bm{k}$ approaching $\mathcal{S}_{\textsc{f};\sigma}^{(0)}$ \cite{Note4}, as has been shown in Ref.~\cite{OS90} for $d=1$ (see appendix B in Ref.~\cite{OS90}, in particular Eq.~(B3)), and in Ref.~\cite{BvdL91} for $d=2$ (see in particular Eq.~(30) in Ref.~\cite{BvdL91} and note that herein $\Delta k \doteq \| \bm{k} - \bm{k}_{\textsc{f};\sigma}\|$, so that $\Delta k \to 0$ corresponds to the right-most parts of Figs.~4-7 in Ref.~\cite{BvdL91}; for an additional detail, see Ref.~\cite{Note5}). If this were not the case, indeed for the $\alpha$ and $\gamma$ in the expressions in Eqs.~(\ref{e342}) and (\ref{e343}) one had $\alpha = 2$ and $\gamma = 0$, appendix \ref{sac}. We note that because of the strict inequalities $0\le \mathsf{n}_{\sigma}(\bm{k}) \le 1$, \textsl{divergence} of $\mathsf{n}_{\sigma}(\bm{k})$ at any $\bm{k}$ signals a fundamental inadequacy of the formalism on the basis of which $\mathsf{n}_{\sigma}(\bm{k})$ has been calculated. It is surprising that in Refs.~\cite{OS90,BvdL91} the \textsl{divergence} of the second-order contribution to $\mathsf{n}_{\sigma}(\bm{k})$ for $\bm{k}$ approaching the underlying Fermi surface has not been explicitly declared as pathological.

The function $\mathsf{n}_{\sigma}^{(2)}(\bm{k})$ being divergent for $d \le 2$ and $\bm{k}$ approaching the underlying Fermi surface $\mathcal{S}_{\textsc{f};\sigma}^{(0)}$, from the considerations in appendix \ref{sac} one immediately infers the asymptotic expressions in Eq.~(\ref{e343}), in which $1 \le \alpha \le 2$, $\gamma \ge 0$, excluding the possibilities of $\alpha=1$, $\gamma >0$, and $\alpha=2$, $\gamma=0$. These are clearly the values with which the exponents $\alpha$ and $\gamma$ in the asymptotic expression in Eq.~(\ref{e342}) are to be identified. It should be noted however that $\mathsf{n}_{\sigma}^{(2)}(\bm{k})$ being proportional to $U^2$, the neighbourhood of $\mathcal{S}_{\textsc{f};\sigma}^{(0)} \equiv \mathcal{S}_{\textsc{f};\sigma}^{\textsc{hf}}$ in which the logarithmic divergence of $\mathsf{n}_{\sigma}^{(2)}(\bm{k})$ becomes noticeable, is dependent on $U$; the smaller the value of $U$, the narrower the latter neighbourhood of $\mathcal{S}_{\textsc{f};\sigma}^{(0)}$. In other words, in the case at hand the processes of effecting the limits of $U/t \to 0$ and $\bm{k} \to \bm{k}_{\textsc{f};\sigma}^{(0)}$, where $\bm{k}_{\textsc{f};\sigma}^{(0)} \in \mathcal{S}_{\textsc{f};\sigma}^{(0)}$, do \textsl{not} commute.

In the light of the above discussions, it is relevant to note that \emph{the expressions in Eqs.~(\ref{e342}) and (\ref{e343}) are specific to the case of the limit of $\bm{k}$ approaching the underlying Fermi surface \textsl{prior to} $U/t$ approaching $0$.} This observation is relevant in that it shows that although the relationship in Eq.~(\ref{e342}) is specific to $\bm{k}$ points \textsl{on} $\mathcal{S}_{\textsc{f};\sigma}$, the functional form of this relationship remains applicable for $\bm{k}$ in a neighbourhood of $\mathcal{S}_{\textsc{f};\sigma}$ whose extent depends on the value of $U/t$. For $\bm{k}$ sufficiently far outside this $U$-dependent neighbourhood of $\mathcal{S}_{\textsc{f};\sigma}$, it is to be expected that one recovers the values $\alpha = 2$, $\gamma = 0$ (see the relevant remarks in Ref.~\cite{Note6}; see also Fig.~\ref{f4} in appendix \ref{sab}, where an interplay is clearly visible between the location of $\bm{k}$ and the value of $U$, represented by respectively $a$ and $u$, in establishing a specific asymptotic behaviour in the underlying function).

That the leading-order term in the asymptotic series expansion \cite{WW62,ETC65,HAL74} of $\Sigma_{\sigma}'(\bm{k};\varepsilon)$ (and similarly as regards $\Sigma_{\sigma}^{(2)}(\bm{k};\varepsilon;[\{G_{\sigma'}\}])$) for $U/t \to 0$ may not in general scale like $U^2$, may be surmised from the expression in Eq.~(46) in conjunction with the data in Fig.~2 of Ref.~\cite{VT97} (for the attractive Hubbard model, Eq.~(8) in conjunction with the data in Fig.~2b of Ref.~\cite{KAT01}). With reference to the data in the above-mentioned Fig.~2 (Fig.~2b), we should emphasize however that the equality $U_{\textrm{sp}} = g_{\uparrow\downarrow}(0)\hspace{0.6pt} U$, where $g_{\uparrow\downarrow}(0) \doteq \langle \h{n}_{\uparrow} \h{n}_{\downarrow}\rangle/(\langle \h{n}_{\uparrow}\rangle \langle \h{n}_{\downarrow}\rangle)$ is the normalized site double occupancy, is merely an \emph{ansatz}. What is significant from the perspective of the considerations of this section, is that the expression for the exact $\Sigma_{\sigma}'(\bm{k};\varepsilon)$ involves a product of the bare on-site energy $U$ and a vertex part, represented in Eq.~(46) of Ref.~\cite{VT97} by the spin-symmetric (or \textsl{charge}) interaction parameter $U_{\textrm{ch}}$ and the spin-antisymmetric (or \textsl{spin}) interaction parameter $U_{\textrm{sp}}$, both of which are distinct from $U$ and do not necessarily to leading order scale like $U$ for $U/t\to 0$.

The reader may also consider Figs.~7 and 8 of Ref.~\cite{VT97}. In the latter figure, one encounters also the prediction of the second-order perturbation theory for $\mathsf{n}_{\sigma}(\bm{k})$, showing \textsl{no} divergence for $\bm{k}$ approaching a point of the underlying Fermi `surface'. This is because the data in Fig.~8 of Ref.~\cite{VT97} correspond to finite lattices, of the sizes $6\times 6$ and $16\times 16$, in addition to a finite temperature. From the expressions in Eqs.~(B1) and (B2) of Ref.~\cite{OS90}, and those in Eq.~(29) of Ref.~\cite{BvdL91}, one clearly observes that the possible divergence of $\mathsf{n}_{\sigma}(\bm{k})$ for a given $\bm{k}$ is due to the energy differences in the denominators of the relevant expressions, corresponding to $\bm{k}$ inside and outside the underlying Fermi sea, becoming vanishing. Owing to the Fermi functions in the numerators of the expressions for $\mathsf{n}_{\sigma}^{(2)}(\bm{k})$, this is only possible when the relevant 1BZ consists of a dense set of points, that is in the thermodynamic limit (unless, for finite systems, a point is counted as being part of both the Fermi sea and its complement with respect to the underlying 1BZ, in which case the divergence of $\mathsf{n}_{\sigma}^{(2)}(\bm{k})$ arising from this point is algebraic, not logarithmic); only in this limit can the above-mentioned denominators become arbitrary small for $\bm{k}$ approaching a point of the underlying Fermi surface, resulting in the aforementioned logarithmic divergence of $\mathsf{n}_{\sigma}^{(2)}(\bm{k})$ in $d \le 2$.

\refstepcounter{dummy}
\label{SinceInD}
Since in $d$ dimensions the Fermi surface corresponding to an $N$-particle metallic GS is a $(d-1)$-dimensional subset of the underlying $d$-dimensional 1BZ, in view of the origin of the divergence of $\mathsf{n}_{\sigma}^{(2)}(\bm{k})$ in $d \le 2$ for $\bm{k}$ approaching the relevant Fermi surface, described above, \emph{this divergence must be a universal characteristic of the $\mathsf{n}_{\sigma}^{(2)}(\bm{k})$ corresponding to the $N$-particle uniform metallic GS of the Hubbard Hamiltonian for arbitrary finite $d$.}

In the light of the above observations, it is interesting to note that the Monte-Carlo calculations by Varney \emph{et al.} \cite{VLBCJS99} on the half-filled Hubbard model in two dimensions reveal that the functional forms in Eq.~(\ref{e343}) for $\mathsf{n}_{\sigma}(\bm{k}_{\textsc{f};\sigma}^-)$ and $\mathsf{n}_{\sigma}(\bm{k}_{\textsc{f};\sigma}^+)$, with $1\le \alpha \le 2$ (in fact, with $\alpha$ far closer to $1$ than $2$, if not $\alpha =1$) and $\gamma\ge 0$, are meaningful for at least $2 \le U/\vert t\vert \le 8$, or, equivalently, $W/4 \le U \le W$, where $W$ denotes the bandwidth in the system under consideration (note the almost linear scaling with $U/t$ of the values of the $\mathsf{n}_{\sigma}(\bm{k})$ in Figs.~1(a) and 1(b) of Ref.~\cite{VLBCJS99} at the $\bm{k}$ points along the $(0,0) - (\pi,\pi)$ direction of the 1BZ nearest to that at which $\mathsf{n}_{\sigma}(\bm{k}) = 1/2$). A similar behaviour is observed in the Monte-Carlo results for the $\mathsf{n}_{\sigma}(\bm{k})$ corresponding to the Hubbard Hamiltonian on a $32$-site ring \cite{LW68-03} away from half-filling and for $0 \le U/\vert t\vert \lesssim 7.5$, or, equivalently, $0 \le U \lesssim 1.9 W$ \cite{vdLMR90} (see Figs.~1 and 2 herein). We note in passing that the GS momentum-distribution functions depicted in Fig.~1 of Ref.~\cite{VLBCJS99} differ considerably from that corresponding to the strong-coupling limit of the Hubbard Hamiltonian. For the latter function, see the expression for $n_{\bm{k}}$ in Eq.~(5.45) of Ref.~\cite{PF03} and note that $n_{\bm{k}}$ is equal to $\mathsf{n}_{\sigma}(\bm{k}) + \mathsf{n}_{\b\sigma}(\bm{k})$, so that $n_{\bm{k}} = 2\hspace{0.4pt} \mathsf{n}_{\sigma}(\bm{k})$, $\forall\sigma$.

Having shown that $\partial\Sigma_{\sigma}'(\bm{k};\varepsilon)/\partial\varepsilon\vert_{\varepsilon = \varepsilon_{\textsc{f}}}$, with $\bm{k} \in \mathcal{S}_{\textsc{f};\sigma}$, does not to leading order scale like $U^2$ as $U/t \to 0$, by continuity we have established that for $\bm{k}$ in a neighbourhood of $\mathcal{S}_{\textsc{f};\sigma}$ and $z$ in a neighbourhood of $\varepsilon_{\textsc{f}}$, $\t{\Sigma}_{\sigma}'(\bm{k};z;[\{G_{\sigma'}\}])$ differs fundamentally from $\t{\Upsigma}_{\sigma}'(\bm{k};z;[\{G_{\sigma'}^{\textsc{mf}}\}])$ in its dependence on $U$. We point out that the conclusion arrived at here contradicts the formal identity in Eq.~(\ref{e315}), however conforms with the observation based on the results in Eqs.~(\ref{e322}) and (\ref{e323}).

% 3.a.5.
\subsubsection{The functional dependence of \texorpdfstring{$\Sigma_{\sigma}'(\bm{k};\varepsilon)$}{} on \texorpdfstring{$U$}{} revisited}
\label{s3a5}
The aim of this section is to uncover the mathematical mechanism to which the deviation of $\t{\Sigma}_{\sigma}'(\bm{k};z;[\{G_{\sigma'}\}])$ from $\t{\Upsigma}_{\sigma}'(\bm{k};z;[\{G_{\sigma'}^{\textsc{mf}}\}])$, for $\bm{k}$ in a neighbourhood of $\mathcal{S}_{\textsc{f};\sigma}$ and $z$ in a neighbourhood of $\varepsilon_{\textsc{f}}$, as described in Sec.~\ref{s3a4}, can be attributed. Knowledge of this mechanism enables one to infer the forms of the leading-order terms in the asymptotic series expansions \cite{WW62,ETC65,HAL74} of the functions on the LHSs of Eqs.~(\ref{e335}) and (\ref{e337}) for $U/t \to 0$ by reliance on the knowledge provided by the expression in Eq.~(\ref{e342}).

The functions $\Sigma_{\sigma}'(\bm{k};\varepsilon;[\{G_{\sigma'}\}])$ and $\Sigma_{\sigma}'(\bm{k};\varepsilon;[\{G_{\sigma}^{\textsc{hf}}\}])$ (to be distinguished from $\Upsigma_{\sigma}'(\bm{k};\varepsilon;[\{G_{\sigma}^{\textsc{hf}}\}])$) are formally related through the following functional series expansion:
\begin{widetext}
\begin{equation}
\t{\Sigma}_{\sigma}'(\bm{k};z;[\{G_{\sigma''}\}]) = \t{\Sigma}_{\sigma}'(\bm{k};z;[\{G_{\sigma''}^{\textsc{hf}}\}]) + \sum_{\bm{k}', \sigma'} \int_{-\infty}^{\infty} \rd\varepsilon' \left. \frac{\delta \t{\Sigma}_{\sigma}'(\bm{k};z;[\{G_{\sigma''}\}])}{\delta G_{\sigma'}(\bm{k}';\varepsilon')}\right|_{\{G_{\sigma''}\} = \{G_{\sigma''}^{\textsc{hf}}\}}\hspace{0.7pt} \hspace{-0.4cm}\big( G_{\sigma'}(\bm{k}';\varepsilon') - G_{\sigma'}^{\textsc{hf}}(\bm{k}';\varepsilon')\big) +
\dots~,
\label{e344}
\end{equation}
\end{widetext}
where, in view of the defining expression in Eq.~(\ref{ea8}), the integration with respect to $\varepsilon'$ over $(-\infty,\infty)$ can be deformed into the complex energy plane. We shall return to this possibility later in this section.

Since $\delta G_{\sigma}(\bm{k};\varepsilon)/\delta G_{\sigma'}(\bm{k}';\varepsilon') = \delta_{\sigma,\sigma'} \delta_{\bm{k},\bm{k}'} \delta(\varepsilon - \varepsilon')$, with $\mathscr{D}_{\nu}^j$ denoting the $j$th $\nu\hspace{0.4pt}$th-order skeleton self-energy diagram contributing to $\Sigma_{\sigma}'$, the second term on the RHS of Eq.~(\ref{e344}) is associated with $\{\mathscr{D}_{\nu}^j \| j, \nu\}$ as follows: for a given $\nu$ and $j$, the total contribution to the second term on the RHS of Eq.~(\ref{e344}) as arising from $\mathscr{D}_{\nu}^j$ consists of the superposition of the contributions of all $2\nu -1$ diagrams deduced from $\mathscr{D}_{\nu}^j$ by successively replacing a single line in $\mathscr{D}_{\nu}^j$ representing $G_{\sigma'}$, by a line representing $G_{\sigma'} - G_{\sigma'}^{\textsc{hf}}$. Prior to the latter replacement, all lines in $\mathscr{D}_{\nu}^j$ representing $G_{\sigma'}$ are to be reinterpreted as representing $G_{\sigma'}^{\textsc{hf}}$, $\forall\sigma'$.

In Sec.~\ref{s3a4} we have established that (specifically in $d \le 2$) for $\bm{k}$ in a neighbourhood of $\mathcal{S}_{\textsc{f};\sigma}$ and $\varepsilon$ in a neighbourhood of $\varepsilon_{\textsc{f}}$ the function $\t{\Sigma}_{\sigma}(\bm{k};z;[\{G_{\sigma'}\}])$ does not to leading order scale like $U^2$, but like a function more dominant than $U^2$ as $U/t \to 0$. This property must evidently be inherent in the expression in Eq.~(\ref{e344}). In the following we shall therefore focus on establishing the relevant mathematical mechanism that gives rise to this property.

On account of the Dyson equation, one has (cf. Eq.~(\ref{e334}))
\begin{widetext}
\begin{equation}
\t{G}_{\sigma}(\bm{k};z) - \t{G}_{\sigma}^{\textsc{hf}}(\bm{k};z) = \t{G}_{\sigma}^{\textsc{hf}}(\bm{k};z) \t{\Sigma}_{\sigma}'(\bm{k};z) \big(1 - \t{G}_{\sigma}^{\textsc{hf}}(\bm{k};z) \t{\Sigma}_{\sigma}'(\bm{k};z)\big)^{-1} \t{G}_{\sigma}^{\textsc{hf}}(\bm{k};z),
\label{e345}
\end{equation}
\end{widetext}
from which one deduces that for $\bm{k}$ sufficiently close to $\mathcal{S}_{\textsc{f};\sigma}^{\textsc{hf}}$ and $\t{\Sigma}_{\sigma}'(\bm{k};\mu^{\textsc{hf}}) \not=0$, one has
\begin{eqnarray}
&&\hspace{0.0cm}\t{G}_{\sigma}(\bm{k};z) - \t{G}_{\sigma}^{\textsc{hf}}(\bm{k};z) \sim - \t{G}_{\sigma}^{\textsc{hf}}(\bm{k};z)-1/\t{\Sigma}_{\sigma}'(\bm{k};\mu^{\textsc{hf}})\nonumber\\
&&\hspace{5.0cm}\text{as}\;\; z \to \mu^{\textsc{hf}}.
\label{e346}
\end{eqnarray}
The function $-\t{G}_{\sigma}^{\textsc{hf}}$ on the RHS of this expression is to be contrasted with the erroneous expression $\t{G}_{\sigma}^{\textsc{hf}} \t{\Sigma}_{\sigma}'[\{G_{\sigma'}\}] \t{G}_{\sigma}^{\textsc{hf}}$ that the formal geometric series expansion of $(1 - \t{G}_{\sigma}^{\textsc{hf}} \t{\Sigma}_{\sigma}'[\{G_{\sigma'}\}])^{-1}$ in powers of $\t{G}_{\sigma}^{\textsc{hf}} \t{\Sigma}_{\sigma}'[\{G_{\sigma'}\}]$ would imply, Eq.~(\ref{e345}). We remark that the asymptotic expression in Eq.~(\ref{e346}) equally applies to $\t{G}_{\sigma}'(\bm{k};z) - \t{G}_{\sigma}^{\textsc{hf}}(\bm{k};z)$, with $G_{\sigma}'$ denoting the function defined in Eq.~(\ref{e332}), provided that $\t{\Sigma}_{\sigma}'(\bm{k};\mu^{\textsc{hf}})$ be replaced by $\t{\Upsigma}_{\sigma}^{(2)}(\bm{k};\mu^{\textsc{hf}};[\{G_{\sigma'}^{\textsc{hf}}\}])$.

Since the functional derivatives on the RHS of Eq.~(\ref{e344}) (of which only one is shown explicitly) are evaluated in terms of $\{G_{\sigma}^{\textsc{hf}}\| \sigma\}$, in the asymptotic region $U/t \to 0$ they qualitatively behave similarly to what one would expect from the self-energy as calculated within the framework of the many-body perturbation theory in terms of $\{G_{\sigma}^{\textsc{hf}}\| \sigma\}$ (note that skeleton self-energy diagrams constitute a proper subset of all connected proper self-energy diagrams). In particular, for $U/t \to 0$ to leading order these derivatives scale like $U^2$ (see later). On the basis of this observation and of the asymptotic expression in Eq.~(\ref{e346}), we conclude that for $\bm{k}$ in a neighbourhood of $\mathcal{S}_{\textsc{f};\sigma}^{\textsc{hf}}$ (cf. Eq.~(\ref{e241})) and $z$ in a neighbourhood of $\varepsilon_{\textsc{f}}^{\textsc{hf}}$ (cf. Eq.~(\ref{e229})) the leading-order term in the asymptotic series expansion of the difference $\t{\Sigma}_{\sigma}'(\bm{k};z;[\{G_{\sigma'}\}] -\t{\Sigma}_{\sigma}'(\bm{k};z;[\{G_{\sigma'}^{\textsc{hf}}\}])$, for $U/t \to 0$, can scale like $U^2$, instead of $U^4$, and even like a function which is more dominant than $U^2$. In this connection, and with reference to appendix \ref{sab}, we note that on denoting the integrand of the integral with respect to $\varepsilon'$ on the RHS of the expression in Eq.~(\ref{e344}) by $f(\varepsilon')$, this integral can be expressed as a contour integral of $\t{f}(z)$ (cf. Eq.~(\ref{ea8})) over $\mathscr{C}(\mu^{\textsc{hf}})$ combined with an integral of $\t{f}(\varepsilon'+i 0^+) - \t{f}(\varepsilon'-i 0^+)$ over the interval $(\mu^{\textsc{hf}},\mu)$, where $\mu$ is the exact chemical potential.

We have thus established the mathematical mechanism that underlies the specific form of the dependence of $\t{\Sigma}_{\sigma}(\bm{k};z)$ on $U$ in the asymptotic region $U/t \to 0$, observed in Sec.~\ref{s3a4}, for $\bm{k}$ and $z$ in a neighbourhood of respectively $\mathcal{S}_{\textsc{f};\sigma}$ and $\varepsilon_{\textsc{f}}$.

Evidently, the equality in Eq.~(\ref{e344}) applies order-by-order, that is it applies for the explicit $\t{\Sigma}_{\sigma}'$ on both sides being replaced by $\t{\Sigma}_{\sigma}^{(\nu)}$, $\nu \ge 1$. Since a $\nu$\hspace{0.4pt}th-order skeleton self-energy diagram consists of $2\nu -1$ distinct Green-function lines, it follows that for any finite $\nu$ the functional series expansion for $\t{\Sigma}_{\sigma}^{(\nu)}[\{G_{\sigma'}\}]$ around $\{G_{\sigma}^{\textsc{hf}}\| \sigma\}$ is terminating; terms corresponding to the $2\nu$\hspace{0.4pt}th- and higher-order functional derivatives of $\t{\Sigma}_{\sigma}^{(\nu)}[\{G_{\sigma'}\}]$ are all identically vanishing. Restricting oneself to the case of $\nu=2$, up to a constant -- independent of $\bm{k}$ and $\varepsilon$ (corresponding to the non-skeleton second-order proper self-energy diagram), the $\t{\Sigma}_{\sigma}^{(2)}[\{G_{\sigma'}^{\textsc{hf}}\}]$ on the RHS of the relevant expression can be replaced by $\t{\Upsigma}_{\sigma}^{(2)}[\{G_{\sigma'}^{\textsc{hf}}\}]$. On account of the asymptotic expression in Eq.~(\ref{e346}) and the subsequent remarks, it follows that aside from the last-mentioned constant, for $U/t\to 0$ the deviation of $\t{\Sigma}_{\sigma}^{(2)}(\bm{k};z;[\{G_{\sigma''}\}])$ from $\t{\Upsigma}_{\sigma}^{(2)}(\bm{k};z;[\{G_{\sigma''}^{\textsc{hf}}\}])$ is more dominant than $U^2$ for $\bm{k}$ and $z$ in a neighbourhood of respectively $\mathcal{S}_{\textsc{f};\sigma}$ and $\varepsilon_{\textsc{f}}$.

By the reasoning of the last but two paragraph, one would be tempted to suppose that in principle an infinity of terms on the RHS of Eq.~(\ref{e344}), denoted by the ellipsis, would contribute to the leading-order term in the asymptotic series expansion of $\t{\Sigma}_{\sigma}'(\bm{k};z;[\{G_{\sigma''}\}])$ for $U/t \to 0$. The details of the previous paragraph reveal however that the coefficient of the term on the RHS of Eq.~(\ref{e344}) associated with, symbolically, $(G_{\sigma'} - G_{\sigma'}^{\textsc{hf}})^p$ and as arising from the contribution of $\t{\Sigma}_{\sigma}^{(\nu)}$ to $\t{\Sigma}_{\sigma}'$, is identically vanishing for $p > 2 \nu -1$, whereby, owing to the direct proportionality of the $p\hspace{0.4pt}$th functional derivative of $\t{\Sigma}_{\sigma}^{(\nu)}[\{G_{\sigma'}\}]$ around $\{G_{\sigma}^{\textsc{hf}} \| \sigma\}$ with $U^{\nu}$, for $\nu = 2$ only three terms (corresponding to $p=1, 2$ and $3$) on the RHS of Eq.~(\ref{e344}) (\textsl{excluding} the first term whose leading-order contribution scales like $U^2$) contribute to the sought-after leading-order asymptotic contribution to $\t{\Sigma}_{\sigma}'[\{G_{\sigma'}\}]$ for $U/t \to 0$. Thus, while the functional form of the latter leading-order asymptotic contribution in its dependence on $U$ is deducible from the second term on the RHS of Eq.~(\ref{e344}), with the $\t{\Sigma}_{\sigma}'[\{G_{\sigma''}\}]$ herein replaced by $\t{\Sigma}_{\sigma}^{(2)}[\{G_{\sigma''}\}]$, calculation of the exact coefficient of this leading-order asymptotic contribution requires also the third and fourth terms on the RHS of Eq.~(\ref{e344}), involving respectively the second and third functional derivatives of $\t{\Sigma}_{\sigma}^{(2)}[\{G_{\sigma''}\}]$ at $\{G_{\sigma}^{\textsc{hf}} \| \sigma\}$, to be taken into account. For a comparable, but not identical, correspondence between the coefficients of the asymptotic series of $\t{\Sigma}_{\sigma}(\bm{k};z)$, in terms of the asymptotic sequence $\{1, 1/z, 1/z^2,\dots\}$ \cite{ETC65,HAL74}, corresponding to $\vert z\vert \to\infty$, and the coefficients of a similar series pertaining to $\t{\Sigma}_{\sigma}^{(\nu)}(\bm{k};z)$, the reader is referred to Sec.~B.7 of Ref.~\cite{BF07a}.

% 3.a.6.
\subsubsection{Summary}
\label{s3a6}
Summarizing, \emph{we have rigorously established that the Fermi surface as calculated on the basis of  $\Upsigma_{\sigma}^{(2)}[\{G_{\sigma'}^{\textsc{hf}}\}]$ suffers from the fundamental deficiency that the number of the $\bm{k}$ points enclosed by it, that is $N_{\textsc{l};\sigma}$, deviates from $N_{\sigma}$, in violation of the Luttinger theorem.} This Fermi surface can therefore \textsl{not} appropriately approximate the exact Fermi surface, for which the Luttinger theorem \cite{LW60,JML60,ID03,BF07a,BF07-12} \textsl{is} well satisfied. With reference to the equality in Eq.~(\ref{e330}) and assuming that $\{G_{\sigma}\| \sigma\}$ are at hand, this is \textsl{not} the case for the Fermi surface calculated on the basis of $\Sigma_{\sigma}^{(2)}[\{G_{\sigma'}\}]$. Similarly for the Fermi surface calculated on the basis of $\Sigma_{\sigma}^{(2)}[\{G_{\sigma'}'\}]$, with $G_{\sigma}'$ determined in terms of $G_{\sigma}^{\textsc{hf}}$ and $\Upsigma_{\sigma}^{(2)}[\{G_{\sigma'}^{\textsc{hf}}\}]$, where $G_{\sigma}^{\textsc{hf}}$ \textsl{self-consistently} corresponds to the single-particle energy dispersion $\varepsilon_{\bm{k}} + \hbar\Upsigma_{\sigma}^{\textsc{hf}}(\bm{k};[\{G_{\sigma'}^{\textsc{hf}}\}]$, $\forall\sigma$, Eqs.~(\ref{e332}) and (\ref{e336}).

On the basis of the fact that the next-to-leading-order term in the formal asymptotic series expansion of $\mathsf{n}_{\sigma}(\bm{k})$ for $U/t \to 0$ in terms of the asymptotic sequence $\{1, U, U^2,\dots\}$ \cite{ETC65,HAL74} is divergent for $\bm{k}$ approaching the underlying Fermi surface \cite{OS90,BvdL91}, on general grounds (appendix \ref{sac}) we have demonstrated that the leading-order term in the asymptotic series expansion of $\t{\Sigma}_{\sigma}(\bm{k};z;[\{G_{\sigma'}\}])$ ($\t{\Sigma}_{\sigma}^{(2)}(\bm{k};z;[\{G_{\sigma'}\}])$) in the region $U/t \to 0$ is more dominant than that of $\t{\Upsigma}_{\sigma}(\bm{k};z;[\{G_{\sigma'}^{\textsc{hf}}\}])$ ($\t{\Upsigma}_{\sigma}^{(2)}(\bm{k};z;[\{G_{\sigma'}^{\textsc{hf}}\}])$) for $\bm{k}$ and $z$ in a neighbourhood of respectively Fermi surface and Fermi energy. The latter term scales like $U^2$ and the former term like $U^{\alpha} \ln^{\gamma}(\vert t\vert/U)$, where $1 \le\alpha \le 2$ and $\gamma \ge 0$, excluding the possibilities $\alpha=0$, $\gamma >0$, and $\alpha =2$, $\gamma=0$, appendix \ref{sac}. Already this observation establishes that in calculating the deviation of $\mathcal{S}_{\textsc{f};\sigma}$ from $\mathcal{S}_{\textsc{f};\sigma}^{(0)}$ in terms of $\Upsigma_{\sigma}^{(2)}[\{G_{\sigma'}^{\textsc{hf}}\}]$ in the region $U/t\to 0$, one in fact neglects the leading asymptotic contribution corresponding to $\Sigma_{\sigma}^{(2)}[\{G_{\sigma'}\}]$, which is missing in $\Upsigma_{\sigma}^{(2)}[\{G_{\sigma'}^{\textsc{hf}}\}]$. We have described the reason for this shortcoming in the paragraph following that containing Eq.~(\ref{e316}), p.~\pageref{ToBeExplicit}.

% 3.b.
\subsection{Analysis}
\label{s3b}

% 3.b.1.
\subsubsection{Overview}
\label{s3b1}
In Refs.~\cite{SG88,HM97} the exact self-energy $\Sigma_{\sigma}'(\bm{k}_{\textsc{f};\sigma}(\varphi);
\varepsilon_{\textsc{f}})$ in the expressions on the RHSs of Eqs.~(\ref{e38}) and (\ref{e314}) is substituted by (Sec.~\ref{s3a2})
\begin{equation}
\Upsigma_{\sigma}^{(2)}(\bm{k}_{\textsc{f};\sigma}^{\textsc{hf}}(\varphi);
\varepsilon_{\textsc{f}}^{(0)}) \equiv \Upsigma_{\sigma}^{(2)}(\bm{k}_{\textsc{f};\sigma}^{\textsc{hf}}(\varphi);
\varepsilon_{\textsc{f}}^{(0)};[\{G_{\sigma'}^{(0)}\}]),
\label{e347}
\end{equation}
resulting in the following equalities, purported to be exact (in the absolute sense) to order $U^2$:
\begin{equation}
\delta k_{\textsc{f};\sigma}^{\textsc{hf}}(\varphi) = \frac{\delta\varepsilon_{\textsc{f}}^{\textsc{hf}} -\hbar \Upsigma_{\sigma}^{(2)}(\bm{k}_{\textsc{f};\sigma}^{\textsc{hf}}
(\varphi);\varepsilon_{\textsc{f}}^{(0)})}
{a_{\sigma}^{\textsc{hf}}(\varphi)},
\label{e348}
\end{equation}
\begin{equation}
\delta\varepsilon_{\textsc{f}}^{\textsc{hf}} = \frac{\int_0^{2\pi} \rd\varphi\, k_{\textsc{f};\sigma}^{\textsc{hf}}(\varphi) \hbar\Upsigma_{\sigma}^{(2)}(\bm{k}_{\textsc{f};\sigma}^{\textsc{hf}}(\varphi);
\varepsilon_{\textsc{f}}^{(0)})/a_{\sigma}^{\textsc{hf}}(\varphi)}{\int_0^{2\pi} \rd\varphi\, k_{\textsc{f};\sigma}^{\textsc{hf}}(\varphi)/a_{\sigma}^{\textsc{hf}}(\varphi)}.
\label{e349}
\end{equation}
The equality in Eq.~(\ref{e348}) directly coincides with that in Eq.~(9) of Ref.~\cite{HM97}. Making use of $\delta(f(x)) = \delta(x)/f'(0)$ for $f(0)= 0$, $f'(0) >0$ and $f(x) \not= 0$ outside $x=0$ in the interval of interest (recall the assumed convexity of the underlying Fermi sea, p.~\pageref{Convex}), from the expression in Eq.~(\ref{e349}) one recovers that in Eq.~(8) of Ref.~\cite{HM97}. For clarity, since for paramagnetic uniform metallic GSs one has $\mathcal{S}_{\textsc{f};\sigma}^{\textsc{hf}} = \mathcal{S}_{\textsc{f};\sigma}^{(0)}$, $a_{\sigma}^{\textsc{hf}}(\varphi)$ is identical to its non-interacting counterpart $a_{\sigma}^{(0)}(\varphi)$, similar to $\bm{k}_{\textsc{f};\sigma}^{\textsc{hf}}(\varphi)$ for which one has $\bm{k}_{\textsc{f};\sigma}^{\textsc{hf}}(\varphi)  \equiv \bm{k}_{\textsc{f};\sigma}^{(0)}(\varphi)$. With reference to the remarks following Eq.~(\ref{e314}), note that also the $\delta k_{\textsc{f};\sigma}^{\textsc{hf}}(\varphi)$ determined on the basis of the expressions in Eqs.~(\ref{e348}) and (\ref{e349}) is invariant under a constant shift, independent of $\varphi$, in $\Upsigma_{\sigma}^{(2)}(\bm{k}_{\textsc{f};\sigma}^{\textsc{hf}}(\varphi);
\varepsilon_{\textsc{f}}^{(0)})$. Thus, using these expressions, the contribution of the second-order non-skeleton self-energy diagram, an \textsl{anomalous} diagram \cite{KL60,LW60,NO98}, can be discarded, it being independent of $\bm{k}$, and thus of $\varphi$.

We note that use of the arguments $\bm{k}_{\textsc{f};\sigma}^{(0)}(\varphi) \equiv \bm{k}_{\textsc{f};\sigma}^{\textsc{hf}}(\varphi)$ and $\varepsilon_{\textsc{f}}^{(0)}$ in Refs.~\cite{SG88,HM97} (see above), and $\varepsilon_{\textsc{f}}^{\textsc{hf}}$ in Ref.~\cite{YY99}, instead of $\bm{k}_{\textsc{f};\sigma}(\varphi)$ and $\varepsilon_{\textsc{f}}$ respectively, is in conformity with the use of the second-order expansion of the exact $\Sigma_{\sigma}'(\bm{k};\varepsilon)$ in powers of $U$. The adopted approximations in Refs.~\cite{SG88,HM97,YY99} would have indeed been consistent with the intended exact determination of $\delta k_{\textsc{f};\sigma}^{\textsc{hf}}(\varphi)$ to order $U^2$, were it not for the fact that $\Upsigma_{\sigma}^{(2)}(\bm{k};\varepsilon;[\{G_{\sigma'}^{\textsc{mf}}\}])$ is specifically for $\bm{k} \in \mathcal{S}_{\textsc{f};\sigma}^{\textsc{mf}}$ and $\varepsilon =\varepsilon_{\textsc{f}}^{\textsc{mf}}$ fundamentally deficient, as summarized above.

We should emphasize that replacing $\bm{k}_{\textsc{f};\sigma}^{\textsc{hf}}(\varphi)$ for $\bm{k}_{\textsc{f};\sigma}(\varphi)$ is based on the assumption that $\Sigma_{\sigma}'(\bm{k};\varepsilon_{\textsc{f}})$ is a continuously differentiable function of $\bm{k}$ in a neighbourhood of $\bm{k} = \bm{k}_{\textsc{f};\sigma}(\varphi)$, and replacing e.g. $\varepsilon_{\textsc{f}}^{(0)}$ for $\varepsilon_{\textsc{f}}$ is based on the assumption that $\Sigma_{\sigma}'(\bm{k}_{\textsc{f};\sigma};\varepsilon)$ is a continuously differentiable function of $\varepsilon$ in a neighbourhood of $\varepsilon = \varepsilon_{\textsc{f}}$. Both of these assumptions are by definition applicable to \textsl{conventional} Fermi liquids \cite{BF99,BF03}. As regards the latter, with reference to the equality in Eq.~(\ref{e341}) one observes that this assumption is invalid in the cases where $Z_{\sigma}(\bm{k}_{\textsc{f};\sigma}) = 0$ (see Secs~\ref{s2c2}, \ref{s4a} and \ref{s4b}). Barring the possible van Hove points of $\varepsilon_{\bm{k}}$ on $\mathcal{S}_{\textsc{f};\sigma}$, Sec.~\ref{s4b}, it is not expected that for $d >1$, $Z_{\sigma}(\bm{k})$ may be vanishing for any $\bm{k} \in \mathcal{S}_{\textsc{f};\sigma}$ in the region $U/t \to 0$. It should be realized that in the event of $\alpha = 1$ in the asymptotic expression in Eq.~(\ref{e342}) (in which case necessarily $\gamma = 0$ -- see the remark following Eq.~(\ref{e343})), for $U/t\to 0$ to leading order the deviation of $\Sigma_{\sigma}'(\bm{k};\varepsilon_{\textsc{f}})$ from $\Sigma_{\sigma}'(\bm{k};\varepsilon_{\textsc{f}}^{(0)})$ scales like $U^2$, since $\varepsilon_{\textsc{f}} -\varepsilon_{\textsc{f}}^{(0)} = O(U)$.

As an aside, the contribution of the second-order anomalous self-energy diagram is determined in terms of a Fermi-surface average \cite[p. 44]{KL60},\cite[p. 166]{NO98}. This specific aspect is apparent in the expression on the RHS of Eq.~(\ref{e349}). Further, for the direct calculation of the `anomalous' part of $\Upsigma_{\sigma}^{(2)}(\bm{k}_{\textsc{f};\sigma}^{\textsc{hf}}(\varphi);
\varepsilon_{\textsc{f}}^{(0)})$, one formally has to employ the finite-temperature many-body formalism \cite[Ch. 7]{FW03} and effect the zero-temperature limit \textsl{after} having effected the thermodynamic limit \cite{KL60,LW60} \cite[\S3.3]{NO98} (on effecting the zero-temperature limit, the first loop from below in Fig.~2b of Ref.~\cite{HM97} gives rise to a $\delta$-function contribution, this on account of the expression in Eq.~(26) of Ref.~\cite{KL60}, as well as that in Eq.~(81) of Ref.~\cite{LW60}, which is to be contrasted with the subsequent expression, in Eq.~(82)). For the calculation of the `anomalous' part of $\Upsigma_{\sigma}^{(2)}(\bm{k}_{\textsc{f};\sigma}^{\textsc{hf}}(\varphi);
\varepsilon_{\textsc{f}}^{(0)})$ under a specific condition, see Sec.~\ref{s3b2}.

For the accuracy of presentation, in Ref.~\cite{SG88} no explicit appeal can be traced to an expression similar to that in Eq.~(\ref{e349}). However, on account of the statement following Eq.~(2) in this reference, namely that ``the chemical potential $\mu(U)$ is determined by fixing the number of electrons to $N$, independently of $U$'', and the fact that the constant $\mu_2$ (the coefficient of $U^2$ in the expansion ``$\mu = \mu_0 + \mu_1 U + \mu_2 U^2 +\dots$'') featuring in Eq.~(4) of Ref.~\cite{SG88}, is the equivalent of the constant $\delta\varepsilon_{\textsc{f}}^{\textsc{hf}}$ on the RHS of Eq.~(\ref{e348}), we conclude that the calculations in Ref.~\cite{SG88} must have relied on an expression similar to that in Eq.~(\ref{e349}). This view is supported by the fact that in Ref.~\cite{SG88}, as in Ref.~\cite{YY99}, no reference has been made to the contribution of the above-mentioned second-order non-skeleton, \textsl{anomalous}, self-energy diagram.

The considerations in Sec.~\ref{s3a} have made explicit that the expressions in Eqs.~(\ref{e348}) and (\ref{e349}) are deficient on two essential grounds. Firstly, in contrast to $\t{\Sigma}_{\sigma}^{(2)}[\{G_{\sigma'}\}]$ and $\t{\Sigma}_{\sigma}^{(2)}[\{G_{\sigma'}'\}]$, where $G_{\sigma}'$ may be one of the two functions defined in Eqs.~(\ref{e332}) and (\ref{e333}), $\t{\Upsigma}_{\sigma}^{(2)}[\{G_{\sigma'}^{\textsc{hf}}\}]$ misses out a term that is more dominant than $U^2$ in the asymptotic region $U/t \to 0$. Secondly, a specific quantity, expressed in terms of $\t{\Upsigma}_{\sigma}^{(2)}[\{G_{\sigma'}^{\textsc{hf}}\}]$ and the function $G_{\sigma}'$ defined in Eq.~(\ref{e332}), fails to satisfy the Luttinger-Ward identity \cite{LW60}, Eq.~(\ref{e337}). In contrast, the same quantity expressed in terms of $\t{\Sigma}_{\sigma}^{(2)}[\{G_{\sigma'}'\}]$ and $\t{G}_{\sigma}'$ does satisfy the Luttinger-Ward identity, Eq.~(\ref{e336}), similarly to the case of the latter quantity being ideally expressed in terms of $\t{\Sigma}_{\sigma}^{(2)}[\{G_{\sigma'}\}]$ and $\t{G}_{\sigma}$, Eq.~(\ref{e330}). Consequently, the Fermi surface defined as the locus of the $\bm{k}$ points at which $1/G_{\sigma}'(\bm{k};\mu^{\textsc{hf}}) = 0$ (see Sec.~\ref{s3b2}), with $G_{\sigma}'$ defined according to the expression in Eq.~(\ref{e332}) in terms of the $G_{\sigma}^{\textsc{hf}}$ presented in Eq.~(\ref{e339}), \textsl{cannot} satisfy the Luttinger theorem. While naively one might expect that the deviation of the relevant $N_{\textsc{l};\sigma}/N_{\textsc{s}}$ from $n_{\sigma} \equiv N_{\sigma}/N_{\textsc{s}}$, Eqs.~(\ref{e12}) and (\ref{e275}), scaled like $U^4$ in the region $U/t \to 0$, we have rigorously shown that in the latter region this deviation scales even more dominantly than $U^2$. For some relevant additional details, see Sec.~\ref{s3b2} and appendix \ref{sab}.

In Ref.~\cite{YY99}, the exact self-energy $\Sigma_{\sigma}'(\bm{k}_{\textsc{f};\sigma}(\varphi);
\varepsilon_{\textsc{f}})$ is substituted by $\Upsigma_{\sigma}^{(2)}(\bm{k}_{\textsc{f};\sigma}(\varphi);
\varepsilon_{\textsc{f}}^{\textsc{hf}};[\{G_{\sigma'}^{\textsc{hf}}\}])$, taking in addition \textsl{no} account of the \textsl{anomalous} second-order self-energy diagram \cite{Note7}, which is to say that the self-energy as employed in Ref.~\cite{YY99} coincides with $\Sigma_{\sigma}^{(2)}(\bm{k}_{\textsc{f};\sigma}(\varphi);
\varepsilon_{\textsc{f}}^{\textsc{hf}};[\{G_{\sigma'}^{\textsc{hf}}\}])$. In contrast to the approaches in Refs.~\cite{SG88,HM97}, the approach in Ref.~\cite{YY99} implicitly relies on the assumption of the validity of the Luttinger theorem in the framework of the second-order perturbation expansion of the self-energy in powers of $U/t$ and thereby avoids use of the expression in Eq.~(\ref{e314}), or that in Eq.~(\ref{e349}). It further avoids use of the expression in Eq.~(\ref{e38}), or that in Eq.~(\ref{e348}), by considering $\delta k_{\textsc{f};\sigma}^{\textsc{hf}}(\varphi)$ as being ``caused by the energy shift resulting from $\re\Sigma_k^{\hspace{0.8pt}\textsc{r}}(0)$''. With reference to the exact identify in Eq.~(\ref{e318}), we conclude that the  $\delta k_{\textsc{f};\sigma}^{\textsc{hf}}(\varphi)$ as calculated in Ref.~\cite{YY99} corresponds to the expression in Eq.~(\ref{e348}) in which $\delta \varepsilon_{\textsc{f}}^{\textsc{hf}}$ is identified with zero. Thus, the working assumption in Ref.~\cite{YY99} may be identified as consisting of the exact equality of the $\delta \varepsilon_{\textsc{f}}^{\textsc{hf}}$ as calculated according to the expression in Eq.~(\ref{e349}), in terms of $\Sigma_{\sigma}^{(2)}(\bm{k}_{\textsc{f};\sigma}(\varphi);
\varepsilon_{\textsc{f}}^{\textsc{hf}};[\{G_{\sigma'}^{\textsc{hf}}\}])$, with the contribution of the second-order non-skeleton self-energy diagram. Leaving aside these details, the exact identity in Eq.~(\ref{e318}) implies that insofar as the self-energy is concerned, the work in Ref.~\cite{YY99} stands on the same footing as, but without being necessarily equivalent to, the works in Refs.~\cite{SG88,HM97}.

The function $\delta k_{\textsc{f};\sigma}^{\textsc{hf}}(\varphi)/U^2$, corresponding to a number of different site-occupation numbers, $n = 2 n_{\sigma}$, $\forall\sigma$, and calculated on the basis of the expressions in Eqs.~(\ref{e348}) and (\ref{e349}), is presented in Fig.~2 of Ref.~\cite{SG88}, in Figs.~2 and 3 of Ref.~\cite{HM97}. The same function as calculated by directly solving the equation for the Fermi surface (see the previous paragraph) is presented in Figs.~6 and 8 of Ref.~\cite{YY99}. These numerical results, if exact to order $U^2$, would have implied that \textsl{any} $U>0$, no matter how small, would result in the violation of the relationship in Eq.~(\ref{e241}).

% 3.b.2.
\subsubsection{An illustrative calculation}
\label{s3b2}
In the light of the above observations, it is instructive to consider calculation of the Fermi surface associated with the total self-energy $\Upsigma_{\sigma}^{\textsc{hf}}[\{G_{\sigma'}^{\textsc{hf}}\}] +\Upsigma_{\sigma}^{(2)}[\{G_{\sigma'}^{\textsc{hf}}\}]$. From the second equality in Eq.~(\ref{e275}), one has
\begin{widetext}
\begin{equation}
N_{\textsc{l};\sigma} = \sum_{\bm{k}} \Theta\big(\varepsilon_{\textsc{f}}^{\textsc{hf}} -[\varepsilon_{\bm{k}} + \hbar \Upsigma_{\sigma}^{\textsc{hf}}(\bm{k};[\{G_{\sigma'}^{\textsc{hf}}\}]) +\hbar \Upsigma_{\sigma}^{(2)}(\bm{k};\varepsilon_{\textsc{f}}^{\textsc{hf}};[\{G_{\sigma'}^{\textsc{hf}}\}])]\big),
\label{e350}
\end{equation}
\end{widetext}
where the argument of the $\Theta$ function is the inverse of the Green function $G_{\sigma}'(\bm{k};\varepsilon)$, defined in Eq.~(\ref{e332}), at $\varepsilon = \varepsilon_{\textsc{f}}^{\textsc{hf}}$. The reason underlying this choice for $\varepsilon$ (whereby the counterpart of the constant $\delta\varepsilon_{\textsc{f}}^{\textsc{hf}}$, Eq.~(\ref{e35}), in the context of the present consideration is identified with zero) is the identity in Eq.~(\ref{e320}); in a neighbourhood of $\varepsilon_{\textsc{f}}^{\textsc{hf}}$, only at $\varepsilon = \varepsilon_{\textsc{f}}^{\textsc{hf}}$ can  $G_{\sigma}'(\bm{k};\varepsilon)$ be unbounded, or $1/G_{\sigma}'(\bm{k};\varepsilon)$ be vanishing. This choice for $\varepsilon$ and the choice of the contour $\mathscr{C}(\mu^{\textsc{hf}})$ in the expression in Eq.~(\ref{e337}) are inextricably connected. This is readily verified by considering the equalities in Eqs.~(\ref{e325}), (\ref{e326}) and (\ref{e327}).

With reference to the expressions in Eqs.~(\ref{e271}), (\ref{e275}) and (\ref{e350}), within the framework of the second-order perturbation theory for the self-energy in terms of $\{G_{\sigma}^{\textsc{hf}}\|\sigma\}$, the Fermi surface corresponding to particles with spin index $\sigma$, to be denoted by $\mathcal{S}_{\textsc{f};\sigma}^{(2)}$, consists of the set of solutions of the following equation:
\begin{equation}
\varepsilon_{\bm{k}} + \hbar \Upsigma_{\sigma}^{\textsc{hf}}(\bm{k};[\{G_{\sigma'}^{\textsc{hf}}\}]) +\hbar \Upsigma_{\sigma}^{(2)}(\bm{k};\varepsilon_{\textsc{f}}^{\textsc{hf}};[\{G_{\sigma'}^{\textsc{hf}}\}]) = \varepsilon_{\textsc{f}}^{\textsc{hf}}.
\label{e351}
\end{equation}
Below we consider the case where $\hbar \Upsigma_{\sigma}^{\textsc{hf}}(\bm{k};[\{G_{\sigma'}^{\textsc{hf}}\}]) \equiv U n_{\b{\sigma}}$, Eq.~(\ref{e224}), independent of $\bm{k}$. Further, in the following $\bm{k}_{\textsc{f};\sigma}$ will denote a general point on $\mathcal{S}_{\textsc{f};\sigma}^{(2)}$. Thus, along the same lines as in Sec.~\ref{s3a} (making use of in particular the equality in Eq.~(\ref{e32})), from the equality in Eq.~(\ref{e351}) one obtains the following expression for $\delta k_{\textsc{f};\sigma}^{\textsc{hf}}(\varphi)$, which is exact (from the perspective of Eq.~(\ref{e351})) to order $U^2$:
\begin{equation}
\delta k_{\textsc{f};\sigma}^{\textsc{hf}}(\varphi) = -\frac{\hbar \Upsigma_{\sigma}^{(2)}(\bm{k}_{\textsc{f};\sigma}^{\textsc{hf}}(\varphi);
\varepsilon_{\textsc{f}}^{\textsc{hf}};[\{G_{\sigma'}^{\textsc{hf}}\}])}{a_{\sigma}^{\textsc{hf}}(\varphi)}.
\label{e352}
\end{equation}
This expression is to be contrasted with that in Eq.~(\ref{e348}), taking into account the exact identity in Eq.~(\ref{e318}) (see Eq.~(\ref{e347})) (see the remark in the previous paragraph concerning the constant $\delta\varepsilon_{\textsc{f}}^{\textsc{hf}}$). On account of the considerations of Sec.~\ref{s3a3}, unless the self-energy in the numerator of the expression on the RHS of Eq.~(\ref{e352}) is identically vanishing for all $\varphi$ (compare with the exact result in Eq.~(\ref{e236})), the equality in Eq.~(\ref{e313}) (under the conditions for which it has been deduced, p.~\pageref{Convex}) is \textsl{not} satisfied to order $U^2$ by the $\delta k_{\textsc{f};\sigma}^{\textsc{hf}}(\varphi)$ in Eq.~(\ref{e352}).

Denoting the contribution of the \textsl{anomalous} second-order self-energy diagram to $\Upsigma_{\sigma}^{(2)}$ by $\Upsigma_{\sigma}^{(2\textrm{a})}$, and the contribution of the second-order \textsl{skeleton} self-energy diagram to $\Upsigma_{\sigma}^{(2)}$ by $\Upsigma_{\sigma}^{(2\textrm{s})}$ (see Eq.~(\ref{e318})), \emph{provided that the expression on the LHS of Eq.~(\ref{e337}) scales to leading order like $U^2$ for $U/t\to \infty$}, the exact $\Upsigma_{\sigma}^{(2\textrm{a})}$ can be calculated as follows.

With $\Omega_{\textsc{1bz}}$ denoting the `volume' of the 1BZ, one has (cf. Eqs.~(\ref{e325}), (\ref{e326}) and (\ref{e327}))
\begin{equation}
\frac{N_{\textsc{l};\sigma}}{N_{\textsc{s}}} \equiv n_{\sigma} - n_{\sigma}'' = \frac{\mathscr{A}_{\sigma}}{\Omega_{\textsc{1bz}}},
\label{e353}
\end{equation}
where
\begin{equation}
n_{\sigma}'' \doteq \lim_{\beta\to\infty} \frac{1}{N_{\textsc{s}}} \sum_{\bm{k}\in\textrm{1BZ}} \b{\upnu}_{\sigma}''(\bm{k}),
\label{e354}
\end{equation}
and $\mathscr{A}_{\sigma}$ is introduced in Eq.~(\ref{e311}). For orientation, for a square lattice with lattice constant $a=1$, one has $\Omega_{\textsc{1bz}} = 4\pi^2$. One trivially obtains that
\begin{equation}
\mathscr{A}_{\sigma} \sim \mathscr{A}_{\sigma}^{\textsc{hf}} + \int_0^{2\pi} \rd\varphi\, k_{\textsc{f};\sigma}^{\textsc{hf}}(\varphi)\hspace{0.6pt} \delta k_{\textsc{f};\sigma}^{\textsc{hf}}(\varphi)\;\; \text{for}\;\; \frac{U}{t}\to 0.
\label{e355}
\end{equation}
Since within the framework of the Hartree-Fock theory the Luttinger-Ward identity is satisfied, one has
\begin{equation}
\frac{\mathscr{A}_{\sigma}^{\textsc{hf}}}{\Omega_{\textsc{1bz}}} = n_{\sigma},
\label{e356}
\end{equation}
so that, following the equality in Eq.~(\ref{e353}), one obtains
\begin{equation}
\frac{1}{\Omega_{\textsc{1bz}}} \int_0^{2\pi} \rd\varphi\, k_{\textsc{f};\sigma}^{\textsc{hf}}(\varphi)\hspace{0.6pt} \delta k_{\textsc{f};\sigma}^{\textsc{hf}}(\varphi) = -n_{\sigma}^{\prime\prime (2)},
\label{e357}
\end{equation}
where $n_{\sigma}^{\prime\prime (2)}$ is the expression on the LHS of Eq.~(\ref{e337}) divided by $N_{\textsc{s}}$, with the $G_{\sigma}'$ in Eq.~(\ref{e337}) denoting the function defined in Eq.~(\ref{e332}) (recall that, here by assumption the expression on the LHS of Eq.~(\ref{e337}) scales to leading order like $U^2$ for $U/t\to 0$). Combining the equalities in Eqs.~(\ref{e352}) and (\ref{e357}), for the constant $\Upsigma_{\sigma}^{(2\textrm{a})}$, independent of $\bm{k}$, one obtains
\begin{widetext}
\begin{equation}
\hbar\Upsigma_{\sigma}^{(2\textrm{a})} = \frac{n_{\sigma}^{\prime\prime (2)} - \Omega_{\textsc{1bz}}^{-1}\int_0^{2\pi} \rd\varphi\, k_{\textsc{f};\sigma}^{\textsc{hf}}(\varphi)\hspace{0.6pt} \hbar\Upsigma_{\sigma}^{(2\textrm{s})}(\bm{k}_{\textsc{f};\sigma}^{\textsc{hf}}(\varphi);
\varepsilon_{\textsc{f}}^{\textsc{hf}})/a_{\sigma}^{\textsc{hf}}(\varphi)}{\Omega_{\textsc{1bz}}^{-1}
\int_0^{2\pi} \rd\varphi\, k_{\textsc{f};\sigma}^{\textsc{hf}}(\varphi)/a_{\sigma}^{\textsc{hf}}(\varphi)}.
\label{e358}
\end{equation}
\end{widetext}
For $n_{\sigma}^{\prime\prime (2))} =0$, the pair of expressions in Eqs.~(\ref{e352}) and (\ref{e358}) yield \textsl{exactly} the same $\delta k_{\textsc{f};\sigma}^{\textsc{hf}}(\varphi)$, $\forall\varphi$, as the pair of expressions in Eqs.~(\ref{e348}) and (\ref{e349}).

% 3.c.
\subsection{Discussion}
\label{s3c}
We have demonstrated that within the framework of the \textsl{non-self-consistent} perturbation theory for the self-energy of a metallic GS, in terms of the single-particle Green functions $\{G_{\sigma}^{\textsc{mf}}\| \sigma\}$ of a mean-field Hamiltonian, the Luttinger theorem is violated to at least the order of the adopted perturbation expansion. Quantitatively, in the case of the $N$-particle uniform metallic GS of the Hubbard Hamiltonian, the deviation of $N_{\textsc{l};\sigma}/N_{\textsc{s}}$ (the ratio of the number of the $\bm{k}$ points comprising the interior of the underlying Fermi sea corresponding to particles with spin index $\sigma$, to the number of lattice sites) from $n_{\sigma} \equiv N_{\sigma}/N_{\textsc{s}}$ scales at least like $U^{\nu}$ as $U/t \to 0$, where $\nu \ge 2$ is the order of the adopted perturbation series for the self-energy (the exclusion of $\nu=1$ is due to the constancy of the self-energy at this order with respect to variations of $\varepsilon$, whereby the Luttinger-Ward identity is trivially satisfied). It follows that the Fermi surfaces $\{\mathcal{S}_{\textsc{f};\sigma}\| \sigma\}$ of an $N$-particle  metallic GS \textsl{cannot} be calculated exactly to $\nu\hspace{0.6pt}$th order in the coupling constant of interaction on the basis of a non-self-consistently-calculated self-energy to $\nu\hspace{0.6pt}$th order. In fact, as a result of this systematic violation of the Luttinger theorem, a non-self-consistently-calculated self-energy fails to capture the non-analytic dependence of the exact self-energy $\Sigma_{\sigma}(\bm{k};\varepsilon)$ on the coupling-constant of interaction for $\bm{k}$ and $\varepsilon$ is a neighbourhood of respectively the Fermi surface and the Fermi energy.

On the basis of the above observations, we have shown that \emph{the extant non-self-consistent second-order calculations \cite{SG88,HM97,YY99}, purporting to show $\mathcal{S}_{\textsc{f};\sigma} \not\subseteq \mathcal{S}_{\textsc{f};\sigma}^{\textsc{hf}}$, are fundamentally deficient, as they fail to take full account of the leading-order contributions to the Fermi surface geometry to order $U^2$.} Therefore, results of these calculations do not constitute evidence against the validity of the relationship $\mathcal{S}_{\textsc{f};\sigma} \subseteq \mathcal{S}_{\textsc{f};\sigma}^{\textsc{hf}}$, Eq.~(\ref{e241}).

Neglecting the above rigorous conclusion, the fact that the calculated $\delta k_{\textsc{f};\sigma}^{\textsc{hf}}(\varphi)/U^2$ are very small, at the largest close to $0.001$ for $n= 0.97$ and $\varphi = 0$ \cite{HM97} ($0.001\hspace{0.6pt} U^2$ is to be compared with $\pi$, for the lattice constant $a=1$, which is approximately the value of $k_{\textsc{f};\sigma}^{(0)}(\varphi=0)$ corresponding to $n \approx 1$) should give one pause in declaring $\mathcal{S}_{\textsc{f};\sigma} \subseteq \mathcal{S}_{\textsc{f};\sigma}^{\textsc{hf}}$ as inexact.

In his section we have emphasised the significance of self-consistency to the calculation of in particular Fermi surface $\mathcal{S}_{\textsc{f};\sigma}$ (insofar as second-order perturbation theory is concerned, see Eqs.~(\ref{e332}), (\ref{e333}) and the subsequent specifications; see also the closing part of the present section, p.~\pageref{InClosing}). The calculations reported in Refs.~\cite{SG88,HM97,YY99,ZEG96} are non-self-consistent. Nojiri \cite{HN99} has performed a \textsl{self-consistent} calculation of $\Sigma_{\sigma}^{(2)}(\bm{k};\varepsilon)$ and on the basis of this calculation has deduced $\delta k_{\textsc{f};\sigma}^{\textsc{hf}}(\varphi) \not\equiv 0$ (see Figs.~6 and 7 in Ref.~\cite{HN99}). Three main remarks concerning the latter and other relevant observations by Nojiri \cite{HN99} are in order.

First, in contrast to the calculations in Refs.~\cite{SG88,HM97,YY99} in which $\delta k_{\textsc{f};\sigma}^{\textsc{hf}}(\varphi)/U^2$ is by design \textsl{fully} independent of $U$, owing to self-consistency and the attendant proliferation of infinite powers of $U$ in the calculated second-order self-energy, the $\delta k_{\textsc{f};\sigma}^{\textsc{hf}}(\varphi)/U^2$ in Ref.~\cite{HN99} \textsl{cannot} be identified with the expansion coefficient of $U^2$ in the series expansion of $\delta k_{\textsc{f};\sigma}^{\textsc{hf}}(\varphi)$ in powers of $U$ (for the sake of argument, here we disregard the fact that, as we have discussed above, the leading-order term in the asymptotic series expansion of $\delta k_{\textsc{f};\sigma}^{\textsc{hf}}(\varphi)$ for $U/t\to 0$ is more dominant than $U^2$). To demonstrate the deviation of this coefficient from zero, Nojiri must have performed a scaling analysis whereby to isolate the true coefficient of $U^2$ in the expansion of $\delta k_{\textsc{f};\sigma}^{\textsc{hf}}(\varphi)$. The significance of such analysis is best appreciated by the fact that the numerical results presented in Ref.~\cite{HN99} correspond to $U/t=4$; it is out of the question that the $\delta k_{\textsc{f};\sigma}^{\textsc{hf}}(\varphi)$ corresponding to such a relatively large value of $U/t$ (in comparison with the bandwidth $W$, for which in the present case one has $W/t = 8$) can in its entirety be attributed to the leading-order contribution to the exact $\delta k_{\textsc{f};\sigma}^{\textsc{hf}}(\varphi)$. In this connection, since the calculations underlying this $\delta k_{\textsc{f};\sigma}^{\textsc{hf}}(\varphi)$ entirely neglect $\Sigma_{\sigma}'(\bm{k};\varepsilon) - \Sigma_{\sigma}^{(2)}(\bm{k};\varepsilon)$, a higher than second-order contribution to the calculated $\delta k_{\textsc{f};\sigma}^{\textsc{hf}}(\varphi)$ is in principle spurious (see also the following paragraph). In spite of these facts, it is interesting to note that by dividing the data displayed in Figs.~6 and 7 of Ref.~\cite{HN99} by $U^2$ (i.e. by $16$), one observes that the results are uniformly by a factor of nearly $2$ \textsl{smaller} than their counterparts as presented in Refs.~\cite{SG88,HM97,YY99}. Related to this, a recent variational study by B\"{u}nemann, Schickling and Gebhard \cite{BSG12} of the Fermi surface of the Hubbard model reveals a noticeably smaller $\delta k_{\textsc{f};\sigma}^{\textsc{hf}}(\varphi)$ than reported in Ref.~\cite{HM97}. For clarity, on dividing by $(U/t)^2$ the $\delta k_{\textsc{f};\sigma}^{\textsc{hf}}(\varphi)$ corresponding to different values of $U/t$ in the inset of Fig.~4 of Ref.~\cite{BSG12} (which correspond to the total band filling $n=0.9$), one observes that to a good approximation all curves collapse into a single universal curve. On account of this observation, one may divide by $(U/t)^2$ the $\delta k_{\textsc{f};\sigma}^{\textsc{hf}}(\varphi)$ displayed in Fig.~4 of Ref.~\cite{BSG12} (corresponding to $U/t =10$) and compare the results with those in Fig.~3 of Ref.~\cite{HM97}. For $n=0.9$ ($0.8$) and $t=1$, the $\delta k_{\textsc{f};\sigma}^{\textsc{hf}}(\varphi)/U^2$ at $\varphi = 0$ in the former figure is approximately equal to $0.0006$ ($0.0001$), while in the latter figure it is approximately equal to $0.0008$ ($0.0005$). It is also interesting to note that the Fermi surface inferred from the quantum-Monte-Carlo results by Moreo \emph{et al.} \cite{MSSWB90}, pertaining to the Hubbard Hamiltonian on a $16 \times 16$ square lattice, with $U/t = 4$ (at $k_{\textsc{b}} T = t/6$ and for $n_{\sigma} = n_{\b\sigma}= 0.435$), almost identically coincides with $\mathcal{S}_{\textsc{f};\sigma}^{(0)} = \mathcal{S}_{\textsc{f};\sigma}^{\textsc{hf}}$.

Interestingly, Nojiri \cite{HN99} presents the calculated value of $1/Z_{\bm{k}_{\textsc{f};\sigma}}$, denoted by $\gamma_{\omega}$ \cite[Eq.~(3.9)]{HN99}, corresponding to $t=1$ and $U=4$, as function of the band filling $n$ in two different directions of the 1BZ,  namely the direction $(0,0) - (0,\pi)$, $\varphi=90^{\circ}$, and the direction $(0,0) - (\pi,\pi)$, $\varphi = 45^{\circ}$ \cite[Fig.~8]{HN99}. From the data in the latter figure, one observes that $1/Z_{\bm{k}_{\textsc{f};\sigma}}-1$ along both directions of the 1BZ are increasing functions of $n$ and specifically for $n$ approaching unity it is of the order of unity, implying that the calculations reported in Ref.~\cite{HN99} indeed do \textsl{not} correspond to a weakly-correlated $N$-particle GS of the Hubbard Hamiltonian. It is noteworthy that the way in which the $\gamma_{\omega}(\varphi)$ corresponding to $\varphi = 90^{\circ}$ exceeds the $\gamma_{\omega}(\varphi)$ corresponding to $\varphi = 45^{\circ}$, for $n$ approaching unity, directly correlates with the magnitudes of $\delta k_{\textsc{f};\sigma}^{\textsc{hf}}(\varphi)$ corresponding to respectively $\varphi = 90^{\circ}$ (which is equivalent to $\varphi=0$) and  $\varphi = 45^{\circ}$ \cite[Fig.~7]{HN99}.

Second, the expression for $\delta k_{\textsc{f};\sigma}^{\textsc{hf}}(\varphi)$ as adopted in Ref.~\cite{HN99} coincides with that in Eq.~(9) of Ref.~\cite{HM97} (our Eq.~(\ref{e348})), which is in principle suited for calculating $\delta k_{\textsc{f};\sigma}^{\textsc{hf}}(\varphi)$ to \textsl{exactly} second-order in $U$ (for the deficiency of Eq.~(\ref{e348}) in non-self-consistent calculations, see Sections~\ref{s3b1} and \ref{s3b2}). It is \textsl{not} suited for the calculations in Ref.~\cite{HN99}, where $U/t=4$; for such a relatively large value of $U/t$, the neglect of $\delta\t{\varepsilon}_{\sigma}^{\,\textsc{hf}}(\varphi)$, Eq.~(\ref{e37}), which in non-self-consistent calculations scales to leading order like $U^4$, is \textsl{not} warranted; Nojiri \cite{HN99} must instead have employed the expression in Eq.~(\ref{e38}) in conjunction with that in Eq.~(\ref{e314}).

Third, and last, the Fourier transforms (from the time- to the frequency-domain, and \emph{vice versa}) underlying the calculations reported in Ref.~\cite{HN99} have been carried out by means of the method of Fast Fourier Transformation (FFT) \cite{PTVF01}. In Ref.~\cite[\S6.3]{BF07a} we have shown the way in which a finite energy cut-off in similar calculations gives rise to violation of the exact property in Eq.~(\ref{e323}) and consequently to violation of the Luttinger theorem. What is noticeable in Ref.~\cite{HN99}, and many similar publications, is the explicit use of the function $\re[\Sigma_{\sigma}(\bm{k};\varepsilon_{\textsc{f}})]$ (the function ``$\re\hspace{-0.6pt}\Sigma_{\bm{k}\sigma}(0)$'' in the notation of Ref.~\cite{HN99}) where $\Sigma_{\sigma}(\bm{k};\varepsilon_{\textsc{f}})$ would suffice. This suggests that in these calculations the imaginary part of the function $\Sigma_{\sigma}(\bm{k};\varepsilon_{\textsc{f}})$ is artificially non-vanishing, implying that the Luttinger theorem cannot be exactly satisfied in these calculations. In the calculations of Ref.~\cite{HN99}, the frequency domain $(-\infty,\infty)$ has been limited to $[-\omega_{\textrm{c}},\omega_{\textrm{c}}]$, where $\omega_{\textrm{c}}/t = 512 \times 0.06 = 30.72$ \cite{HN99}, to be contrasted with $W/t = 8$, or $U/t = 4$.

\emph{We conclude that the extant numerical results concerning $\delta k_{\textsc{f};\sigma}^{\textsc{hf}}(\varphi)$ are deficient and thus \textsl{not} capable of negating the validity of the expression in Eq.~(\ref{e241}).}

\refstepcounter{dummy}
\label{InClosing}
In closing, we shed some additional light on the main aspects discussed in this section, Sec.~\ref{s3}, by considering a perturbation expansion in terms of the mean-field Green functions $\{G_{\sigma}^{\textsc{mf}}\| \sigma\}$, where
\begin{equation}
\t{G}_{\sigma}^{\textsc{mf}}(\bm{k};z) = \frac{\hbar}{z - [\varepsilon_{\bm{k}} + \hbar \Sigma_{\sigma}(\bm{k};\varepsilon_{\textsc{f}})]},
\label{e359}
\end{equation}
in which the real-valued function $\Sigma_{\sigma}(\bm{k};\varepsilon_{\textsc{f}})$ is the exact self-energy $\t{\Sigma}_{\sigma}(\bm{k};z)$ evaluated at the exact Fermi energy $\varepsilon_{\textsc{f}}$. One can convince oneself that the following considerations equally apply when $\t{\Sigma}_{\sigma}(\bm{k};z)$ is substituted by its self-consistently-calculated counterpart (to any arbitrary order in the coupling constant of the interaction potential) and $\varepsilon_{\textsc{f}}$ by the self-consistent value for the Fermi energy that conforms with the requirement of the Luttinger theorem \cite{BF07a}.

The expressions in Eqs.~(\ref{e234}) and (\ref{e236}) signify the fundamental difference between the mean-field Green function in Eq.~(\ref{e359}) and that in Eq.~(\ref{e339}). The function in Eq.~(\ref{e359}) also differs from the Green function encountered in Ref.~\cite[p.~1425]{LW60} (see Table I in Ref.~\cite{BF07a} for the relevant notational conventions), defined in terms of $\Sigma_{\sigma}(\bm{k}_{\textsc{f};\sigma};\varepsilon_{\textsc{f}})$, where $\bm{k}_{\textsc{f};\sigma}$ denotes the Fermi wave vector in the direction of $\bm{k}$, and explicitly referred to in Ref.~\cite[p.~1322]{VT97}. The function in Eq.~(\ref{e359}) is more general than the latter Green function the results corresponding to which we shall also discuss below. With reference to the result in Eq.~(\ref{e236}), we note that the function $\t{G}_{\sigma}^{\textsc{mf}}(\bm{k};z)$ defined in terms of $\Sigma_{\sigma}(\bm{k}_{\textsc{f};\sigma};\varepsilon_{\textsc{f}})$ coincides with that in Eq.~(\ref{e339}) on replacing the $\{G_{\sigma'}^{\textsc{hf}}\}$ on the RHS of this equation by $\{G_{\sigma'}\}$ (see however Eq.~(\ref{e224})). We note in passing that for a general Fermi sea, there can be more than one Fermi wave vector in the direction of $\bm{k}$, a possibility that we neglect here for transparency (this neglect amounts to assuming the underlying Fermi sea to be \textsl{convex}, Sec.~\ref{s3}, p.~\pageref{Convex}).

For the Green function $\t{G}_{\sigma}^{\textsc{mf}}$ in Eq.~(\ref{e359}), from the Dyson equation one has
\begin{widetext}
\begin{equation}
\t{G}_{\sigma}(\bm{k};z) = \big(1 - \t{G}_{\sigma}^{\textsc{mf}}(\bm{k};z) [\hbar\t{\Sigma}_{\sigma}(\bm{k};z) - \hbar\t{\Sigma}_{\sigma}(\bm{k};\varepsilon_{\textsc{f}})]\big)^{-1} \t{G}_{\sigma}^{\textsc{mf}}(\bm{k};z).
\label{e360}
\end{equation}
For $z\to \varepsilon_{\textsc{f}}$ one can employ the following general equality (cf. Eq.~(\ref{e341})) \cite{BF99}
\begin{equation}
\hbar\t{\Sigma}_{\sigma}(\bm{k};z) - \hbar\t{\Sigma}_{\sigma}(\bm{k};\varepsilon_{\textsc{f}}) = \Big(1 - \frac{1}{Z_{\sigma}(\bm{k})}\Big) (z - \varepsilon_{\textsc{f}}) + o(z - \varepsilon_{\textsc{f}}),\;\, \forall \bm{k},
\label{e361}
\end{equation}
where $o$  is the small-$o$ order symbol, to be distinguished from the large-$O$ order symbol \cite[\S2.11]{WW62}. Since for $U/t \to 0$ the Landau quasi-particle weight $Z_{\sigma}(\bm{k}) \to 1$ (cf. Eqs.~(\ref{e341}) and (\ref{e342})), it follows that the coefficient of the $(z-\varepsilon_{\textsc{f}})$ on the RHS of Eq.~(\ref{e361}) approaches zero for $U/t \to 0$. With reference to the defining expression in Eq.~(\ref{e232}), one observes that for $z \to \varepsilon_{\textsc{f}}$ and $\bm{k}\to \bm{k}_{\textsc{f};\sigma} \in\mathcal{S}_{\textsc{f};\sigma}$ one has
\begin{equation}
\t{G}_{\sigma}^{\textsc{mf}}(\bm{k};z) [\hbar\t{\Sigma}_{\sigma}(\bm{k};z) - \hbar\t{\Sigma}_{\sigma}(\bm{k};\varepsilon_{\textsc{f}})] \sim \frac{\hbar \big(1-1/Z_{\sigma}(\bm{k}_{\textsc{f};\sigma})\big)}{1 - \big([\bm{\nabla}\varepsilon_{\bm{k}} + \hbar\bm{\nabla}\Sigma_{\sigma}(\bm{k};\varepsilon_{\textsc{f}})]\vert_{\bm{k} = \bm{k}_{\textsc{f};\sigma}}\big) \cdot (\bm{k}-\bm{k}_{\textsc{f};\sigma})/(z-\varepsilon_{\textsc{f}})},
\label{e362}
\end{equation}
\end{widetext}
making explicit that in the case at hand the limiting processes $z \to \varepsilon_{\textsc{f}}$ and $\bm{k}\to \bm{k}_{\textsc{f};\sigma}$ do \textsl{not} commute; whereas on taking the limit $z\to \varepsilon_{\textsc{f}}$ first, one obtains zero for a fixed $\bm{k} \not=\bm{k}_{\textsc{f};\sigma}$, on taking the limit $\bm{k}\to \bm{k}_{\textsc{f};\sigma}$ first, one obtains $\hbar (1-1/Z_{\sigma}(\bm{k}_{\textsc{f};\sigma}))$ for a fixed $z\not= \varepsilon_{\textsc{f}}$, which, according to the expressions in Eqs.~(\ref{e341}) and (\ref{e342}), approaches zero for $U/t\to 0$. From the perspective of the present considerations, it is interesting that for sufficiently small values of $U/\vert t\vert$ (a necessary condition \textsl{only} for the case of $\bm{k}$ approaching $\mathcal{S}_{\textsc{f};\sigma}$ for a fixed $z \not= \varepsilon_{\textsc{f}}$), the function $\t{G}_{\sigma}(\bm{k};z)$ in Eq.~(\ref{e360}) can be expanded in powers of $\t{G}_{\sigma}^{\textsc{mf}}(\bm{k};z) [\hbar\t{\Sigma}_{\sigma}(\bm{k};z) - \hbar\t{\Sigma}_{\sigma}(\bm{k};\varepsilon_{\textsc{f}})]$ for $\bm{k}$ and $z$ in a neighbourhood of respectively $\mathcal{S}_{\textsc{f};\sigma}$ and $\varepsilon_{\textsc{f}}$, irrespective of the order in which $\bm{k}$ and $z$ may approach $\mathcal{S}_{\textsc{f};\sigma}$ and $\varepsilon_{\textsc{f}}$. The result that one thus obtains is in stark contrast with the asymptotic results in Eqs.~(\ref{e340}) and (\ref{e346}). The significant implication of the present result for the Luttinger-Ward identity in the framework of a finite-order perturbation expansion for the self-energy should be evident.

Following the same approach as above, for the relevant expression corresponding to the case where $\t{G}_{\sigma}^{\textsc{mf}}(\bm{k};z)$ is defined in terms of $\Sigma_{\sigma}(\bm{k}_{\textsc{f};\sigma};\varepsilon_{\textsc{f}})$, with $\bm{k}_{\textsc{f};\sigma}$ denoting the Fermi wave vector in the direction of $\bm{k}$, for $z \to \varepsilon_{\textsc{f}}$ and $\bm{k}\to \bm{k}_{\textsc{f};\sigma} \in\mathcal{S}_{\textsc{f};\sigma}$ one obtains
\begin{widetext}
\begin{equation}
\t{G}_{\sigma}^{\textsc{mf}}(\bm{k};z) [\hbar\t{\Sigma}_{\sigma}(\bm{k};z) - \hbar\t{\Sigma}_{\sigma}(\bm{k}_{\textsc{f};\sigma};\varepsilon_{\textsc{f}})] \sim \frac{\hbar \big(1-1/Z_{\sigma}(\bm{k}_{\textsc{f};\sigma})\big) +\hbar (\bm{\nabla}\Sigma_{\sigma}(\bm{k};\varepsilon_{\textsc{f}})\vert_{\bm{k}=\bm{k}_{\textsc{f};\sigma}}) \cdot (\bm{k}-\bm{k}_{\textsc{f};\sigma})/(z-\varepsilon_{\textsc{f}})}{1 - \big(\bm{\nabla}\varepsilon_{\bm{k}}\vert_{\bm{k} = \bm{k}_{\textsc{f};\sigma}}\big) \cdot (\bm{k}-\bm{k}_{\textsc{f};\sigma})/(z-\varepsilon_{\textsc{f}})},
\label{e363}
\end{equation}
where to leading order in $(z-\varepsilon_{\textsc{f}})$ we have replaced $\bm{\nabla}\t{\Sigma}_{\sigma}(\bm{k};z)$ with $\bm{\nabla}\Sigma_{\sigma}(\bm{k};\varepsilon_{\textsc{f}})$. On effecting the limit $\bm{k} \to \bm{k}_{\textsc{f};\sigma}$ for a fixed $z \not= \varepsilon_{\textsc{f}}$, for the LHS of the expression in Eq.~(\ref{e363}) one obtains the same result as in the case considered above, namely $\hbar (1-1/Z_{\sigma}(\bm{k}_{\textsc{f};\sigma}))$. In contrast with the case considered above however, here on taking the limit $z\to \varepsilon_{\textsc{f}}$ for a fixed $\bm{k} \not= \bm{k}_{\textsc{f};\sigma}$, and provided that $\big(\bm{\nabla}\varepsilon_{\bm{k}}\vert_{\bm{k} = \bm{k}_{\textsc{f};\sigma}}\big) \cdot (\bm{k}-\bm{k}_{\textsc{f};\sigma}) \not= 0$, instead of zero one obtains
\begin{equation}
\lim_{z\to \varepsilon_{\textsc{f}}} \t{G}_{\sigma}^{\textsc{mf}}(\bm{k};z) [\hbar\t{\Sigma}_{\sigma}(\bm{k};z) - \hbar\t{\Sigma}_{\sigma}(\bm{k}_{\textsc{f};\sigma};\varepsilon_{\textsc{f}})] \sim -\frac{\hbar (\bm{\nabla}\Sigma_{\sigma}(\bm{k};\varepsilon_{\textsc{f}})\vert_{\bm{k}=\bm{k}_{\textsc{f};\sigma}}) \cdot (\bm{k}-\bm{k}_{\textsc{f};\sigma})}{\big(\bm{\nabla}\varepsilon_{\bm{k}}\vert_{\bm{k} = \bm{k}_{\textsc{f};\sigma}}\big) \cdot (\bm{k}-\bm{k}_{\textsc{f};\sigma})}.
\label{e364}
\end{equation}
\end{widetext}
In the case of the $N$-particle uniform GS of the Hubbard Hamiltonian where $\Sigma_{\sigma}^{\textsc{hf}}(\bm{k})$ is independent of $\bm{k}$, Eq.~(\ref{e224}), the function $\bm{\nabla}\Sigma_{\sigma}(\bm{k};\varepsilon_{\textsc{f}})$ can be replaced by $\bm{\nabla}\Sigma_{\sigma}'(\bm{k};\varepsilon_{\textsc{f}}) \equiv \bm{\nabla}S_{\sigma}(\bm{k})$, Eq.~(\ref{e234}) (see the last remark following Eq.~(\ref{e238}) above). The dependence of this function on $U/\vert t\vert$ in the region $U/t \to 0$ is \textsl{similar} to the expression on the RHS of Eq.~(\ref{e342}). With reference to the expression in Eq.~(\ref{e341}), we thus conclude that insofar as the functional dependence of the function on the LHS of Eq.~(\ref{e363}) on $U/\vert t\vert$, in the region $U/t \to 0$, is concerned, the order of the limiting processes $z\to \varepsilon_{\textsc{f}}$ and $\bm{k} \to \bm{k}_{\textsc{f};\sigma} \in \mathcal{S}_{\textsc{f};\sigma}$ is immaterial (the limiting \textsl{values} for a given non-vanishing $U/\vert t\vert$ are in general not the same however). Note that the expression in Eq.~(\ref{e364}) breaks down at the van Hove points of the non-interacting energy dispersion that may be located on $\mathcal{S}_{\textsc{f};\sigma}$ (see Sec.~\ref{s4b}).

The above considerations establish that from the perspective of the Luttinger theorem, a self-consistent calculation based on the Green function in Eq.~(\ref{e359}) is qualitatively similar to a self-consistent calculation based on the Green function derived from the latter function through replacing the $\Sigma_{\sigma}(\bm{k};\varepsilon_{\textsc{f}})$ herein by $\Sigma_{\sigma}(\bm{k}_{\textsc{f};\sigma};\varepsilon_{\textsc{f}})$, where $\bm{k}_{\textsc{f};\sigma}$ denotes the Fermi wave vector in the direction of $\bm{k}$.

% 4.
\section{Behaviour of \texorpdfstring{$\mathsf{n}_{\sigma}(\bm{k})$}{} for \texorpdfstring{$\bm{k}$}{} close to \texorpdfstring{$\mathcal{S}_{\SC{\textsc{f}};\sigma}$}{}}
\label{s4}
With reference to the expressions in Eqs.~(\ref{e218}) and (\ref{e220}), we introduce the auxiliary function $\zeta_{\bm{k};\sigma}$ and write
\begin{equation}
\beta_{\bm{k};\sigma}^{<} \equiv \mathsf{n}_{\sigma}(\bm{k})\, (n_{\b\sigma} + \zeta_{\bm{k};\sigma}), \label{e41}
\end{equation}
where, in the light of the expressions in Eqs.~(\ref{e221}) and (\ref{e240}) and following the result in Eq.~(\ref{e222}),
\begin{equation}
\zeta_{\bm{k};\sigma} \sim 0\;\; \text{for}\;\; \bm{k} \to \bm{k}_{\textsc{f};\sigma} \in \mathcal{S}_{\textsc{f};\sigma}.
\label{e42}
\end{equation}
One thus has (see Eqs.~(\ref{e218}) and (\ref{e220}))
\begin{eqnarray}
\varepsilon_{\bm{k};\sigma}^{<} &=& \varepsilon_{\bm{k}} + U n_{\b\sigma} + U \zeta_{\bm{k};\sigma}, \label{e43}\\
\varepsilon_{\bm{k};\sigma}^{>} &=& \varepsilon_{\bm{k}} + U n_{\b\sigma} - U \Lambda_{\sigma}(\bm{k})\hspace{0.7pt} \zeta_{\bm{k};\sigma},
\label{e44}
\end{eqnarray}
where
\begin{equation}
\Lambda_{\sigma}(\bm{k}) \doteq \frac{\mathsf{n}_{\sigma}(\bm{k})}{1-\mathsf{n}_{\sigma}(\bm{k})} \ge 0. \label{e45}
\end{equation}
In view of the condition in Eq.~(\ref{e42}), one observes that $\varepsilon_{\bm{k};\sigma}^{<}$ and $\varepsilon_{\bm{k};\sigma}^{>}$ indeed take the same value for $\bm{k} = \bm{k}_{\textsc{f};\sigma}$ and that this value conforms with the result in Eq.~(\ref{e223}). Note that, since by assumption $\mathsf{n}_{\sigma}(\bm{k}) \not= 1$ \cite{Note1}, the combination $\Lambda_{\sigma}(\bm{k}) \hspace{0.7pt} \zeta_{\bm{k};\sigma}$ in the above expression for $\varepsilon_{\bm{k};\sigma}^{>}$ does not impose a stricter condition on $\zeta_{\bm{k};\sigma}$ than merely $\zeta_{\bm{k};\sigma}\sim 0$ for $\bm{k} \to \bm{k}_{\textsc{f};\sigma}$. For later use, we point out that following the definition in Eq.~(\ref{e45}) one has
\begin{equation}
\mathsf{n}_{\sigma}(\bm{k}) = \frac{\Lambda_{\sigma}(\bm{k})}{1+\Lambda_{\sigma}(\bm{k})}.
\label{e46}
\end{equation}

By defining
\begin{equation}
\Gamma_{\sigma}(\bm{k}) \doteq \frac{\mu - \varepsilon_{\bm{k};\sigma}^{<}}{\varepsilon_{\bm{k};\sigma}^{>}-\mu},
\label{e47}
\end{equation}
on the basis of the expressions in Eqs.~(\ref{e240}) and (\ref{e284}) one verifies that \cite{BF03}
\begin{equation}
\mathcal{S}_{\textsc{f};\sigma} = \big\{ \bm{k}\,\|\, \Lambda_{\sigma}(\bm{k}) = \Gamma_{\sigma}(\bm{k}) \big\}. \label{e48}
\end{equation}
Expanding $\varepsilon_{\bm{k}}$ around $\bm{k} = \bm{k}_{\textsc{f};\sigma} \in \mathcal{S}_{\textsc{f};\sigma}$, from the results in Eqs.~(\ref{e223}), (\ref{e43}) and (\ref{e44}) one deduces that
\begin{equation}
\Gamma_{\sigma}(\bm{k}) \sim \frac{-\bm{a}_{\sigma} \cdot (\bm{k}-\bm{k}_{\textsc{f};\sigma}) - U \zeta_{\bm{k};\sigma}}{\bm{a}_{\sigma} \cdot (\bm{k}-\bm{k}_{\textsc{f};\sigma}) - U \Lambda_{\sigma}(\bm{k})\, \zeta_{\bm{k};\sigma}}\;\;\text{for}\;\; \bm{k} \to \bm{k}_{\textsc{f};\sigma}, \label{e49}
\end{equation}
where
\begin{equation}
\bm{a}_{\sigma} \equiv \bm{a}(\bm{k}_{\textsc{f};\sigma}) \doteq \left. \bm{\nabla} \varepsilon_{\bm{k}}\right|_{\bm{k} = \bm{k}_{\textsc{f};\sigma}}.
\label{e410}
\end{equation}

% 4.a.
\subsection{Case 1}
\label{s4a}
An interesting case corresponds to $\vert\zeta_{\bm{k};\sigma}\vert$ approaching zero faster than $\|\bm{k}-\bm{k}_{\textsc{f};\sigma}\|$ for $\bm{k} \to \bm{k}_{\textsc{f};\sigma}^{\pm}$, as is the case when, for instance \cite{BF03,BF04a}
\begin{equation}
\vert\zeta_{\bm{k};\sigma}\vert \sim \vert b_{\sigma}(\bm{k}_{\textsc{f};\sigma}^{\pm})\vert\, \|\bm{k}-\bm{k}_{\textsc{f};\sigma}\|^{\gamma^{\pm}}\;\; \text{with}\;\; 0 < \gamma^{\pm} <1,
\label{e411}
\end{equation}
where the superscript $-$ ($+$) corresponds to $\bm{k}$ approaching $\mathcal{S}_{\textsc{f};\sigma}$ from inside (outside) the Fermi sea. In such and similar cases \cite{BF03,BF04a}, to leading order the expression in Eq.~(\ref{e49}) takes the following simplified form:
\begin{equation}
\Gamma_{\sigma}(\bm{k}) \sim \frac{1}{\Lambda_{\sigma}(\bm{k})}\;\; \text{for}\;\; \bm{k} \to \bm{k}_{\textsc{f};\sigma}.
\label{e412}
\end{equation}
From this and the expression in Eq.~(\ref{e48}) one obtains $\Lambda_{\sigma}^2(\bm{k}) \sim 1$ for $\bm{k} \to \bm{k}_{\textsc{f};\sigma}$, which, on account of $\Lambda_{\sigma}(\bm{k}) \ge 0$, Eq.~(\ref{e45}), can be written as
\begin{equation}
\Lambda_{\sigma}(\bm{k}) \sim 1 \;\; \text{for}\;\; \bm{k} \to \bm{k}_{\textsc{f};\sigma}.
\label{e413}
\end{equation}
In view of the equality in Eq.~(\ref{e46}), this implies that
\begin{equation}
\mathsf{n}_{\sigma}(\bm{k}) \sim \frac{1}{2} \;\; \text{for}\;\; \bm{k} \to \bm{k}_{\textsc{f};\sigma}. \label{e414}
\end{equation}
Since the validity of this result is independent of whether $\bm{k}$ approaches $\bm{k}_{\textsc{f};\sigma}$ from inside or outside the underlying Fermi sea, it follows that in the case at hand $\mathsf{n}_{\sigma}(\bm{k})$ is \textsl{continuous} at $\bm{k} = \bm{k}_{\textsc{f};\sigma}$ and takes the value $1/2$ at $\bm{k} = \bm{k}_{\textsc{f};\sigma}$. Thus, following the Migdal theorem, Eq.~(\ref{e255}), in the present case \cite{BF03,BF04a}
\begin{equation}
Z_{\bm{k}_{\textsc{f};\sigma}} = 0.
\label{e415}
\end{equation}

We note that in the case of the one-dimensional Tomonaga-Luttinger model for spin-less fermions, one has \cite{ML65,JV94}
\begin{equation}
\mathsf{n}_r(k) \sim \frac{1}{2} - C \sgn(k- r k_{\textsc{f}})\, \vert k - r k_{\textsc{f}}\vert^{\alpha}\;\; \text{for}\;\; k \to r k_{\textsc{f}},
\label{e416}
\end{equation}
where $C >0$ is a dimensional constant and $\alpha \in (0,1)$ \cite{ML65}. Here $\mathsf{n}_r(k)$ stands for the GS momentum-distribution function corresponding to the $r$-moving fermions, where $r=+$ signifies `right' and $r=-$, `left'. The expression in Eq.~(\ref{e414}) is seen to be in conformity with that in Eq.~(\ref{e416}). In addition, following the asymptotic expression in Eq.~(\ref{e416}), for $\Lambda_{\sigma}(\bm{k})$, Eq.~(\ref{e45}), one obtains
\begin{equation}
\Lambda_r(k) \sim 1 - 4 C \sgn(k- r k_{\textsc{f}})\, \vert k- r k_{\textsc{f}}\vert^{\alpha}\;\;\text{for}\;\; k \to r k_{\textsc{f}},
\label{e417}
\end{equation}
which is in conformity with the leading-order asymptotic result in Eq.~(\ref{e413}). Since the equality of $\Lambda_{\sigma}(\bm{k})$ with $\Gamma_{\sigma}(\bm{k})$ applies only for $\bm{k}$ \textsl{on}  $\mathcal{S}_{\textsc{f};\sigma}$, Eq.~(\ref{e48}), it is not possible to infer the next-to-leading-order term in the asymptotic series expansion \cite{WW62,ETC65,HAL74} of $\Gamma_r(k)$ corresponding to $k \to r k_{\textsc{f}}$ from the expression in Eq.~(\ref{e417}). This term can however be deduced from the expression in Eq.~(\ref{e49}). In the cases where $\alpha + \gamma^{\pm} < 1$, from this expression one trivially obtains that
\begin{equation}
\Gamma_r(k) \sim 1 + 4 C \sgn(k- r k_{\textsc{f}})\, \vert k - r k_{\textsc{f}}\vert^{\alpha}\;\;\text{for}\;\; k \to r k_{\textsc{f}}.
\label{e418}
\end{equation}
This and the expression in Eq.~(\ref{e417}) reveal the way in which in the case at hand $\Lambda_r(k)$ and $\Gamma_r(k)$ approach the common value of $1$ at $k= r k_{\textsc{f}}$ (cf. Eq.~(\ref{e48})) for $k$ approaching $r k_{\textsc{f}}$.

Under the conditions for which the asymptotic expression in Eq.~(\ref{e412}) applies (e.g. for $0 <\gamma^{\pm} < 1$ -- see Refs.~\cite{BF03,BF04a}), for $\bm{k}$ sufficiently close to $\mathcal{S}_{\textsc{f};\sigma}$ the expression in Eq.~(\ref{e280}) can be presented as
\begin{equation}
\mathsf{n}_{\sigma}(\bm{k}) \gtrless \frac{1}{1+ \Lambda_{\sigma}(\bm{k})} \sim \frac{1}{2} + \frac{1-\Lambda_{\sigma}(\bm{k})}{4}\;\; \text{for}\;\; \bm{k} \in \left\{\begin{array}{l}  \mathrm{FS}_{\sigma}, \\ \\ \ol{\mathrm{FS}}_{\sigma}. \end{array} \right.
\label{e419}
\end{equation}
Specialising to the case of the Tomonaga-Luttinger model for spin-less fermions, making use of the expression in Eq.~(\ref{e417}), for $k$ sufficiently close to $r k_{\textsc{f}}$ the result in Eq.~(\ref{e419}) can be written as
\begin{equation}
\mathsf{n}_r(k) \gtrless \frac{1}{2} \mp C\, \vert k - r k_{\textsc{f}}\vert^{\alpha}\;\;\text{for}\;\; k \lessgtr r k_{\textsc{f}}.
\label{e420}
\end{equation}
Since $C>0$,  in the light of the asymptotic expression in Eq.~(\ref{e416}) one observes that the inequalities in Eq.~(\ref{e420}) amount to true statements. Hereby is, insofar as the one-dimensional Tomonaga-Luttinger model for spin-less fermions is concerned, the validity of the inequalities in Eq.~(\ref{e280}) explicitly demonstrated for a finite neighbourhood of $r k_{\textsc{f}}$, $r=+, -$.

% 4.b.
\subsection{Case 2}
\label{s4b}
Here we consider a case corresponding to a class of uniform metallic GSs to which the \textsl{conventional} Fermi-liquid state \cite{BF03,BF99} belongs. For other cases, we refer the reader to Refs.~\cite{BF03,BF04a}. This case corresponds to the asymptotic relationship (cf. Eq.~(\ref{e411})) \cite{BF03}
\begin{equation}
\zeta_{\bm{k};\sigma} \sim \bm{b}_{\sigma}^{\pm} \cdot (\bm{k}-\bm{k}_{\textsc{f};\sigma})\;\; \text{for}\;\; \bm{k} \to \bm{k}_{\textsc{f};\sigma} \in \mathcal{S}_{\textsc{f};\sigma},
\label{e421}
\end{equation}
where (cf. Eq.~(\ref{e410}))
\begin{equation}
\bm{b}_{\sigma}^{\pm} \equiv \bm{b}_{\sigma}(\bm{k}_{\textsc{f};\sigma}^{\pm}) \doteq \left. \bm{\nabla} \zeta_{\bm{k};\sigma}\right|_{\bm{k}=\bm{k}_{\textsc{f};\sigma}^{\pm}},
\label{e422}
\end{equation}
in which the superscript $-$ ($+$) signifies the \textsl{limit} of $\bm{\nabla} \zeta_{\bm{k};\sigma}$ for $\bm{k}$ approaching $\mathcal{S}_{\textsc{f};\sigma}$ from inside (outside) the underlying Fermi sea. For the reasons that we have indicated in Sec.~\ref{s2d}, $\bm{b}_{\sigma}^{-}$ and $\bm{b}_{\sigma}^{+}$ cannot be equal, a fact best illustrated by the inequalities in Eq.~(\ref{e428}) below.

For the following considerations it will be convenient to introduce, for a given $\bm{k}_{\textsc{f};\sigma}$, the \textsl{outward} unit vector $\h{\bm{n}}(\bm{k}_{\textsc{f};\sigma})$ normal to $\mathcal{S}_{\textsc{f};\sigma}$ at $\bm{k}_{\textsc{f};\sigma}$ \cite{Note8}, and without loss of generality assume that vector $\bm{k}$ is defined according to
\begin{equation}
\bm{k} = \bm{k}_{\textsc{f};\sigma} + \kappa\, \h{\bm{n}}(\bm{k}_{\textsc{f};\sigma}),
\label{e423}
\end{equation}
where, for sufficiently small $\vert\kappa\vert$, $\kappa <0$ ($\kappa >0$) corresponds to $\bm{k}$ located inside (outside) the underlying Fermi sea. For definiteness, the adjective \textsl{outward}, referred to above, signifies that $\h{\bm{n}}(\bm{k}_{\textsc{f};\sigma})$ points in the direction \textsl{away} from the Fermi sea at $\bm{k}_{\textsc{f};\sigma}$.

For $\bm{k}$ as defined according to the equality in Eq.~(\ref{e423}), and $\kappa \to 0$, the expression in Eq.~(\ref{e49}) takes the following simplified form:
\begin{equation}
\Gamma_{\sigma}(\bm{k}) \sim \frac{-(a_{\sigma} + U b_{\sigma}^{\pm})}{a_{\sigma} - U \Lambda_{\sigma}^{\pm}\, b_{\sigma}^{\pm}}\;\; \text{for}\;\; \kappa \gtrless 0,
\label{e424}
\end{equation}
where
\begin{equation}
a_{\sigma} \doteq \bm{a}_{\sigma} \cdot \h{\bm{n}}(\bm{k}_{\textsc{f};\sigma}),\;\; b_{\sigma}^{\pm} \doteq \bm{b}_{\sigma}^{\pm} \cdot \h{\bm{n}}(\bm{k}_{\textsc{f};\sigma}),
\label{e425}
\end{equation}
and
\begin{equation}
\Lambda_{\sigma}^{\pm} \doteq \Lambda_{\sigma}(\bm{k}_{\textsc{f};\sigma}^{\pm}).
\label{e426}
\end{equation}
With reference to the defining expression in Eq.~(\ref{e47}), we note that the numerator of the expression on the RHS of Eq.~(\ref{e424}) arises from $\mu-\varepsilon_{\bm{k};\sigma}^{<}$ and the denominator from $\varepsilon_{\bm{k};\sigma}^{>}-\mu$. More explicitly, we have \textsl{not} multiplied the expression in Eq.~(\ref{e47}) by $\frac{-1}{-1}$ in arriving at the expression in Eq.~(\ref{e424}). In view of the inequalities in Eq.~(\ref{e217}), $b_{\sigma}^{-}$ and $b_{\sigma}^{+}$ are thus bound to satisfy the following inequalities \cite{BF03}:
\begin{equation}
b_{\sigma}^- > \frac{a_{\sigma}}{U \Lambda_{\sigma}^-},\;\;\; b_{\sigma}^+ < -\frac{a_{\sigma}}{U}. \label{e427}
\end{equation}
Since $\Lambda_{\sigma}^- \ge 0$, Eq.~(\ref{e45}), and $a_{\sigma} \ge 0$ (as a consequence of $\h{\bm{n}}(\bm{k}_{\textsc{f};\sigma})$ being the \textsl{outward} unit vector normal to $\mathcal{S}_{\textsc{f};\sigma}$ at $\bm{k}_{\textsc{f};\sigma}$, Eq.~(\ref{e425})), it follows that for $U >0$, which we consider in this paper, one must have
\begin{equation}
b_{\sigma}^- > 0,\;\;\; b_{\sigma}^{+} < 0.
\label{e428}
\end{equation}
Further, since $\mathsf{n}_{\sigma}(\bm{k}_{\textsc{f};\sigma}^-) \uparrow 1$ and $\mathsf{n}_{\sigma}(\bm{k}_{\textsc{f};\sigma}^+) \downarrow 0$ for $U/\vert t\vert \to 0$ (cf. Eq.~(\ref{e343})), from the expression in Eq.~(\ref{e45}) one infers that $\Lambda_{\sigma}^{-} \uparrow +\infty$ and $\Lambda_{\sigma}^{+} \downarrow 0$ for $U/\vert t\vert \to 0$. If to leading order $\Lambda_{\sigma}^-$ diverges like $\vert t\vert/U$ for $U/\vert t\vert \to 0$, then the $1/(U \Lambda_{\sigma}^-)$ in Eq.~(\ref{e427}) approaches a \textsl{finite} constant for $U/\vert t\vert \to 0$. The significance of this observation lies in the fact that unlike $b_{\sigma}^+$, which, following the right-most inequality in Eq.~(\ref{e427}), has to decrease indefinitely towards $-\infty$ for $U/\vert t\vert \to 0$, $b_{\sigma}^-$ is \textsl{not} bound necessarily to increase indefinitely towards $+\infty$ for $U/\vert t\vert \to 0$. We shall return to behaviours of $b_{\sigma}^{\pm}$ as functions of $U/\vert t\vert$ in Sec.~\ref{s4b1} below.

For $\bm{k} \in \mathcal{S}_{\textsc{f};\sigma}$, making use of the expression in Eq.~(\ref{e424}), from the equality in Eq.~(\ref{e48}) one obtains the following quadratic equation for $\Lambda_{\sigma}^{\pm}$ \cite{BF03}:
\begin{equation}
U b_{\sigma}^{\pm} (\Lambda_{\sigma}^{\pm})^2 - a_{\sigma} \Lambda_{\sigma}^{\pm} - (a_{\sigma} + U b_{\sigma}^{\pm})  = 0.
\label{e429}
\end{equation}
Taking into account that $\Lambda_{\sigma}^{\pm} \ge 0$, Eq.~(\ref{e45}), from the above equation one obtains
\begin{equation}
\Lambda_{\sigma}^{\pm} = 1 + \frac{a_{\sigma}}{U b_{\sigma}^{\pm}} \lessgtr 1,
\label{e430}
\end{equation}
where the inequalities follow from those in Eq.~(\ref{e428}). On account of the results in Eq.~(\ref{e430}), the expression in Eq.~(\ref{e46}) can be written as \cite{BF03}
\begin{equation}
\mathsf{n}_{\sigma}(\bm{k}_{\textsc{f};\sigma}^{\pm}) = \frac{a_{\sigma} + U b_{\sigma}^{\pm}}{a_{\sigma} + 2 U b_{\sigma}^{\pm}} \lessgtr \frac{1}{2},
\label{e431}
\end{equation}
where the inequalities follow from those in Eq.~(\ref{e427}). From the above equality one immediately infers that
\begin{equation}
\mathsf{n}_{\sigma}(\bm{k}_{\textsc{f};\sigma}^{-}) = \mathsf{n}_{\sigma}(\bm{k}_{\textsc{f};\sigma}^{+}) = \frac{1}{2}\;\; \text{for}\;\; a_{\sigma} =0,
\label{e432}
\end{equation}
that is, $\mathsf{n}_{\sigma}(\bm{k})$ is continuous, and takes the value of $1/2$, at those points of $\mathcal{S}_{\textsc{f};\sigma}$ that coincide with the van Hove points of the non-interacting energy dispersion $\varepsilon_{\bm{k}}$. With reference to the Migdal theorem, Eq.~(\ref{e255}), at these points $Z_{\bm{k}_{\textsc{f};\sigma}} = 0$ \cite{BF03}, in conformity with the observations by other authors \cite{IED96,GGV96-7,KK04}.

\begin{figure}[t!]
\includegraphics[angle=0, width=0.43\textwidth]{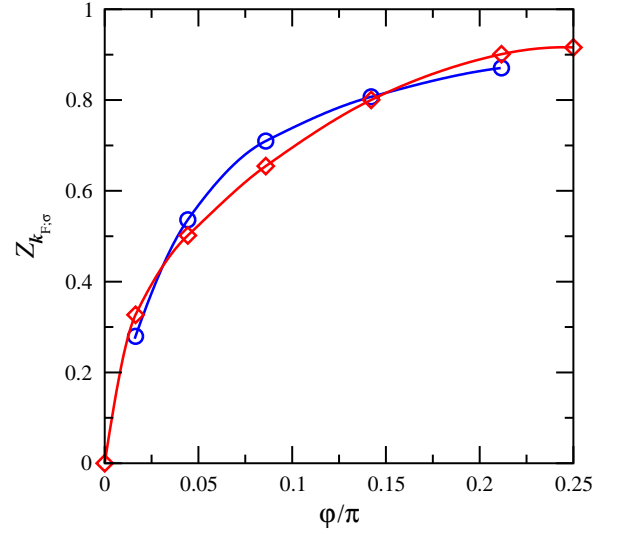}
\caption{(Colour) $Z_{\bm{k}_{\textsc{f};\sigma}}$ versus $\upvarphi \doteq \widehat{\bm{k}_{\textsc{f};\sigma}, (\pi,0)}$ as calculated by Katanin and Kampf \protect\cite{KK04} ($\circ$) and on the basis of the expression in Eq.~(\protect\ref{e255}) in conjunction with the expressions in Eq.~(\protect\ref{e431}), using $b_{\sigma}^-=0.0912$, $b_{\sigma}^+ = -1.4158$ ($\diamond$). The results correspond to $U/t=2$ and the van Hove filling associated with $t'/t=0.1$, i.e. $n=0.918\,022\,\dots$ \protect\cite{BF04c}. The point $\upvarphi =0$ corresponds to a van Hove point of the energy dispersion $\varepsilon_{\bm{k}}$ on the Fermi surface, where $\bm{a}(\bm{k}_{\textsc{f};\sigma}) = \bm{0}$ (cf. Eq.~(\protect\ref{e432})). The apparent difference between the two curves reflects the isotropy of the $b_{\sigma}^{\mp}$ employed in the calculations. Note that the values of these constants conform with the inequalities in Eq.~(\protect\ref{e428}). Further, with reference to the remarks following Eq.~(\protect\ref{e428}), it is notable that the ratio $b_{\sigma}^+/b_{\sigma}^-$ of these constants is of the order of $10$. }\label{f2}
\end{figure}

In Fig.~\ref{f2} we present the $Z_{\bm{k}_{\textsc{f};\sigma}}$ as calculated on the basis of the expressions in Eq.~(\ref{e431}) for the Hubbard model on the square lattice with lattice constant $a=1$ and corresponding to the energy dispersion
\begin{equation}
\varepsilon_{\bm{k}} = -2t \big(\hspace{-0.6pt}\cos(k_x) + \cos(k_y)\big) + 4 t' \cos(k_x) \cos(k_y),
\label{e433}
\end{equation}
where $k_x$ and $k_y$ are the Cartesian coordinates of $\bm{k}$. The data in Fig.~\ref{f2} are specific to $t=1$, $t'=t/10$ and $U/t=2$. The site occupation number $n=0.918\,022\,\dots$ to which the data in Fig.~\ref{f2} correspond, is the so-called van Hove filling, named thus by the fact that at this filling the saddle points of the energy dispersion in Eq.~(\ref{e433}) at $(\pm\pi,0)$ and $(0,\pm\pi)$ are located on $\mathcal{S}_{\textsc{f}}^{(0)} = \mathcal{S}_{\textsc{f};\sigma}^{\textsc{hf}}$. In Fig.~\ref{f2}, $\upvarphi$ denotes the angle between the positive $k_x$ axis and $\bm{k}_{\textsc{f};\sigma}$, so that $\upvarphi=0$ corresponds to the van Hove point at $(\pi,0)$. In Fig.~\ref{f2} we compare our results \cite{BF04c}, calculated on the basis of the expressions in Eq.~(\ref{e431}), with those calculated by Katanin and Kampf \cite{KK04} on the basis of a functional renormalization-group formalism. Since we have not calculated $b_{\sigma}^{\pm}$ independently, this comparison only tests the validity of the functional forms for $\mathsf{n}_{\sigma}(\bm{k}_{\textsc{f};\sigma}^{\pm})$ as presented in Eq.~(\ref{e431}). The apparent deviation between the two sets of data in Fig.~\ref{f2} is essentially attributable to the dependence of $b_{\sigma}^{\pm}$ on $\upvarphi$ that has \textsl{not} been taken into account in our calculations.

% 4.b.1.
\subsubsection{\texorpdfstring{The asymptotic expressions for $\mathsf{n}_{\sigma}(\bm{k}_{\textsc{f};\sigma}^{\mp})$}{} corresponding to \texorpdfstring{$U/t\to 0$}{} revisited}
\label{s4b1}
Here we establish direct relationships between the coefficients $b_{\sigma}^{\pm}$, Eqs.~(\ref{e421}) and (\ref{e425}), in the asymptotic region $U/\vert t\vert \ll 1$ and the coefficients $\upalpha_{\sigma}^{\pm}(\bm{k}_{\textsc{f};\sigma})$ in the asymptotic expressions in Eq.~(\ref{e343}). By doing so, we gain insight into the behaviours to be expected of $b_{\sigma}^{\pm}$ as functions of $U$. \emph{Having already dealt with the condition $a_{\sigma} = 0$, Eq.~(\ref{e432}), unless we indicate otherwise, in the following $a_{\sigma} > 0$ \cite{Note8}.}

With reference to the first asymptotic expression in Eq.~(\ref{e343}), we posit that
\begin{equation}
b_{\sigma}^- \sim b_{\sigma;0}^- \hspace{0.7pt}\Big(\frac{U}{\vert t\vert}\Big)^{\alpha-1}\ln^{\gamma}\hspace{-0.1cm}\Big(\frac{\vert t\vert}{U}\Big)\;\; \text{for}\;\; \frac{U}{t} \to 0,
\label{e434}
\end{equation}
where $b_{\sigma;0}^-$ is a finite positive constant (see the discussions following Eq.~(\ref{e428}) above and note that according to the left-most inequality in the latter equation, $b_{\sigma}^-$ is strictly positive, and so must be $b_{\sigma;0}^-$). From the expression in Eq.~(\ref{e431}) one obtains
\begin{equation}
\mathsf{n}_{\sigma}(\bm{k}_{\sigma}^-) \sim 1 - \frac{b_{\sigma;0}^-  \vert t\vert}{a_{\sigma}}\hspace{0.7pt} \Big(\frac{U}{\vert t\vert}\Big)^{\alpha}\ln^{\gamma}\hspace{-0.1cm}\Big(\frac{\vert t\vert}{U}\Big) \;\;\text{as}\;\; \frac{U}{t}\to 0.
\label{e435}
\end{equation}
Comparing this result with the first expression in Eq.~(\ref{e343}), one infers that
\begin{equation}
\upalpha_{\sigma}^- \equiv \frac{\vert t\vert}{a_{\sigma}}\hspace{0.7pt} b_{\sigma;0}^-,
\label{e436}
\end{equation}
where $\upalpha_{\sigma}^- \equiv \upalpha_{\sigma}^-(\bm{k}_{\textsc{f};\sigma})$. For the ratio $\vert t\vert/a_{\sigma}$, the reader is referred to Eqs.~(\ref{e410}) and (\ref{e425}), and for the energy dispersion $\varepsilon_{\bm{k}}$ in the former equation, to Eq.~(\ref{e433}). Since $\upalpha_{\sigma}^-$ is finite, the result in Eq.~(\ref{e436}) amounts to an \textsl{a posteriori} justification for the above assumption regarding the finiteness of $b_{\sigma;0}^-$.

On account of the remarks with regard to the behaviours of $\mathsf{n}_{\sigma}(\bm{k}_{\textsc{f};\sigma}^+)$ and $b_{\sigma}^+$ following Eq.~(\ref{e428}) above, we express $b_{\sigma}^+$ as follows:
\begin{equation}
b_{\sigma}^+ \equiv -\frac{a_{\sigma}}{U} - \delta b_{\sigma}^+,
\label{e437}
\end{equation}
where, in the light of the second asymptotic expression in Eq.~(\ref{e343}) (cf. Eq.~(\ref{e434})),
\begin{equation}
\delta b_{\sigma}^+ \sim \delta  b_{\sigma;0}^+\hspace{0.7pt}\Big(\frac{U}{\vert t\vert}\Big)^{\alpha-1}\ln^{\gamma}\hspace{-0.1cm}\Big(\frac{\vert t\vert}{U}\Big)\;\; \text{for}\;\; \frac{U}{t} \to 0,
\label{e438}
\end{equation}
in which $\delta b_{\sigma;0}^+$ is a finite positive constant. Making use of the expression in Eq.~(\ref{e431}), one readily obtains that
\begin{equation}
\mathsf{n}_{\sigma}(\bm{k}_{\textsc{f};\sigma}^+) \sim \frac{\delta b_{\sigma;0}^+ \vert t\vert}{a_{\sigma}}\hspace{0.7pt}\Big(\frac{U}{\vert t\vert}\Big)^{\alpha}\ln^{\gamma}\hspace{-0.1cm}\Big(\frac{\vert t\vert}{U}\Big)
\;\; \text{as}\;\; \frac{U}{t} \to 0.
\label{e439}
\end{equation}
Comparing this result with the second expression in Eq.~(\ref{e343}), one infers that (cf. Eq.~(\ref{e436}))
\begin{equation}
\upalpha_{\sigma}^+ \equiv \frac{\vert t\vert}{a_{\sigma}}\hspace{0.7pt}\delta b_{\sigma;0}^+,
\label{e440}
\end{equation}
where $\upalpha_{\sigma}^+ \equiv \upalpha_{\sigma}^+(\bm{k}_{\textsc{f};\sigma})$. Since $\upalpha_{\sigma}^+$ is finite, the result in Eq.~(\ref{e440}) amounts to an \emph{a posteriori} justification of the above assumption regarding the finiteness of $\delta  b_{\sigma;0}^+$. Note that following the expressions in Eqs.~(\ref{e436}) and (\ref{e440}), $\upalpha_{\sigma}^- = \upalpha_{\sigma}^+$ implies $b_{\sigma;0}^- = \delta b_{\sigma;0}^+$.

% 5.
\section{On ferromagnetism}
\label{s5}
The existence or lack of existence of ferromagnetic regions in the zero-temperature phase diagram of the Hubbard Hamiltonian is a long-standing theoretical question, for an overview of which we refer the reader to chapter 8 of Ref.~\cite{PF03}. Restricting oneself to Hubbard Hamiltonians defined on bipartite lattices, the established trend turns out to be that the more accurately correlation effects are taken account of, the smaller the ferromagnetic regions of the phase diagram become. Particularly, the Lieb-Mattis theorem \cite{LM62,EMH95} rules out ferromagnetic GSs for $d=1$, $\varepsilon_k = -2 t \cos(k)$ and $U <\infty$. Monte-Carlo simulations by Hirsch \cite{JEH85} for the Hubbard Hamiltonian on a square lattice and involving nearest-neighbour hopping of non-interacting particles, reveal that for at least $0 \le U/t\le 10$ and away from half-filling the GS of the system is a paramagnetic metal, an observation that accords with that by Rudin and Mattis \cite{RM85} in a subsequent study. Here we show that the exact property $\mathcal{S}_{\textsc{f};\sigma} \subseteq \mathcal{S}_{\textsc{f};\sigma}^{\textsc{hf}}$, Eq.~(\ref{e241}), amounts to a kinematic constraint that in particular in the case of the Hubbard Hamiltonian, and barring a limited range of values of $U$ \cite{BF03}, renders ferromagnetic uniform metallic states as unviable GSs; even in the just-mentioned limited range, which turns out to shrink for decreasing values of $n$, the GS need not be ferromagnetic, however.

For $n_{\sigma}, n_{\b\sigma} >0$, following the definition in Eq.~(\ref{e226}) and the relationship in Eq.~(\ref{e241}), one has \cite[\S8]{BF03}
\begin{equation}
\varepsilon_{\bm{k}_{\textsc{f};\sigma}} + \hbar \Sigma_{\sigma}^{\textsc{hf}}(\bm{k}_{\textsc{f};\sigma}) = \varepsilon_{\bm{k}_{\textsc{f};\b\sigma}} + \hbar \Sigma_{\b\sigma}^{\textsc{hf}}(\bm{k}_{\textsc{f};\b\sigma}),
\label{e51}
\end{equation}
which in the case of the Hubbard Hamiltonian, for which the Hartree-Fock self-energy $\Sigma_{\sigma}^{\textsc{hf}}(\bm{k})$ has the form presented in Eq.~(\ref{e224}), can be expressed as \cite[Fig.~2]{BF03}
\begin{equation}
n_{\sigma} - n_{\b\sigma} = \frac{1}{U} (\varepsilon_{\bm{k}_{\textsc{f};\sigma}} -\varepsilon_{\bm{k}_{\textsc{f};\b\sigma}}).
\label{e52}
\end{equation}
Clearly, this equality is identically satisfied for $n_{\sigma} = n_{\b\sigma}$. In contrast, it amounts to a restriction on the realization of metallic GSs with $n_{\sigma} \not= n_{\b\sigma}$ for a predetermined value of $n \equiv n_{\sigma} + n_{\b\sigma}$. The condition $n_{\sigma} \not= n_{\b\sigma}$ with $n_{\sigma}, n_{\b\sigma} >0$ corresponds to \textsl{partial} ferromagnetism, and, with either $n_{\sigma}$ or $n_{\b\sigma}$ vanishing, to \textsl{saturated} ferromagnetism. Evidently, the expressions in Eqs.~(\ref{e51}) and (\ref{e52}) become meaningless for saturated ferromagnetic GSs, for which one of the two sets $\mathcal{S}_{\textsc{f};\sigma}$ and $\mathcal{S}_{\textsc{f};\b\sigma}$ is empty.

\begin{figure}[t!]
\includegraphics[angle=0, width=0.43\textwidth]{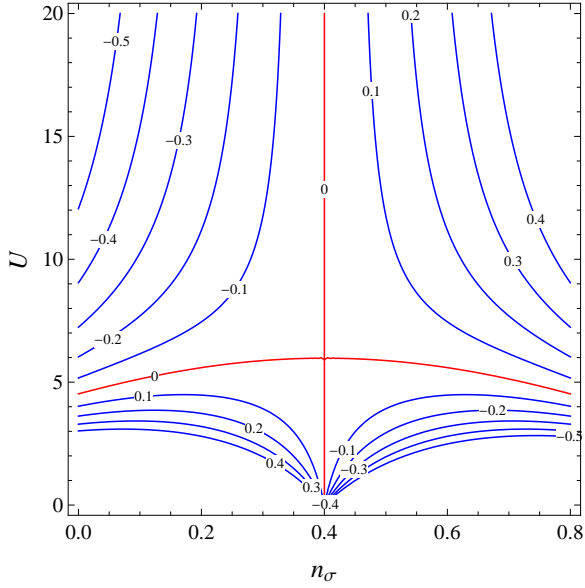}
\caption{(Colour) Contour plot of $(n_{\sigma}-n_{\b\sigma}) + 2t [\cos(\pi n_{\sigma}) - \cos(\pi n_{\b\sigma})]/U$, Eq.~(\protect\ref{e55}), in the $n_{\sigma}$-$U$ plane for $n=0.8$ and $t=1$. Whereas uniform paramagnetic metallic GSs, characterized by $n_{\sigma}=n_{\b\sigma} = n/2$, are permitted for \textsl{all} $U \ge 0$, a partially ferromagnetic uniform metallic GS is potentially feasible only for a restricted range of values of $U$ (here, for $4.5225 \lesssim U  \lesssim 5.9757$). The width of this window of $U$ decreases for decreasing values of $n$.} \label{f3}
\end{figure}

To examine the effectiveness of the kinematic constraint in Eq.~(\ref{e52}) in preventing ferromagnetic uniform metallic GSs from being realized, we consider the Hubbard Hamiltonian in $d=1$ \cite{LW68-03}, defined on a regular lattice with lattice constant $a=1$ and in terms of the tight-binding energy dispersion
\begin{equation}
\varepsilon_k = -2 t \cos(k).
\label{e53}
\end{equation}
For this energy dispersion, taking into account the relationship in Eq.~(\ref{e241}), one has the following equality, which, since $\textrm{1BZ} = [-\pi,\pi)$, applies for all physical band fillings:
\begin{equation}
k_{\textsc{f};\sigma} = \pi\hspace{0.15pt} n_{\sigma}.
\label{e54}
\end{equation}
Making use of this equality, that in Eq.~(\ref{e52}) transforms into the following explicit relationship between $n_{\sigma}$ and $n_{\b\sigma}$ \cite[\S8]{BF03}:
\begin{equation}
n_{\sigma} - n_{\b\sigma} = -\frac{2 t}{U} \big(\!\cos(\pi\hspace{0.15pt} n_{\sigma}) - \cos(\pi\hspace{0.15pt} n_{\b\sigma})\big),
\label{e55}
\end{equation}
in which $\{ n_{\sigma}, n_{\b\sigma}\}$ are constrained by $n_{\sigma}, n_{\b\sigma} > 0$ (see the previous paragraph) \textsl{and} $n_{\sigma} + n_{\b\sigma} = n$ for a prescribed value of $n$. In Fig.~\ref{f3} we present the contour plot of the deviation of the LHS of the equality in Eq.~(\ref{e55}) from its RHS in the $n_{\sigma}$-$U$ plane for $n=0.8$. The trivial zero at $n_{\sigma} = n/2 = 0.4$ of this difference for \textsl{all} $U$ is apparent. The interesting information revealed by this contour plot is that this difference can become zero for $n_{\sigma} \not= n/2$ \textsl{only} for $U$ inside a finite interval. As we have indicated above, whether for $U$ in the latter interval the metallic GS is ferromagnetic or paramagnetic, is a question that is to be settled on the basis of energetic considerations. Following the Lieb-Mattis theorem \cite{LM62,EMH95}, no ferromagnetic GS is realized for $U$ in this interval; outside this interval, and barring saturated ferromagnetic metallic GSs, the kinematic constraint in Eq.~(\ref{e55}) leaves no room for uniform ferromagnetic metallic GSs to be viable.

We have carried out similar calculations as the one described above for $d=2$ and the non-interacting energy dispersion presented in Eq.~(\ref{e433}), and observed a similar trend as corresponding to $d=1$. Importantly, the limited range of values of $U$ over which the equality in Eq.~(\ref{e52}) is satisfied for $n_{\sigma} \not= n_{\b\sigma}$ remains; the only difference with the case of $d=1$ turns out to be the increase in the different combinations of unequal $n_{\sigma}$ and $n_{\b\sigma}$, subject to $n_{\sigma} + n_{\b\sigma} = n$, for which the equation in Eq.~(\ref{e52}) is satisfied in the latter range of values of $U$.

% 6.
\section{On the excited states}
\label{s6}
For a detailed discussion of the relationship between in particular $\varepsilon_{\bm{k};\sigma}^{<}$ and the single-particle excitation energy dispersions \textsl{as measured} by means of the angle-resolved photoemission spectroscopy (ARPES) \cite{DHS03}, we refer the reader to Ref.~\cite{BF04b}. Here we mainly focus on a central element of the latter work (see also Ref.~\cite[\S6.4]{BF07a}).

The functions (cf. Eq.~(\ref{e251}))
\begin{equation}
P_{\sigma}^{<}(\bm{k};\varepsilon) \doteq \frac{1}{\hbar} \frac{A_{\sigma}(\bm{k};\varepsilon)}{\mathsf{n}_{\sigma}(\bm{k})},\;\; P_{\sigma}^{>}(\bm{k};\varepsilon) \doteq \frac{1}{\hbar} \frac{A_{\sigma}(\bm{k};\varepsilon)}{1-\mathsf{n}_{\sigma}(\bm{k})},
\label{e61}
\end{equation}
are, for all $\bm{k}$, normalized probability distribution functions for respectively $\varepsilon <\mu$ and $\varepsilon >\mu$: they are non-negative, and yield $1$, for all $\bm{k}$, on being integrated with respect to $\varepsilon$ over respectively $(-\infty,\mu]$ and $[\mu,+\infty)$, Eqs.~(\ref{e249}) and (\ref{e250}). For definiteness, although $A_{\sigma}(\bm{k};\varepsilon)$ is defined for all $\varepsilon \in (-\infty,+\infty)$, \emph{unless we indicate otherwise, in what follows we consider $P_{\sigma}^{<}(\bm{k};\varepsilon)$ and $P_{\sigma}^{>}(\bm{k};\varepsilon)$ as being defined over respectively $(-\infty,\mu]$ and $[\mu,+\infty)$.}

On account of $P_{\sigma}^{<}(\bm{k};\varepsilon)$ and $P_{\sigma}^{>}(\bm{k};\varepsilon)$ being probability distribution functions, and in view of the expressions in Eq.~(\ref{e251}), for an arbitrary $\bm{k} \in \textrm{1BZ}$, $\varepsilon_{\bm{k};\sigma}^{<}$ and $\varepsilon_{\bm{k};\sigma}^{>}$ are the \textsl{mean values} of $\varepsilon$ as distributed according to respectively $P_{\sigma}^{<}(\bm{k};\varepsilon)$ and $P_{\sigma}^{>}(\bm{k};\varepsilon)$. It follows that in the regions of the $\bm{k}$ space where  $P_{\sigma}^{<}(\bm{k};\varepsilon)$ ($P_{\sigma}^{>}(\bm{k};\varepsilon)$) consists of a single dominant peak, $\varepsilon_{\bm{k};\sigma}^{<}$ ($\varepsilon_{\bm{k};\sigma}^{>}$) must to a good approximation coincide with the energy $\varepsilon$ at which $P_{\sigma}^{<}(\bm{k};\varepsilon)$ ($P_{\sigma}^{>}(\bm{k};\varepsilon)$), and thus $A_{\sigma}(\bm{k};\varepsilon)$, is dominantly peaked (see specifically Sec.~\ref{s2c1}). In other words, in the latter regions of the $\bm{k}$ space, there exists an intimate approximate relationship, a relationship that becomes exact in the non-interacting limit, between $\varepsilon_{\bm{k};\sigma}^{\lessgtr}$ and the dispersion of the single-particle excitation energies \cite{BF04b}.

With reference to the inequalities on the RHS of the $\Leftrightarrow$ in Eq.~(\ref{e253}) (see also Fig.~\ref{f1}), one can appreciate the properties $\varepsilon_{\bm{k};\sigma}^{<} \le \varepsilon_{\bm{k};\sigma}^{\textsc{hf}}$ for $\bm{k}$ inside the relevant Fermi sea and $\varepsilon_{\bm{k};\sigma}^{>} > \varepsilon_{\bm{k};\sigma}^{\textsc{hf}}$ for $\bm{k}$ outside, by realizing that for any non-vanishing coupling constant of interaction, $P_{\sigma}^{\lessgtr}(\bm{k};\varepsilon)$ are \textsl{not} symmetric functions of $\varepsilon$ with respect to $\varepsilon_{\bm{k};\sigma}^{\lessgtr}$, each distribution function being comprised of a high-energy tail (and, in general, of some non-negligible satellite peaks in the early parts of the tail \cite{HL69,HL67}), which in the case of the Hubbard Hamiltonian sets in to decay at least exponentially \cite[\S\hspace{0.0pt}B.3]{BF07a} for $\vert \varepsilon-\mu\vert$ in excess of the largest energy parameter of the Hamiltonian; these properties cause the energy $\varepsilon_{\bm{k};\sigma}^{<}$ for $\bm{k}$ inside the Fermi sea, and the energy $\varepsilon_{\bm{k};\sigma}^{>}$ for $\bm{k}$ outside, to be displaced towards higher \textsl{binding energies} in comparison with the single-particle excitation energy $\varepsilon_{\bm{k};\sigma}^{\textsc{hf}}$ within the framework of the exact Hartree-Fock theory, where the self-energy is independent of $\varepsilon$ and the associated single-particle spectral function is equal to $A_{\sigma}^{\textsc{hf}}(\bm{k};\varepsilon) \doteq \hbar\, \delta(\varepsilon - \varepsilon_{\bm{k};\sigma}^{\textsc{hf}})$.

On replacing the lower bound $-\infty$ and the upper bound $+\infty$ of the energy integrals, such as those in Eqs.~(\ref{e249}) and (\ref{e250}), by respectively $-E$ and $+E$, where $E$ is in principle an arbitrary energy parameter satisfying $-E < \mu < E$, one obtains a generalized momentum-distribution function $\mathsf{n}_{\sigma}(\bm{k};E)$, and its complementary part $\int_{-E}^{E} \rd\varepsilon\, A_{\sigma}(\bm{k};\varepsilon) - \mathsf{n}_{\sigma}(\bm{k};E)$, in terms of which one defines the normalized probability distribution functions $P_{\sigma}^{<}(\bm{k};E,\varepsilon)$ and $P_{\sigma}^{>}(\bm{k};E,\varepsilon)$, for $\varepsilon$ over  respectively $[-E,\mu]$ and $[\mu,E]$ according to expressions similar to those in Eq.~(\ref{e61}) \cite{BF04b}. These distribution functions define energy dispersions $\varepsilon_{\bm{k};\sigma}^{<}(E)$ and $\varepsilon_{\bm{k};\sigma}^{>}(E)$ (in the same manner that $P_{\sigma}^{<}(\bm{k};\varepsilon)$ and $P_{\sigma}^{>}(\bm{k};\varepsilon)$ define $\varepsilon_{\bm{k};\sigma}^{<}$ and $\varepsilon_{\bm{k};\sigma}^{>}$), and, for any \textsl{finite} $E$, one can calculate the expectation value of $(\varepsilon- \varepsilon_{\bm{k};\sigma}^{\lessgtr}(E))^2$ with respect to $P_{\sigma}^{\lessgtr}(\bm{k};E,\varepsilon)$, which, in the regions of $\bm{k}$ space where $P_{\sigma}^{<}(\bm{k};E,\varepsilon)$ ($P_{\sigma}^{>}(\bm{k};E,\varepsilon)$) consists of a single dominant peak over $[-E,\mu]$ ($[\mu,E]$), amounts to an approximate value for the width of this peak \cite{BF04b}. We point out that introduction of the finite energy cut-off $E$ is in general necessary both mathematically and experimentally; mathematically, because the second moments of $P_{\sigma}^{<}(\bm{k};E,\varepsilon)$ and $P_{\sigma}^{>}(\bm{k};E,\varepsilon)$ are in general unbounded for $E \to \infty$ (this is not the case as regards a lattice Hamiltonian such as the Hubbard Hamiltonian in Eq.~(\ref{e11}) in which both $\varepsilon_{\bm k}$ and the interaction potential are bounded over the bounded $\mathrm{1BZ}$, however it can well be the case for the more general Hamiltonian in Eq.~(\ref{ea1})) \cite{Note9}, and experimentally, because, firstly, $A_{\sigma}(\bm{k};\varepsilon)$ is in practice measured only over a finite energy range and, secondly, for a real system the single-band approximation of the underlying Hamiltonian fails when considering the single-particle spectral function over an unreasonably extended range of energies.

Considering only the ARPES (to be contrasted with the \textsl{inverse} photoemission spectroscopy, with which $\varepsilon_{\bm{k};\sigma}^{>}$ and $\varepsilon_{\bm{k};\sigma}^{>}(E)$ are associated) \cite{DHS03}, below we restrict ourselves to dealing with $\varepsilon_{\bm{k};\sigma}^{<}$ and $\varepsilon_{\bm{k};\sigma}^{<}(E)$.

On the basis of the above observations, in Ref.~\cite{BF04b} we have related the experimentally-observed `kink' in the single-particle energy dispersions and the abrupt change of the width of the peak in the measured $A_{\sigma}(\bm{k};\varepsilon)$ at the `kink' wave vector $\bm{k}_{\star}$ (see, e.g., Ref.~\cite{XJZ03}), to an abrupt change in $\mathsf{n}_{\sigma}(\bm{k})$, or $\mathsf{n}_{\sigma}(\bm{k};E)$, at $\bm{k} = \bm{k}_{\star}$. A moment's reflection reveals that for establishing this relationship, there is \textsl{no} need for identically equating either $\varepsilon_{\bm{k};\sigma}^{<}$, or $\varepsilon_{\bm{k};\sigma}^{<}(E)$, with the single-particle energy dispersion as inferred \textsl{directly} from the measured $A_{\sigma}(\bm{k};\varepsilon)$ \cite{BF04b}. By defining (cf. Eq.~(\ref{e255}))
\begin{equation}
Z_{\bm{k}_{\star};\sigma} \doteq \mathsf{n}_{\sigma}(\bm{k}_{\star}^-) - \mathsf{n}_{\sigma}(\bm{k}_{\star}^+),
\label{e62}
\end{equation}
we have for $\bm{v}_{\bm{k};\sigma}^{<} \doteq \frac{1}{\hbar} \bm{\nabla} \varepsilon_{\bm{k};\sigma}^{<}$ in Ref.~\cite{BF04b} (see Eq.~(67) and Fig.~2 herein) obtained
\begin{equation}
\bm{v}_{\bm{k}_{\star}^-;\sigma}^{<} = (1 + \lambda_{\star})\, \bm{v}_{\bm{k}_{\star}^+;\sigma}^{<}, \label{e63}
\end{equation}
where \cite[Eqs.~(64) and (68)]{BF04b}
\begin{equation}
\lambda_{\star} \approx \frac{Z_{\bm{k}_{\star};\sigma}}{\mathsf{n}_{\sigma}(\bm{k}_{\star}^-)} \le 1. \label{e64}
\end{equation}
The result $1 < 1 + \lambda_{\star} \lesssim 2$ is in full conformity with experimental observations (see, e.g., Ref.~\cite[Fig.~4]{AL01}). In the above expressions, $\bm{k}_{\star}^{\pm}$ are defined similarly to $\bm{k}_{\textsc{f};\sigma}^{\pm}$ \cite{Note10}.

The above-mentioned abrupt change in $\mathsf{n}_{\sigma}(\bm{k})$, or $\mathsf{n}_{\sigma}(\bm{k};E)$, at $\bm{k} = \bm{k}_{\star}$, can be directly attributed, through the expression for $\mathsf{n}_{\sigma}(\bm{k})$ in Eq.~(\ref{e249}), to the existence of a sharp peak in $A_{\sigma}(\bm{k}_{\star};\varepsilon)$ inside the region $\varepsilon <\mu$ (as regards $\mathsf{n}_{\sigma}(\bm{k};E)$, inside $[-E,\mu]$). In this way, in Ref.~\cite{BF04b} we have obtained the functional relationship between the slopes of the measured energy dispersion at both sides of $\bm{k} = \bm{k}_{\star}$ (in the radial direction of the relevant 1BZ) and the amount of change in $\mathsf{n}_{\sigma}(\bm{k})$ at $\bm{k} = \bm{k}_{\star}$, described by the expressions in Eqs.~(\ref{e62}) -- (\ref{e64}).

Lastly, we remark that the behaviour of the self-energy $\Sigma_{\sigma}(\bm{k};\varepsilon)$ as a function of $\varepsilon$ as inferred on the basis of the considerations in Ref.~\cite{BF04b} (see Fig.~1 herein), is in remarkably good agreement with that inferred by Hwang \emph{et al.} \cite{HTG04} from the infrared spectra of the high-$T_{\textrm{c}}$ compound Bi$_2$Sr$_2$CaCu$_2$O$_{8+\delta}$ \cite{Note11}. Later calculations of $\Sigma_{\sigma}(\bm{k};\varepsilon)$, by Byczuk \emph{et al.} \cite{KB07} and Macridin \emph{et al.} \cite{AM07}, also reproduce the $\Sigma_{\sigma}(\bm{k};\varepsilon)$ as deduced in Ref.~\cite{BF04b} (compare the diagrams in Fig.~2 of Ref.~\cite{KB07} as well as those in Figs.~2 and 3 of Ref.~\cite{AM07} with those in Fig.~1 of Ref.~\cite{BF04b}).

% 7.
\section{Summary and concluding remarks}
\label{s7}
In this paper we have reproduced and reviewed some of the main results of Refs.~\cite{BF03,BF04a,BF04b}. In doing so, we have bypassed use of the abstract one-to-one mappings $\bm{\Phi}_{\sigma}^{<}(\bm{k})$ and $\bm{\Phi}_{\sigma}^{>}(\bm{k})$ that are central directly to the considerations of Ref.~\cite{BF03}, and indirectly to those of Refs.~\cite{BF04a,BF04b}. We hope hereby to have made references \cite{BF03,BF04a,BF04b} more accessible.

The most significant finding of Refs.~\cite{BF03,BF04a} to which we have paid considerable attention in this paper, Sec.~\ref{s2}, is that the exact self-energy $\Sigma_{\sigma}(\bm{k};\varepsilon)$ pertaining to the $N$-particle uniform metallic GS of the single-band Hubbard Hamiltonian in Eq.~(\ref{e11}) \cite{BF03}, and the more general single-band Hamiltonian in Eq.~(\ref{ea1}) \cite{BF04a}, reduces to the \textsl{exact} Hartree-Fock self-energy $\Sigma_{\sigma}^{\textsc{hf}}(\bm{k})$ for $\varepsilon = \varepsilon_{\textsc{f}}$ \textsl{and} $\bm{k} \in \mathcal{S}_{\textsc{f};\sigma}$, Eqs.~(\ref{e234}) and (\ref{e236}) , where $\varepsilon_{\textsc{f}}$ denotes the \textsl{exact} Fermi energy and $\mathcal{S}_{\textsc{f};\sigma}$ the \textsl{exact} Fermi surface corresponding to particles with spin index $\sigma$ in the $N$-particle GS under consideration. As a consequence, one has $\mathcal{S}_{\textsc{f};\sigma} \subseteq \mathcal{S}_{\textsc{f};\sigma}^{\textsc{hf}}$, Eq.~(\ref{e241}), where $\mathcal{S}_{\textsc{f};\sigma}^{\textsc{hf}}$, defined in terms of $\varepsilon_{\textsc{f}}$, denotes the counterpart of $\mathcal{S}_{\textsc{f};\sigma}$ within the framework of the \textsl{exact} Hartree-Fock theory, Eq.~(\ref{e226}). The \textsl{pseudogap} region of the Fermi surface of the interacting $N$-particle metallic GS is the difference set $\mathcal{S}_{\textsc{f};\sigma}^{\textsc{hf}}\backslash \mathcal{S}_{\textsc{f};\sigma}$ \cite{BF03,BF04a}, which may be empty, depending on the value of the band filling $n$, the form of the non-interacting energy dispersion $\varepsilon_{\bm{k}}$, and the strength of the on-site interaction energy $U$.

In Sec.~\ref{s3} we have rigorously demonstrated that the extant second-order computational results purporting to show $\mathcal{S}_{\textsc{f};\sigma} \not\subseteq \mathcal{S}_{\textsc{f};\sigma}^{\textsc{hf}}$, are fundamentally deficient, this on account of the breakdown of the Luttinger theorem in the underlying calculations at best at the second order in $U$. The considerations of this section have made explicit that in employing non-self-consistent many-body perturbation theory for uniform metallic GSs, it is essential that the Fermi \textsl{energy} of the adopted mean-field (MF) theory identically coincide with the exact Fermi energy. This is additional to the well-known requirement that for anisotropic metallic GSs the Fermi surface $\mathcal{S}_{\textsc{f};\sigma}^{\textsc{mf}}$ not be deformed with respect to $\mathcal{S}_{\textsc{f};\sigma}$, $\forall\sigma$ \cite[\S5.7]{PN64}.

In Sec.~\ref{s3} we have further shown that even in the cases where $\mathcal{S}_{\textsc{f};\sigma}^{\textsc{mf}} \subseteq \mathcal{S}_{\textsc{f};\sigma}$, $\forall\sigma$, and $\varepsilon_{\textsc{f}}^{\textsc{mf}} = \varepsilon_{\textsc{f}}$, a \textsl{non-self-consistent} perturbation expansion for the self-energy fails correctly to reproduce the exact $\Sigma_{\sigma}(\bm{k};\varepsilon)$ for $\bm{k}$ in a neighbourhood of $\mathcal{S}_{\textsc{f};\sigma}$ and $\varepsilon$ in a neighbourhood of $\varepsilon_{\textsc{f}}$. This in consequence of the fact that within the framework of a $\nu\hspace{0.6pt}$th-order non-self-consistent perturbation expansion of the self-energy, for any finite value of $\nu \ge 2$, the Luttinger theorem is at best quantitatively violated at order $U^{\nu}$ (see above). Such systematic violation prevents a coherence effect from developing, an effect brought about by the mechanism that embodies the Luttinger theorem, according to which \textsl{all} points of $\mathcal{S}_{\textsc{f};\sigma}$ are put into direct contact with $\varepsilon_{\textsc{f}}$. We should emphasize, as we have done in Sec.~\ref{s3a4}, that the extent of the last-mentioned neighbourhood of  $\mathcal{S}_{\textsc{f};\sigma}$ is not fixed, but is dependent on the value of $U/t$, diminishing towards zero with $U/t \to 0$.

Two energy dispersions, $\varepsilon_{\bm{k};\sigma}^{<}$ and $\varepsilon_{\bm{k};\sigma}^{>}$, Eqs.~(\ref{e215}) and (\ref{e216}), feature very prominently in the considerations of both Refs.~\cite{BF03,BF04a,BF04b} and the present paper. These energies are \textsl{variational} single-particle excitation energies, satisfying $\varepsilon_{\bm{k};\sigma}^{<} < \mu < \varepsilon_{\bm{k};\sigma}^{>}$, $\forall\bm{k}$, where $\mu$ is the chemical potential corresponding to the $N$-particle uniform GS of the system under consideration. These energy dispersions, which are defined for both metallic and insulating uniform GSs, have the remarkable property that for metallic GSs they coincide, up to deviations of the order of $1/N$, with $\mu$ for $\bm{k} \in \mathcal{S}_{\textsc{f};\sigma}$. On the basis of this property, in Sec.~\ref{s4} we have deduced the functional forms of the GS momentum-distribution function $\textsf{n}_{\sigma}(\bm{k})$ specific to a variety of uniform metallic GSs for $\bm{k}$ infinitesimally away from the underlying Fermi surface $\mathcal{S}_{\textsc{f};\sigma}$, expressed in terms of two GS correlation functions that characterize the leading-order terms in the asymptotic series expansions of $\mu-\varepsilon_{\bm{k};\sigma}^{<}$ and $\varepsilon_{\bm{k};\sigma}^{>}-\mu$ for $\bm{k}$ approaching $\mathcal{S}_{\textsc{f};\sigma}$ \cite{BF03,BF04a}. Amongst others, we have deduced the functional forms of the aforementioned correlation functions in the asymptotic region $U/t \to 0$.

In Sec.~\ref{s5} we have made explicit the way in which the relationship $\mathcal{S}_{\textsc{f};\sigma} \subseteq \mathcal{S}_{\textsc{f};\sigma}^{\textsc{hf}}$, Eq.~(\ref{e241}), amounts to a kinematic constraint for ferromagnetic uniform metallic states to qualify as potential GSs of the single-pand Hamiltonians considered in this paper, Eqs.~(\ref{e11}) and (\ref{ea1}).

On the basis of similar considerations as indicated above, and by establishing a well-defined correspondence between in particular $\varepsilon_{\bm{k};\sigma}^{<}$ and the single-particle energy dispersions as measured by means of the ARPES \cite{DHS03}, in Sec.~\ref{s6} we have established a direct relationship between the `kinks' in the experimentally-observed single-particle energy dispersions in a number of materials and the behaviour of the underlying $\textsf{n}_{\sigma}(\bm{k})$ for $\bm{k}$ in the vicinity of the `kink' wave vector $\bm{k}_{\star}$ \cite{BF04b}. We hope hereby to have succeeded in drawing the attention of in particular experimentalists to the relationship that exists between the experimentally-measured single-particle energy dispersions and $\varepsilon_{\bm{k};\sigma}^{<}$ and $\varepsilon_{\bm{k};\sigma}^{>}$. The existence of this relationship is of considerable practical significance, in particular because the exact identity in Eq.~(\ref{e252}) enables one to gain quantitative insight into the dispersion of the single-particle excitation energies \textsl{above} $\mu$, which is related to the energy dispersion $\varepsilon_{\bm{k};\sigma}^{>}$, on the basis of the measurements concerning the single-particle excitation energies \textsl{below} $\mu$, which is related to the energy dispersion $\varepsilon_{\bm{k};\sigma}^{<}$. In this connection and with reference to the expression in Eq.~(\ref{e249}), one should note that $\mathsf{n}_{\sigma}(\bm{k})$ is determined through the knowledge of $A_{\sigma}(\bm{k};\varepsilon)$ for $\varepsilon$ over $(-\infty,\mu]$, and that, according to the expressions in Eqs.~(\ref{ea3}) -- (\ref{ea5}), knowledge of $\mathsf{n}_{\sigma}(\bm{k})$ and the bare interaction potential $\t{v}$ suffices to calculate $\Sigma_{\sigma}^{\textsc{hf}}(\bm{k})$. If for no other purpose, the experimentally-determined energy dispersions $\varepsilon_{\bm{k};\sigma}^{<}$ and $\varepsilon_{\bm{k};\sigma}^{>}$ (better, $\varepsilon_{\bm{k};\sigma}^{<}(E)$ and $\varepsilon_{\bm{k};\sigma}^{>}(E)$) can be fruitfully used for determining the Fermi surface of the $N$-particle uniform metallic GS of any system whose low-energy properties are accurately described by a single-band Hamiltonian. For clarity, an experimentally-determined $\varepsilon_{\bm{k};\sigma}^{<}$ is the energy dispersion determined on the basis of an experimentally-measured $A_{\sigma}(\bm{k};\varepsilon)$ \cite{DHS03} and the left-most expression in Eq.~(\ref{e251}). In Sec.~\ref{s2e}, we have speculated on a way in which also the Luttinger surface \cite[\S2.4]{BF07a} of \textsl{insulating} $N$-particle uniform GSs may be determined with the aid of $\varepsilon_{\bm{k};\sigma}^{<}$, $\varepsilon_{\bm{k};\sigma}^{>}$ and $\mathsf{n}_{\sigma}(\bm{k})$.

% Appendix A.
\begin{appendix}
\section{Some aspects concerning a more general single-band Hamiltonian than the single-band Hubbard Hamiltonian}
\label{saa}
The considerations of this paper are focused on the $N$-particle uniform metallic GS of the single-band Hubbard Hamiltonian $\wh{\mathcal{H}}$ in Eq.~(\ref{e11}). With reference to Ref.~\cite{BF04a}, in the main text we have emphasized that the results presented in this paper are with some appropriate modifications also applicable to the $N$-particle uniform metallic GS of the following, more general, single-band Hamiltonian \cite{BF04a}:
\begin{eqnarray}
&&\hspace{-1.0cm} \wh{H} = \!\sum_{\bm{k},\sigma}\! \varepsilon_{\bm{k}}\,
\h{a}_{\bm{k};\sigma}^{\dag} \h{a}_{\bm{k};\sigma} \nonumber\\
&&\hspace{-0.5cm} + \frac{1}{2 \Omega}
\sum_{\sigma,\sigma'} \sum_{\bm{k}, \bm{p}, \bm{q}} \t{v}(\|\bm{q}\|)\,
\h{a}_{\bm{k}+\bm{q};\sigma}^{\dag} \h{a}_{\bm{p}-\bm{q};\sigma'}^{\dag}
\h{a}_{\bm{p};\sigma'} \h{a}_{\bm{k};\sigma},
\label{ea1}
\end{eqnarray}
where $\Omega$ is the macroscopic volume of the system under consideration (`macroscopic' because we explicitly deal with \textsl{metallic} GSs), and $\t{v}(\|\bm{q}\|)$ the Fourier transform of the isotropic two-body potential, $v(\|\bm{r}-\bm{r}'\|)$ or $v(\|\bm{R}_j-\bm{R}_{j'}\|)$ (depending on whether the system under consideration is defined in a continuum subset of $\mathds{R}^d$, or on the lattice $\{\bm{R}_j\}$ embedded in $\mathds{R}^d$, appendix \ref{sad}), through which the particles in the system interact \cite{Note12}. The sums with respect to $\bm{k}$, $\bm{p}$ and $\bm{q}$ in Eq.~(\ref{ea1}) are over the following discrete sets of points in the wave-vector space, which conform with the box boundary condition (that is, the distance between two adjacent points along the $j$th principal axis of this space is equal to $2\pi/L_j$, $j=1,\dots, d$, where $L_j$ is the macroscopic length of the system along the $j$th principal axis): (1) in the cases where the system under consideration is defined on the lattice $\{\bm{R}_j \| j=1,2,\dots, N_{\textsc{s}}\}$, which we assume to be a Bravais lattice, the relevant set consists of the points of the first Brillouin zone, $\mathrm{1BZ}$, specific to $\{\bm{R}_j\}$; (2) in the cases where the system is defined in a continuum subset of $\mathds{R}^d$ and the Bravais lattice $\{\bm{R}_j\}$ signifies the position of `atoms' (assuming mono-atomic crystals -- see appendix \ref{sad}), the relevant set consists of the points $\bm{k}$ describable as $\bm{k} = \bm{\kappa} + \bm{K}$, where $\bm{\kappa} \in \textrm{1BZ}$ and $\bm{K} \in \{\bm{K}_j \| j = 1,2,\dots,N_{\textsc{s}}\}$, the latter being the set of vectors spanning the lattice reciprocal to $\{\bm{R}_j\}$.

The Hubbard Hamiltonian in Eq.~(\ref{e11}) is recovered from the Hamiltonian in Eq.~(\ref{ea1}) by effecting the following transformation:
\begin{equation}
\t{v}(\|\bm{q}\|) \rightharpoonup \frac{\Omega\, U}{N_{\textsc{s}}},
\label{ea2}
\end{equation}
and confining the wave-vector sums in Eq.~(\ref{ea1}) to the points of the relevant $\mathrm{1BZ}$; in doing so, such vector as for instance $\bm{k}+\bm{q}$ is to be identified with $\bm{k} +\bm{q} +\bm{K}_0$, where $\bm{K}_0$ is the reciprocal-lattice vector, corresponding to the underlying direct lattice $\{ \bm{R}_j\}$, for which $\bm{k} +\bm{q} +\bm{K}_0 \in \mathrm{1BZ}$.

% A.1.
\subsection{The exact Hartree-Fock self-energy}
\label{saa1}
Using the rules for calculating the contributions of the self-energy diagrams \cite[\S3.9]{FW03}, for the Hartree-Fock self-energy $\Sigma_{\sigma}^{\textsc{hf}}(\bm{k}) $ as expressed in terms of the \textsl{exact} Green functions $\{ G_{\sigma} \| \sigma \}$, i.e. $\Sigma_{\sigma}^{\textsc{hf}}(\bm{k};[\{G_{\sigma'}\}])$, Sec.~\ref{s3a2}, one obtains \cite[\S4.10]{FW03}
\begin{equation}
\Sigma_{\sigma}^{\textsc{hf}}(\bm{k}) = \Sigma^{\textsc{h}}(\bm{k}) + \Sigma_{\sigma}^{\textsc{f}}(\bm{k}), \label{ea3}
\end{equation}
where (see Eqs.~(\ref{e13}) and (\ref{e230}))
\begin{equation}
\Sigma^{\textsc{h}}(\bm{k}) = \frac{1}{\hbar\, \Omega}\, \t{v}(0) \sum_{\bm{k}',\sigma'} \mathsf{n}_{\sigma'}(\bm{k}') \equiv \t{v}(0)\, \frac{N}{\hbar\,\Omega},
\label{ea4}
\end{equation}
\begin{equation}
\Sigma_{\sigma}^{\textsc{f}}(\bm{k}) = -\frac{1}{\hbar\, \Omega}\, \sum_{\bm{k}'} \t{v}(\|\bm{k}-\bm{k}'\|)\, \mathsf{n}_{\sigma}(\bm{k}').
\label{ea5}
\end{equation}
In the cases where $\t{v}(0)$ is unbounded, $\Sigma^{\textsc{h}}(\bm{k})$ is discarded on account of its cancellation against the self-energy contribution arising from the interaction of particles with a positively-charged uniform background; by the same consideration, in these cases the point $\bm{k}' = \bm{k}$ is excluded from the set of points over which the summation on the RHS of Eq.~(\ref{ea5}) is carried out \cite[\S2.1]{BF02}.

Making use of the transformation in Eq.~(\ref{ea2}), for the Hubbard Hamiltonian one obtains (see Eqs.~(\ref{e12}) and (\ref{e13}))
\begin{equation}
\Sigma^{\textsc{h}}(\bm{k}) = \frac{1}{\hbar} U n,\;\; \Sigma_{\sigma}^{\textsc{f}}(\bm{k}) = -\frac{1}{\hbar} U n_{\sigma},
\label{ea6}
\end{equation}
leading, on account of the equality in Eq.~(\ref{ea3}), to the expression in Eq.~(\ref{e224}). In the above expressions, $\mathsf{n}_{\sigma}(\bm{k})$ originates from \cite[Eq.~(10.7)]{FW03}
\begin{equation}
\mathsf{n}_{\sigma}(\bm{k}) = \frac{1}{\hbar} \int_{-\infty}^{\infty} \frac{\rd\varepsilon}{2\pi i}\, \e^{i\varepsilon 0^+/\hbar} G_{\sigma}(\bm{k};\varepsilon),
\label{ea7}
\end{equation}
which is equivalent to the expression in Eq.~(\ref{e249}). This equivalence is established by realizing that (see Eq.~(\ref{e246}))
\begin{equation}
G_{\sigma}(\bm{k};\varepsilon) \doteq \lim_{\eta\downarrow 0} \t{G}_{\sigma}(\bm{k};\varepsilon \pm i \eta),\;\; \varepsilon \gtrless \mu,
\label{ea8}
\end{equation}
where $\t{G}_{\sigma}(\bm{k};z)$, $z\in \mathds{C}$, is the single-particle Green function in Eq.~(\ref{e268}). Such explicit definition for the `physical' single-particle Green function $G_{\sigma}(\bm{k};\varepsilon)$, $\varepsilon \in \mathds{R}$, is necessary owing to $\t{G}_{\sigma}(\bm{k};z)$ being discontinuous across (some parts of) the real $\varepsilon$ axis; explicitly, on account of the spectral representation in Eq.~(\ref{e268}), those parts at which, for the given $\bm{k}$, $A_{\sigma}(\bm{k};\varepsilon) \not= 0$. Since $\t{G}_{\sigma}(\bm{k};z)$ is analytic everywhere in $\mathds{C}$ outside the real axis \cite{JML61,BF07a}, and since $\t{G}_{\sigma}(\bm{k};z) \sim \hbar/z$ for $\vert z\vert \to \infty$ \cite{JML61,BF07a}, through deforming the integration contour over $[\mu,+\infty)$ into the upper-half of the complex $\varepsilon$ plane (note the converging factor $\e^{i\varepsilon 0^+/\hbar}$) and making use of the definition in Eq.~(\ref{e246}), from the expression in Eq.~(\ref{ea7}) one arrives at the one in Eq.~(\ref{e249}).

From the perspective of the considerations of this paper, the significance of the above standard expressions lies in the fact that they reveal that the $\varepsilon_{\textsc{f}}$ (or $\mu$, Eq.~(\ref{e27})) implicit in the expression for the \textsl{exact} Hartree-Fock self-energy, coincides with the exact Fermi energy $\varepsilon_{\textsc{f}}$, apparent from the equivalence of the expressions in Eqs.~(\ref{e249}) and (\ref{ea7}). Violation of the equality in Eq.~(\ref{e229}) would therefore be tantamount to an internal inconsistency in the \textsl{exact} Hartree-Fock theory: the Fermi energy on which the \textsl{exact} $\Sigma_{\sigma}^{\textsc{hf}}(\bm{k})$ implicitly depends, i.e. the exact Fermi energy $\varepsilon_{\textsc{f}}$, would differ from the Fermi energy $\varepsilon_{\textsc{f}}^{\textsc{hf}}$ pertaining to the $N$-particle GS of this theory as determined through solving the equation in Eq.~(\ref{e228}). Although such inconsistency cannot \emph{a priori} be ruled out, in Sec.~\ref{s2f} we demonstrate that the equality in Eq.~(\ref{e229}) is indeed valid. The considerations in Sec.~4 of Ref.~\cite{BF04a} shed additional light on some relevant properties of the \textsl{exact} $\Sigma_{\sigma}^{\textsc{hf}}(\bm{k})$.

% Appendix B.
\section{Simple examples illustrative of the way in which the Luttinger-Ward identity breaks down}
\label{sab}
In Sec.~\ref{s3a3} we have argued that despite the equality in Eq.~(\ref{e338}), in the asymptotic region $U/t \to 0$ the leading-order term in the asymptotic series expansion \cite{WW62,ETC65,HAL74} of the expression on the LHS of Eq.~(\ref{e337}) does \textsl{not} scale like $U^4$, in contradiction with what a formal geometric series expansion of the function $\t{G}_{\sigma}'$ in Eq.~(\ref{e332}) in powers of $\t{G}_{\sigma}^{\textsc{hf}} \t{\Upsigma}_{\sigma}[\{G_{\sigma'}^{\textsc{hf}}\}]$ would suggest. In this appendix we show the validity of the argument in Sec.~\ref{s3a3} by considering two simple mathematical models that shed light on some relevant processes.

We begin by considering the functions
\begin{equation}
\upgamma_a^0(z) \doteq \frac{1}{z - a}\;\;\text{and}\;\; \upzeta_b(z) \doteq \frac{u^2}{z - b},\;\, a, b, u \in \mathds{R},
\label{eb1}
\end{equation}
as representing respectively $\t{G}_{\sigma}^{\textsc{hf}}(\bm{k};z)$, Eq.~(\ref{e339}), and  $\t{\Upsigma}_{\sigma}^{(2)}(\bm{k};z;[\{G_{\sigma'}^{\textsc{hf}}\}])$ for a given $\bm{k} \in \textrm{1BZ}$. By identifying $\mu^{\textsc{hf}}$ with $1$, the choice $a=1$ would correspond to $\bm{k}$ being a vector on the Fermi surface $\mathcal{S}_{\textsc{f};\sigma}^{\textsc{hf}}$. We note that similar to the latter functions, those in Eq.~(\ref{eb1}) to leading order decay like $1/z$ for $\vert z\vert \to \infty$ \cite[Eqs.~(B.65), (B.103)]{BF07a}. Thus
\begin{equation}
\upgamma_{a,b}(z) \doteq \frac{\upgamma_a^0(z)}{1 - \upgamma_a^0(z) \upzeta_b(z)}
\label{eb2}
\end{equation}
may be viewed as representing the function $\t{G}_{\sigma}'(\bm{k};z)$ in Eq.~(\ref{e332}).

In analogy with the expression on the LHS of Eq.~(\ref{e337}), we introduce the following function:
\begin{equation}
f_{a,b}(u) \doteq \int_{1-i\infty}^{1+i\infty} \frac{\rd z}{2\pi i}\, \big(\upgamma_{a,b}(z) - \upgamma_a^0(z)\big) \frac{\partial}{\partial z} \upzeta_b(z),
\label{eb3}
\end{equation}
where the choice of the contour of integration (more specifically, the point at which it passes through the real axis, Eq.~(\ref{e328})) is dictated by the above-mentioned identification of $\mu^{\textsc{hf}}$ with $1$.

We note that since the function $\upzeta_{b}(z)$, in contrast to $\t{\Upsigma}_{\sigma}^{(2)}(\bm{k};z;[\{G_{\sigma'}^{\textsc{hf}}\}])$, does \textsl{not} correspond to second-order skeleton self-energy diagrams evaluated in terms of $\upgamma_a^0(z)$, the function
\begin{equation}
h_{a,b}(u) \doteq \int_{1-i\infty}^{1+i\infty} \frac{\rd z}{2\pi i}\, \upgamma_a^0(z) \frac{\partial}{\partial z} \upzeta_b(z)
\label{eb4}
\end{equation}
is not necessarily vanishing (cf. Eq.~(\ref{e338})). It is for this reason that we have defined the function $f_{a,b}(u)$ in terms of the difference $\upgamma_{a,b}(z) - \upgamma_a^0(z)$, instead of $\upgamma_{a,b}(z)$ alone.

While $\upgamma_{a,b}(z)$ is bounded at $z =1$, $\forall a, b \in \mathds{R}$ and $u \not=0$, $\upgamma_a^0(z)$ has a pole at $z=1$ for $a = 1$. This implies that for $a = 1$ the integral in Eq.~(\ref{eb4}) must be viewed as a Cauchy principal-value integral \cite[\S4.5]{WW62}. With this in mind and assuming that $b > 1$, one trivially obtains that
\begin{equation}
h_{a,b}(u) = -\frac{\Theta(1-a) u^2}{(b-a)^2},
\label{eb5}
\end{equation}
where $\Theta(0) = 1/2$ by definition. The peculiarity of the case of $a=1$ is due to the fact that in this case only one-half the residue of the integrand of $h_{a,b}(u)$ corresponding to the simple pole of $\upgamma_a^0(z)$ at $z=1$ is taken account of by the Cauchy principal-value integration \cite[\S4.5]{WW62}.

For $a \le 1$ and $b> 1$ one obtains
\begin{equation}
f_{a,b}(u) = -\frac{1}{2} + \frac{b -a}{2 \sqrt{4 u^2 + (b-a)^2}} + \frac{\Theta(1-a) u^2}{(b-a)^2},
\label{eb6}
\end{equation}
from which one deduces the following asymptotic series expansions \cite{WW62,ETC65,HAL74} corresponding to $u \to 0$:
\begin{equation}
f_{1,b}(u) \sim -\frac{u^2}{2 (b-1)^2} + \frac{3 u^4}{(b-1)^4} -\frac{10 u^6}{(b-1)^6} +\dots,
\label{eb7}
\end{equation}
\begin{equation}
f_{a,b}(u) \sim \frac{3 u^4}{(b-a)^4} -\frac{10 u^6}{(b-a)^6} + \dots,\; a < 1.
\label{eb8}
\end{equation}
One observes that in contrast to the case of $a <1$, the function $f_{a,b}(u)$ for $a=1$ to leading order scales like $u^2$ for $u\to 0$. Below we show that the case of $a=1$ is relevant to the considerations in Sec.~\ref{s3a3}, even though, in the light of the sum with respect to $\bm{k}$ on the LHS of Eq.~(\ref{e337}) and the fact that in the thermodynamic limit Fermi surface amounts to a subset of measure zero of the underlying 1BZ (see the remark following Eq.~(\ref{eb1}) above), the case of $a=1$ may not seem relevant from the perspective of the validity or failure of the Luttinger-Ward identity \cite{LW60} at order $U^2$. We note in passing that the asymptotic series in Eq.~(\ref{eb7}) (Eq.~(\ref{eb8})) is the Taylor series of $f_{1,b}(u)$ ($f_{a,b}(u)$) around $u=0$ for $\vert u\vert < (b-1)/2$ ($\vert u\vert < (b-a)/2$).

For $a<1$, in contrast to the case of $a=1$, the contour of integration with respect to $z$ in the defining expression for $f_{a,b}(z)$, Eq.~(\ref{eb3}), can be deformed, thereby accounting for the \textsl{full} residue of the integrand corresponding to the pole of $\upgamma_a^0(z)$ at $z = a$. The possibility of such contour deformation is rooted in the fact that the function $\upzeta_b(z)$ under consideration, Eq.~(\ref{eb1}), and therefore its derivative $\partial \upzeta_b(z)/\partial z$, is analytic in a finite neighbourhood of $z=1$. This is not the case for the self-energy $\t{\Upsigma}_{\sigma}^{(2)}(\bm{k};z;[\{G_{\sigma'}^{\textsc{hf}}\}])$ corresponding to an $N$-particle \textsl{metallic} GS that the function $\upzeta_b(z)$ is intended to represent. For this self-energy, the point $z = \mu^{\textsc{hf}}$ (where $\mu^{\textsc{hf}} = \varepsilon_{\textsc{f}}^{\textsc{hf}}$ up to a deviation of the order of $1/N$) is a branch-point singularity \cite[\S5.7]{WW62} for \textsl{all} $\bm{k} \in \textrm{1BZ}$. This fact is easily established by considering the identity in Eq.~(\ref{e320}) and taking into account that with $z \equiv \varepsilon + i \eta$, where $\varepsilon, \eta \in \mathds{R}$, the same self-energy is discontinuous at $\eta = 0$ for $\varepsilon$ in a neighbourhood of $\varepsilon_{\textsc{f}}^{\textsc{mf}}$ (see the remark following Eq.~(\ref{ea8}) as well as that following Eq.~(B.59) in Ref.~\cite{BF07a}).

Considering the exact $\t{\Sigma}_{\sigma}(\bm{k};z)$ corresponding to an $N$-particle GS, for this function the energies $z = \mu_{N;\sigma}^-$ and $z = \mu_{N;\sigma}^+$, Eqs.~(\ref{e22}) and (\ref{e23}), turn into \textsl{limit} or \textsl{accumulation} points \cite[\S2.21]{WW62} of the sets $\{\varepsilon_{s;\sigma}^- \| s\}$ and $\{\varepsilon_{s;\sigma}^+ \| s\}$ of the single-particle excitation energies \cite[Eq.~(B.3)]{BF07a} in the limit of $N=\infty$ \cite{Note13}. As $N\to \infty$, $z=\mu_{N;\sigma}^-$ and $z=\mu_{N;\sigma}^+$ become in fact branch-point singularities \cite[\S5.7]{WW62} of $\t{\Sigma}_{\sigma}(\bm{k};z)$, coinciding up to a deviation of the order of $1/N$ in the case of \textsl{metallic} $N$-particle GSs, Eqs.~(\ref{e26}) and (\ref{e27}). In the case of metallic GSs, the contour of integration $\mathscr{C}(\mu)$ in the expression for the Luttinger-Ward function, Eq.~(\ref{e327}), is therefore `pinched' \cite[\S6.3.1]{IZ80} by the branch-point singularities $z = \mu_{N;\sigma}^-$ and $z = \mu_{N;\sigma}^+$, that is, the crossing point of $\mathscr{C}(\mu)$ with the real axis of the $z$ plane becomes immovable. By identifying $\upzeta_b(z)$ with a function whose branch-point singularities similarly to those of the exact $\t{\Sigma}_{\sigma}(\bm{k};z)$ corresponding to metallic GSs `pinch' the contour of integration $\mathscr{C}(\mu)$ (here $\mathscr{C}(\mu^{\textsc{hf}})$) on the real energy axis, it is to be expected that for $a \uparrow 1$ the corresponding function $f_{a,b}(u)$ behaves, in its dependence on $u$ in the region $u\to 0$, similarly to the function $f_{1,b}(u)$ considered earlier in this appendix.

In the light of the above observations, we now consider the function
\begin{equation}
\upzeta_b(z) \doteq \frac{u^2}{z - b} \Big(2 - \frac{\sqrt{1 - z}}{\sqrt{b-z}}\Big),
\label{eb9}
\end{equation}
instead of that presented in Eq.~(\ref{eb1}). For $b >1$, the function in Eq.~(\ref{eb9}) has two branch points \cite[\S5.7]{WW62}, at $z=1$ and $z=b$. The choice for the function $\upzeta_b(z)$ in Eq.~(\ref{eb9}) is motivated by its simplicity, the fact that for $b \not= 1$ the point $z=1$ (representing the chemical potential $\mu^{\textsc{hf}}$ in this appendix)  is its branch point, and that for $\vert z\vert \to \infty$ to leading order it coincides with the function $\upzeta_b(z)$ in Eq.~(\ref{eb1}). We note that in contrast to the above-mentioned self-energy $\t{\Sigma}_{\sigma}(\bm{k};z)$ whose two branch points $\mu_{N;\sigma}^-$ and $\mu_{N;\sigma}^+$ `pinch' \cite[\S6.3.1]{IZ80} the contour $\mathscr{C}(\mu)$ on the real energy axis of the $z$ plane in the case of metallic $N$-particle GSs, Eqs.~(\ref{e26}) and (\ref{e27}), the branch point $z=1$ of the function $\upzeta_b(z)$ in Eq.~(\ref{eb9}) prohibits displacement of the crossing point of the contour of integration in the defining expression for $f_{a,b}(u)$, Eq.~(\ref{eb3}), \textsl{only} to the right of $z=1$; for $a <1$, the last-mentioned crossing point can be freely displaced to the left of $z=1$, so long as it remains to the right of $a$. In the model under consideration, the `pinching' is therefore effected by the limiting process $a\uparrow 1$, increasingly narrowing the interval $[a,1]$.

Although the function $f_{a,b}(u)$ corresponding to the function $\upzeta_b(z)$ in Eq.~(\ref{eb9}) may be expressible in closed form, we have made \textsl{no} serious attempt to this end. Instead, we have studied this function wholly numerically.

\begin{figure}[t!]
\includegraphics[angle=0, width=0.43\textwidth]{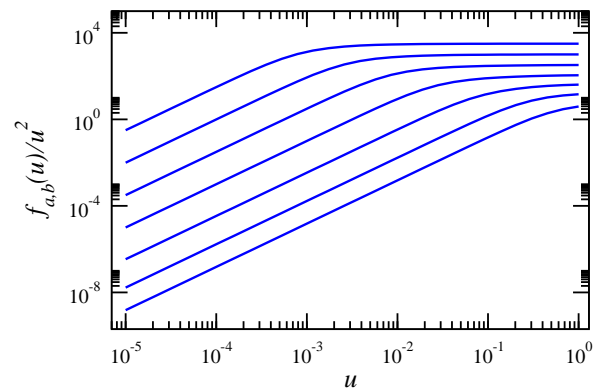}
\caption{(Colour) The log-log plot of the function $f_{a,b}(u)/u^2$ versus $u$ for $b=2$ and seven different values of $a$, with $f_{a,b}(u)/u^2$ increasing for increasing values of $a$. The function $f_{a,b}(u)$ is defined in Eq.~(\protect\ref{eb3}), determined in the terms of the function $\upzeta_b(z)$ introduced in Eq.~(\protect\ref{eb9}). The values $a_i$ assigned to $a$ are: $a_1 = 0.5$ (lower-most curve), $a_2= 0.9$, $a_3 = 0.99$, $a_4 = 0.999$, $a_5 = 0.999\,9$, $a_6 = 0.999\,99$ and $a_7 = 0.999\,999$ (upper-most curve). For orientation, up to a logarithmic correction the function $f_{a_1,2}(u)/u^2$ for $10^{-5} \le u \le 10^{-1}$ is described by $A u^{\upalpha}$, where $A = 10.383$ and $\upalpha = 1.9532$. From the behaviour of $f_{a,2}(u)/u^2$ as a function of $a$, as $a\uparrow 1$, one observes that up to a possible logarithmic correction, to leading order $f_{a,2}(u)$ should diverge like $1/(1-a)^{\upkappa_u}$, where $\upkappa_u$ is a positive function of $u$, independent of $a$. Although the correspondence between $f_{a,2}(u)$ and $1/(1-a)^{\upkappa_u}$ is \textsl{not} a rigorous one (in particular, $\lim_{a\uparrow 1} (1-a)^{\upkappa_u} f_{a,2}(u)$ appears not to exist -- a numerical observation), nonetheless numerical results for $f_{a,2}(1)$ suggest $\upkappa_{1.0} \approx 0.534$ as very reasonable. } \label{f4}
\end{figure}

In Fig.~\ref{f4} we present $f_{a,b}(u)/u^2$, with $b=2$, as a function of $u$ for different values of the parameter $a$. One observes that up to a possible logarithmic correction, the most dominant contribution to $f_{a,2}(u)$ scales like $u^{\upalpha}$, where $\upalpha \approx 4$ for `small' and $\upalpha \approx 2$ for `large' values of $u$. The boundary region separating the `small' and `large' values of $u$ is clearly seen to be determined by the value of $a$, with the region corresponding to `large' values of $u$ increasingly expanding, at the expense of that corresponding to `small' values of $u$, for $a$ approaching $1$, the value representing the chemical potential $\mu^{\textsc{hf}}$ in the considerations of the present appendix. We have therefore established that for the simpler model described in terms of the function $\upzeta_{b}(z)$ introduced in Eq.~(\ref{eb1}), indeed the case of $a=1$ is representative not only of $\bm{k}$ being \textsl{on} the underlying Fermi surface, but also of $\bm{k}$ being located in a finite neighbourhood of this surface.

With reference to the apparent divergence of $f_{a,2}(u)$ for $a\uparrow 1$,  Fig.~\ref{f4}, we note that although the number of $\bm{k}$ points in the neighbourhood of the Fermi surface is a relatively small fraction of the total number of points of which the underlying 1BZ is comprised, quantitatively the contribution to the sum on the LHS of Eq.~(\ref{e337}) of each of the points in the close neighbourhood of the Fermi surface is considerably larger than that of a point of the 1BZ outside this neighbourhood.

% Appendix C.
\section{On the asymptotic series with unbounded expansion coefficients}
\label{sac}
Let $f(u)$ be a well-defined bounded function in some neighbourhood of $u=0$ and
\begin{equation}
f(u) \sim a_0 + a_1 u + a_2 u^2 + \dots
\label{ec1}
\end{equation}
its \textsl{formal} asymptotic series expansion corresponding to $u\to 0$ in terms of the asymptotic sequence $\{1, u, u^2,\dots\}$ \cite{WW62,ETC65,HAL74}. When in this formal series the coefficients $\{a_j \| j=0,\dots, m\}$ are bounded but $a_{m+1}$ is unbounded, the leading-order term in the asymptotic series expansion of $f(u) - (a_0 + \dots + a_m u^m)$, for $u\to 0$, is more dominant than $u^{m+1}$ and, provided that $a_m \not= 0$ (see later), less dominant than $u^{m}$ \cite[\S2.2]{BF02} (in the region $u\to 0$, the function $\xi(u)$ is more [less] dominant than $u^{p}$ if $u^p/\xi(u) \sim 0$ [$\xi(u)/u^p \sim 0$] as $u\to 0$). This leading-order term is deduced through evaluating the infinite sum $\sum_{j=m+1}^{\infty} a_j^{\textrm{s}} u^j$, where $a_j^{\textrm{s}}$ is the unbounded (singular) part of $a_j$, $j > m$. We note that since by assumption $f(u)$ is well-defined, the unboundedness of $a_{m+1}$ implies that besides $a_{m+1}$, an infinite subset of the infinite set $\{a_j \| j > m\}$ must be unbounded \cite[\S2.2]{BF02}.

Here we focus on the cases where the coefficients $\{a_j \| j=0,1,2,\dots\}$ are defined in terms of integrals, whereby an unbounded $a_j$, $j > m$, corresponds to a non-existent integral and $a_j^{\textrm{s}}$ to that part of the relevant integrand whose integral is unbounded. In this light, $\sum_{j=m+1}^{\infty} a_j^{\textrm{s}} u^j$ stands for the function that one obtains on exchanging the orders of $\sum_{j=m+1}^{\infty}$ and the integral associated with $a_j^{\textrm{s}}$ (it may in general be necessary to redefine integrands so as to create a region of integration common to all $a_j$, $j> m$). This exchange of the orders of summation and integration and the subsequent evaluation of the sum with respect to $j$ result in a bounded function of $u$ whose leading-order term in the asymptotic series expansion for $u\to 0$ is more dominant than $u^{m+1}$, however less dominant than $u^{m}$, provided that $a_m \not= 0$ \cite[\S2.2]{BF02}.

The unboundedness of $a_{m+1}$ implying $\partial^j f(u)/\partial u^j$ to be unbounded at $u=0$ for $j=m+1$, it follows that viewed as a function of a complex variable, $f(u)$ is non-analytic at $u=0$. More explicitly, since $f(u)$ is by assumption well-defined and bounded, $f(u)$ is non-analytically singular \cite[\S4.2]{ECT52} at $u=0$, whereby $f(u) - (a_0 + \dots + a_m u^m)$ cannot be described by a single asymptotic series that is valid \textsl{uniformly} for all $\arg(u)$ as $\vert u\vert \to 0$ \cite{ETC65,HAL74}. Thus, restricting the considerations to $u \in \mathds{R}$, an unbounded $a_{m+1}$, for any finite value of $m$, implies that the leading asymptotic contribution to $f(u) - (a_0 + \dots + a_m u^m)$ for $u\to 0$ must depend on the sign of $u$. This term must therefore be expressible as $c\hspace{0.7pt} \vert u\vert^{\alpha} \ln^{\gamma}(1/\vert u\vert)$, where the values of the constants $c$, $\alpha$ and $\gamma$ in general depend on the sign of $u$. Here, $m < \alpha \le m+1$ and $\gamma \ge 0$, where $\alpha$ and $\gamma$ cannot simultaneously be equal to respectively $m+1$ and $0$. For $a_m =0$, the possibility of $\alpha = m$, to be contrasted with $m < \alpha$, cannot \emph{a priori} be ruled out. What is evident however, is that  in the cases where $\alpha = m$, one must have $\gamma = 0$, for  $\alpha=m$ together with $\gamma >0$ would imply an unbounded $a_{m-1}$, in violation of the assumption of $a_{m+1}$ being the first in the sequence $\{a_0, a_1, a_2,\dots\}$ to be unbounded (this is similar to the case where $a_m \not=0$, for which the relevant $\alpha$ and $\gamma$ cannot simultaneously take the values $m+1$ and $\gamma = 0$, respectively). We shall discuss the former aspect below. For now, we note that unless $f(0) \not= \lim_{u\to 0} f(u)$, the possibility of $\alpha = m$ when $m=0$ and $a_0 \equiv f(0) = 0$, is \emph{a priori} ruled out.

For illustration, we first consider the function
\begin{equation}
f(u) \doteq \int_{0}^{\infty} \rd x\, \frac{\sin(u x)}{x^2 +1},\;\; u \in \mathds{R},
\label{ec2}
\end{equation}
for which one has
\begin{equation}
f(u) = \mathrm{Shi}(u) \cosh(u) - \mathrm{Chi}(\vert u\vert)\sinh(u),
\label{ec3}
\end{equation}
where $\mathrm{Shi}$ is the hyperbolic sine integral function \cite[5.2.3]{AS72} and $\mathrm{Chi}$ the hyperbolic cosine integral function \cite[5.2.4]{AS72}. From the expression in Eq.~(\ref{ec3}), one readily obtains
\begin{equation}
f(u) \sim -u \ln(\pm u) + (1 - \upgamma)\hspace{0.4pt} u + \dots\;\; \text{for}\;\; u \to 0^{\pm},
\label{ec4}
\end{equation}
where $\upgamma = 0.57721\,56649\dots$ is the Euler constant \cite[6.1.3]{AS72}, and $u \to 0^{+}$ ($u\to 0^{-}$) denotes $u \downarrow 0$ ($u \uparrow 0$) (see above). With $\sin(x) = x - x^3/6 +\dots$, in the light of the fact that the integral $\int_0^{\infty} \rd x\, x/(x^2+1)$ (coinciding with the coefficient of $u$ in the \textsl{formal} asymptotic series expansion of $f(u)$ corresponding to $u\to 0$) is unbounded, the form of the asymptotic expression in Eq.~(\ref{ec4}) is in conformity with the above observations. According to these, because of the unboundedness of the latter integral, the asymptotic term proportional to $u$ must be superseded by a term more dominant than $u$ and less dominant than $u^0$, which is indeed the case (note that $f(u)$ is a continuous function of $u$ at $u=0$ and $f(0)=0$).

We now consider the function
\begin{equation}
f(u) \doteq \int_0^{\infty} \rd x\, \frac{\cos(u x)}{x^2 +1},\;\; u \in \mathds{R},
\label{ec5}
\end{equation}
for which one has
\begin{equation}
f (u) = \frac{\pi}{2} \e^{-\vert u\vert}.
\label{ec6}
\end{equation}
Clearly, $f(u)$ is continuous but not differentiable at $u=0$, it being cusped. One has
\begin{equation}
f(u) \sim \frac{\pi}{2} \big(1 \mp u + \frac{1}{2!} u^2 \mp \frac{1}{3!} u^3 + \dots \big)\;\;\text{for}\;\; u \to 0^{\pm}.
\label{ec7}
\end{equation}
Two aspects of these asymptotic series are worthy of note.

Firstly, $f(u)-a_0$, where $a_0 \equiv f(0) = \pi/2$, does not have a \textsl{uniform} asymptotic series expansion \cite{ETC65,HAL74} for $u \to 0$ in terms of the asymptotic sequence $\{u, u^2,\dots\}$, which here is to say that in the asymptotic region $u\to 0$, $f(u)-a_0$ \textsl{cannot} be described by a single (`single' \emph{qua} form) asymptotic series in terms of the latter asymptotic sequence, but by \textsl{two} distinct ones, one specific to the region $u <0$ and one specific to the region $u > 0$ (note that here by assumption $u \in \mathds{R}$). This non-uniformity is related to the non-analyticity of $f(u)$ at $u=0$.

Secondly, whereas the Taylor series expansion $\cos(x) =  1 - x^2/2 + x^4/24 -\dots$ contains only \textsl{even} powers of $x$, the two asymptotic series in Eq.~(\ref{ec7}) contain both even and \textsl{odd} powers of $u$. To appreciate this feature, we refer the reader to our above explicit references to the condition $a_m \not= 0$ (here $m=1$) in asserting the condition $m <\alpha$, to be distinguished from $m \le \alpha$. In the present case, where the \textsl{formal} asymptotic series expansion of $f(u)$ for $u\to 0$ in terms of the asymptotic sequence $\{1,u, u^2,\dots\}$ is of the form $f(u) \sim a_0 + a_2 u^2 + a_4 u^4 + \dots$, one has $a_{2j-1} = 0$ for all $j \in \mathds{N}$. The unboundedness of the integral $\int_0^{\infty} \rd x\, x^m/(x^2 + 1)$ for amongst others $m = 2j$, with $j \in \mathds{N}$, implies that $a_{2j}$ is unbounded for all $j \in \mathds{N}$. Thus, while on general grounds the leading order asymptotic contribution to $f(u) - a_0$ for e.g. $u\downarrow 0$, where $a_0 \equiv f(0) = \pi/2$, is of the form $c\hspace{0.7pt} u^{\alpha} \ln^{\gamma}(1/u)$ (see above), with $a_1 = 0$ one \textsl{cannot} conclude that either $1 < \alpha < 2$, $\gamma \ge 0$, or $\alpha=2$, $\gamma> 0$; one \textsl{can} also have $\alpha = 1$, $\gamma=0$, as is indeed attested by the relevant explicit expression in Eq.~(\ref{ec7}). Similarly for the case of $u\uparrow 0$.

The above observations are of direct relevance to the case of the expressions in Eqs.~(\ref{e342}) and (\ref{e343}), where in particular the question arises as to whether one may have $\alpha = 1$ for the exponent $\alpha$ in these expressions. Above we have indirectly shown that insofar as the expressions in Eqs.~(\ref{e342}) and (\ref{e343}) are concerned, the combination $\alpha = 1$, $\gamma=0$ cannot \emph{a priori} be ruled out.

% Appendix D.
\section{On the question of whether or not the GS number-density distribution function of the Hubbard Hamiltonian is non-interacting \texorpdfstring{$v$}{}-representable}
\label{sad}
The problem that we consider in this appendix is not directly related to the main subjects dealt with in this paper. It is however of relevance to the question with regard to the applicability of a specific formulation of the zero-temperature many-body perturbation theory, presented in Refs.~\cite{BF97a,BF97b,BF99b}, to the Hubbard Hamiltonian, when the latter is considered in a specific way to be described below. In this formulation, which explicitly relies on the assumption of the non-interacting $v$-representability of the GS number-density of the interacting system, a self-consistent mean-field Hamiltonian (coinciding with the Kohn-Sham Hamiltonian \cite{KS65}) is generated that with considerable advantage can be used in many-body calculations.

There are two distinct ways in which the Hubbard Hamiltonian may be considered, one in which the direct space consists of the lattice sites $\{ {\bm R}_j\}$ (we assume $\{ {\bm R}_j\}$ to be a Bravais lattice) embedded within $\mathds{R}^d$ (I), and the other in which the direct space consists of a continuum subset of $\mathds{R}^d$ in which the lattice $\{ {\bm R}_j\}$ signifies positions of `atoms' (II) (here we consider mono-atomic crystals and systems subject to periodic boundary condition). Almost all theoretical treatments of the Hubbard Hamiltonian view this Hamiltonian from the former perspective, however the latter perspective is the one adopted in the original systematic derivation of the Hubbard Hamiltonian in Ref.~\cite{JH63}.

In description I, the single-particle Hilbert space of the Hubbard Hamiltonian is spanned by $\{\psi_{\bm{k}}^{(\textsc{i})}(\bm{R}_j)  \| \bm{k}\in \textrm{1BZ}\}$, where
\begin{equation}
\psi_{\bm{k}}^{(\textsc{i})}(\bm{R}_j) \doteq \frac{1}{\sqrt{N_{\textsc{s}}}}\, \e^{i \bm{k}\cdot \bm{R}_j},\;\; j \in \{1,2,\dots,N_{\textsc{s}}\},
\label{ed1}
\end{equation}
and in description II, this space is spanned by the Bloch functions $\{\psi_{\bm{k}}^{(\textsc{ii})}({\bm r}) \| \bm{k}\in \textrm{1BZ}\}$, where \cite[Eq.~(4)]{JH63}
\begin{equation}
\psi_{\bm{k}}^{(\textsc{ii})}({\bm r}) \doteq \frac{1}{\sqrt{N_{\textsc{s}}}} \sum_{j=1}^{N_{\textsc{s}}} \e^{i \bm{k}\cdot \bm{R}_j} \phi({\bm r}-\bm{R}_j),\;\; \bm{r} \in \mathds{R}^d,
\label{ed2}
\end{equation}
in which $\phi(\bm{r})$ is an atomic orbital centred at $\bm{r} = \bm{0}$. One has
\begin{equation}
\sum_{j=1}^{N_{\textsc{s}}} \psi_{\bm{k}}^{(\textsc{i}) *}(\bm{R}_j) \psi_{\bm{k}'}^{(\textsc{i})}(\bm{R}_j) = \delta_{\bm{k},\bm{k}'},
\label{ed3}
\end{equation}
and, provided that
\begin{equation}
\int \mathrm{d}^dr\; \phi^*(\bm{r}-\bm{R}_j) \phi(\bm{r}-\bm{R}_{j'}) = \delta_{j,j'},
\label{ed4}
\end{equation}
\begin{equation}
\int \mathrm{d}^dr\; \psi_{\bm{k}}^{(\textsc{ii}) *}({\bm r}) \psi_{\bm{k}'}^{(\textsc{ii})}({\bm r}) = \delta_{\bm{k},\bm{k}'}.
\label{ed5}
\end{equation}
Irrespective of the nature of the orbital $\phi(\bm{r})$, for $\bm{k}\not= \bm{k}'$ the validity of the equality in Eq.~(\ref{ed5}) is guaranteed on account of $\psi_{\bm{k}}^{(\textsc{ii})}({\bm r})$ and $\psi_{\bm{k}'}^{(\textsc{ii})}({\bm r})$, constructed according to the Bloch sum in Eq.~(\ref{ed2}), belonging to different irreducible representations of the translation group associated with $\{ \bm{R}_i\}$ \cite{JFC84}. The condition in Eq.~(\ref{ed4}) is only necessary in order for $\psi_{\bm{k}}^{(\textsc{ii})}({\bm r})$ to be normalized to unity for all $\bm{k}$.

With $\h{a}_{\bm{k};\sigma}$ denoting the canonical annihilation operator corresponding to wave vector $\bm{k}$ and spin index $\sigma$, for the annihilation field operators one has \cite[Eq.~(2.1)]{FW03}
\begin{eqnarray}
\h{\psi}_{\sigma}^{(\textsc{i})}(\bm{R}_j) &=& \sum_{\bm{k}\in \mathrm{1BZ}} \h{a}_{\bm{k};\sigma}\, \psi_{\bm{k}}^{(\textsc{i})}(\bm{R}_j) \doteq \h{c}_{j;\sigma},
\label{ed6} \\
\h{\psi}_{\sigma}^{(\textsc{ii})}(\bm{r}) &=& \sum_{\bm{k}\in \mathrm{1BZ}} \h{a}_{\bm{k};\sigma}\, \psi_{\bm{k}}^{(\textsc{ii})}(\bm{r}).
\label{ed7}
\end{eqnarray}
One can explicitly show that $\{\h{c}_{j;\sigma}\| j, \sigma\}$ are canonical site annihilation operators, satisfying $[\h{c}_{j;\sigma}^{\phantom{\dag}},\h{c}_{j';\sigma'}^{\dag}]_+ = \delta_{j,j'}\hspace{0.6pt} \delta_{\sigma,\sigma'}$. Further, on replacing the $\psi_{\bm{k}}^{(\textsc{ii})}(\bm{r})$ on the RHS of Eq.~(\ref{ed7}) with the expression for this function as presented in Eq.~(\ref{ed2}), making use of the defining expression for $\h{c}_{j;\sigma}$ in Eq.~(\ref{ed6}), one readily verifies that $\h{\psi}_{\sigma}^{(\textsc{ii})}(\bm{r}) = \sum_{j=1}^{N_{\textsc{s}}} \phi(\bm{r}-\bm{R}_j)\hspace{0.6pt} \h{c}_{j;\sigma}$, establishing that indeed $\h{c}_{j;\sigma}$ annihilates a fermion with spin index $\sigma$ in the orbital state $\phi(\bm{r}-\bm{R}_j)$ \cite[p.~242]{JH63}.

On the basis of the expressions in Eqs.~(\ref{ed6}) and (\ref{ed7}), for the GS \textsl{site-occupation numbers} $\{n_{\sigma}^{(\textsc{i})}(\bm{R}_{j})\}$ in the $N$-particle \textsl{uniform} GS of $\wh{\mathcal{H}}$ one has
\begin{eqnarray}
n_{\sigma}^{(\textsc{i})}(\bm{R}_{j}) &\doteq& \langle\Psi_{N;0}\vert\h{\psi}_{\sigma}^{(\textsc{i})\dag}(\bm{R}_j) \h{\psi}_{\sigma}^{(\textsc{i})}(\bm{R}_j)\vert\Psi_{N;0}\rangle \nonumber\\
&\equiv& \frac{1}{N_{\textsc{s}}} \sum_{\bm{k}\in \textrm{1BZ}} \mathsf{n}_{\sigma}(\bm{k}) = n_{\sigma},\;\;\forall j,
\label{ed8}
\end{eqnarray}
where in arriving at the last equality we have used the expressions in Eqs.~(\ref{e230}) and (\ref{e12}). The result in Eq.~(\ref{ed8}) could be directly inferred on the basis of the observation that $\h{\psi}_{\sigma}^{(\textsc{i})\dag}(\bm{R}_j) \h{\psi}_{\sigma}^{(\textsc{i})}(\bm{R}_j) \equiv \h{c}_{j;\sigma}^{\dag} \h{c}_{j;\sigma}^{\phantom{\dag}}$, Eq.~(\ref{ed6}). On the other hand, for the \textsl{number density} $n_{\sigma}^{(\textsc{ii})}(\bm{r})$ in the \textsl{uniform} GS one has
\begin{eqnarray}
n_{\sigma}^{(\textsc{ii})}(\bm{r}) &\doteq& \langle\Psi_{N;0}\vert\h{\psi}_{\sigma}^{(\textsc{ii})\dag}(\bm{r}) \h{\psi}_{\sigma}^{(\textsc{ii})}(\bm{r})\vert\Psi_{N;0}\rangle \nonumber\\
&\equiv& \sum_{\bm{k}\in \textrm{1BZ}} \mathsf{n}_{\sigma}(\bm{k})\, \vert\psi_{\bm{k}}^{(\textsc{ii})}({\bm r})\vert^2,
\label{ed9}
\end{eqnarray}
where in arriving at the last expression we have used the equality $\langle\Psi_{N;0}\vert \h{a}_{\bm{k};\sigma}^{\dag} \h{a}_{\bm{k}';\sigma}^{\phantom{\dag}}\vert \Psi_{N;0}\rangle = \mathsf{n}_{\sigma}(\bm{k})\hspace{0.6pt}\delta_{\bm{k},\bm{k}'}$, Eq.~(\ref{e28}) (note that eigenstates of the total-momentum operator $\h{P}$ \cite[Eq.~(7.50)]{FW03} corresponding to different eigenvalues are orthogonal). One observes that whereas the form of $\mathsf{n}_{\sigma}(\bm{k})$, so long as it yields the exact value for $N_{\sigma}$ through the expression in Eq.~(\ref{e230}), plays \textsl{no} role on the behaviour of $n_{\sigma}^{(\textsc{i})}(\bm{R}_{j})$ as a function of $j$, the form of $\mathsf{n}_{\sigma}(\bm{k})$ is of direct consequence for the behaviour of $n_{\sigma}^{(\textsc{ii})}(\bm{r})$ as a function of $\bm{r}$.

The question with regard to the non-interacting $v$-representability of $n_{\sigma}^{(\textsc{ii})}(\bm{r})$ can be expressed as follows: is there an $\mathsf{n}_{0;\sigma}(\bm{k}) \in \{0,1\}$, $\forall \bm{k}\in \textrm{1BZ}$, for which
\begin{equation}
\sum_{\bm{k}\in \textrm{1BZ}} \mathsf{n}_{0;\sigma}(\bm{k})\, \vert\psi_{\bm{k}}^{(\textsc{ii})}({\bm r})\vert^2 \equiv n_{\sigma}^{(\textsc{ii})}(\bm{r})\,?
\label{ed10}
\end{equation}
\emph{The answer to this question is in the negative.} The function $\mathsf{n}_{\sigma}(\bm{k})$ not being in general identically vanishing for any $\bm{k} \in \textrm{1BZ}$, according to Eq.~(\ref{ed9}) $n_{\sigma}^{(\textsc{ii})}(\bm{r})$ has contributions corresponding to the entire single-particle Hilbert space of the problem, spanned by the \textsl{complete} set $\{\psi_{\bm{k}}^{(\textsc{ii})}(\bm{r}) \| \bm{k}\}$ whose cardinal number is $N_{\textsc{s}}$. Unless all sites (`atoms') are either singly \textsl{or} doubly occupied (assuming the spin index of all particles in the former case to be $\sigma$), this is however not the case with $n_{0;\sigma}^{(\textsc{ii})}(\bm{r})$ (the function on the LHS of Eq.~(\ref{ed10})), which owing to $\mathsf{n}_{0;\sigma}(\bm{k}) \in \{0,1\}$, $\forall \bm{k}\in \textrm{1BZ}$, has contributions corresponding to only $N_{\sigma}$ (where $N_{\sigma} < N_{\textsc{s}}$) basis functions from the set $\{\psi_{\bm{k}}^{(\textsc{ii})}(\bm{r}) \| \bm{k}\}$.

Alternatively, the validity of the identity in Eq.~(\ref{ed10}) would imply
\begin{equation}
\sum_{\bm{k}\in \textrm{1BZ}} \big( \mathsf{n}_{\sigma}(\bm{k}) - \mathsf{n}_{0;\sigma}(\bm{k})\big)\, \vert\psi_{\bm{k}}^{(\textsc{ii})}({\bm r})\vert^2 = 0,\;\; \forall \bm{r}.
\label{ed11}
\end{equation}
Since the set 1BZ is \textsl{countable} (consisting of $N_{\textsc{s}}$ distinct points), it is in principle possible that an $\mathsf{n}_{0;\sigma}(\bm{k}) \in \{0,1\}$ may exist for which the equality in Eq.~(\ref{ed11}) applies for $\bm{r}$ varying over a countable set, however this \textsl{cannot} be the case for $\bm{r}$ varying over a set consisting of a \textsl{continuum} of points. For clarity, let $(\mathbb{A})_{l,l'} \doteq \vert\psi_{\bm{k}_{l'}}^{(\textsc{ii})}({\bm r}_l)\vert^2$, where $\{ \bm{k}_l \| l =1,2,\dots,N_{\textsc{s}}\} \equiv \textrm{1BZ}$. Thus, for, e.g., $\bm{r}_l \in \{\bm{R}_j\| j=1,2,\dots, N_{\textsc{s}}\}$, $\mathbb{A}$ is an $N_{\textsc{s}} \times N_{\textsc{s}}$ matrix and the equality in Eq.~(\ref{ed11}) signifies $\{\mathsf{n}_{\sigma}(\bm{k}_l) - \mathsf{n}_{0;\sigma}(\bm{k}_l)\| l\}$ as being the set of the components of an eigenvector of $\mathbb{A}$ corresponding to zero eigenvalue. Since it cannot \emph{a priori} be ruled out that $\mathbb{A}$ is singular, for $\bm{r}_l \in \{\bm{R}_j\}$ it is in principle possible that the latter eigenvector may be non-trivial, i.e. while $\mathsf{n}_{\sigma}(\bm{k}) \not\equiv \mathsf{n}_{0;\sigma}(\bm{k})$, the equality in Eq.~(\ref{ed11}) may in principle be satisfied. This \textsl{cannot} be the case however when $\bm{r}$ varies over a \textsl{continuum} set, in which case the equality in Eq.~(\ref{ed11}) amounts to an uncountably large number of conditions to be satisfied by the countable set of variables $\{\mathsf{n}_{\sigma}(\bm{k}_l) - \mathsf{n}_{0;\sigma}(\bm{k}_l)\| l\}$. The same applies for $\bm{r}$ varying over a countable set whose cardinal number is greater than $N_{\textsc{s}}$.

To appreciate the peculiarity of the Hubbard model when viewed from perspective II, one should realize that the single-particle Hilbert space of this model is rigid, it being spanned by the \textsl{fixed} set of functions $\{\psi_{\bm{k}}^{(\textsc{ii})}(\bm{r}) \| \bm{k}\}$, Eq.~(\ref{ed2}), independent of the coupling constant $\lambda$ of interaction. The very crucial aspect of the $\lambda$-dependent \textsl{self-consistent} Kohn-Sham orbitals \cite{KS65} that are encountered in the conventional applications of the density-functional theory to \textsl{inhomogeneous} GSs, is missing here entirely, a fact reflected in the expression on the RHS of Eq.~(\ref{ed9}) and that on the LHS of Eq.~(\ref{ed10}); for any arbitrary non-negative value of $\lambda$, $n_{\sigma}^{(\textsc{ii})}(\bm{r})$ is determined fully by $\mathsf{n}_{\sigma}(\bm{k})$, the `orbitals' $\{\psi_{\bm{k}}^{(\textsc{ii})}(\bm{r}) \| \bm{k}\}$ being entirely independent of $\lambda$. It is interesting to note that $n_{\sigma}^{(\textsc{ii})}(\bm{r})$ is the diagonal part of the single-particle density matrix $\varrho_{\sigma}^{(\textsc{ii})}(\bm{r},\bm{r}')$, described by an expression similar to that on the RHS of Eq.~(\ref{ed9}) in which  $\psi_{\bm{k}}^{(\textsc{ii})}({\bm r}) \psi_{\bm{k}}^{(\textsc{ii})\hspace{0.4pt}*}({\bm r}')$ takes the place of
$\vert\psi_{\bm{k}}^{(\textsc{ii})}({\bm r})\vert^2$. The fact that for interacting GSs $\mathsf{n}_{\sigma}(\bm{k}) \not\in\{0,1\}$, whereby $\mathsf{n}_{\sigma}^2(\bm{k}) \not\equiv \mathsf{n}_{\sigma}(\bm{k})$, accounts for the non-idempotency of $\varrho_{\sigma}^{(\textsc{ii})}(\bm{r},\bm{r}')$ for these GSs.

We have thus demonstrated that for the Hubbard Hamiltonian the \textsl{number densities} $\{ n_{\sigma}^{(\textsc{ii})}({\bm r})\}$, to be distinguished from the \textsl{occupation numbers} $\{ n_{\sigma}^{(\textsc{i})}(\bm{R}_j)\}$, are \textsl{not} non-interacting $v$-representable. This is the result arrived at by Schindlmayr and Godby \cite{SG95} through their numerical calculations on a number of one-dimensional Hubbard chains. Although in a subsequent paper Sch\"onhammer \emph{et al.} \cite{SGN95} have pointed out the distinction between the ``\textsl{site occupation function(al) theory}'' and the \textsl{density functional theory}, we believe that the details in this appendix constitute the first explicit demonstration of the exactness of the numerical finding by Schindlmayr and Godby \cite{SG95} regarding the failure of the interacting $\{ n_{\sigma}^{(\textsc{ii})}({\bm r})\}$ to be non-interacting $v$-representable.

Before closing, one remark is in order. From Eqs.~(\ref{ed6}) and (\ref{ed1}) one deduces that
\begin{equation}
\h{a}_{\bm{k};\sigma} = \frac{1}{\sqrt{N_{\textsc{s}}}} \sum_{j=1}^{N_{\textsc{s}}} \h{c}_{j;\sigma}\, \e^{-i \bm{k}\cdot \bm{R}_j},
\label{ed12}
\end{equation}
whereby one can write (cf. Eq.~(\ref{ed9}))
\begin{eqnarray}
&&\hspace{-1.1cm}\h{n}_{\sigma}^{(\textsc{ii})}(\bm{r}) \doteq \h{\psi}_{\sigma}^{(\textsc{ii})\dag}(\bm{r}) \h{\psi}_{\sigma}^{(\textsc{ii})}(\bm{r})\nonumber\\
&&\hspace{0.05cm}=\sum_{j,j'} \phi^*(\bm{r}-\bm{R}_j) \phi(\bm{r}-\bm{R}_{j'})\, \h{c}_{j;\sigma}^{\dag} \h{c}_{j';\sigma}.
\label{ed13}
\end{eqnarray}
With $\h{n}^{(\textsc{ii})}(\bm{r}) \doteq \sum_{\sigma} \h{n}_{\sigma}^{(\textsc{ii})}(\bm{r})$, and barring one minor difference, the expression for $\h{n}^{(\textsc{ii})}(\bm{r})$ coincides with the expression for $\h{n}(\bm{r})$ employed in Ref.~\cite{SG95} (see Eq.~(5) herein), so that our above statement regarding the non-interacting $v$-representability of $\h{n}_{\sigma}^{(\textsc{ii})}(\bm{r})$ has indeed direct bearing on the above-mentioned numerical finding by Schindlmayr and Godby. The difference to which we have just referred, corresponds to the restriction of the double summation $\sum_{j,j'}$ as encountered in the expression in Eq.~(\ref{ed13}) to a summation over pairs of nearest neighbours in Eq.~(5) of Ref.~\cite{SG95}, signified by $\sum_{<i,j>}$. This is not a fundamental difference, since change of $\sum_{j,j'}$ into $\sum_{<i,j>}$ does not lead to a change of the form of the expression in Eq.~(\ref{ed9}) above, rather to a \textsl{modification} of the behaviour of the $\vert\psi_{\bm{k}}^{(\textsc{ii})}(\bm{r})\vert^2$ herein as a function of $\bm{r}$ and $\bm{k}$. Quantitatively, this modification is relatively small for $\phi(\bm{r})$ a sufficiently localised orbital centred at $\bm{r}=\bm{0}$.
$\hfill\Box$

\end{appendix}
%_______________________________________________________________

%\vspace{0.6cm}

\bibliographystyle{apsrev}

\begin{thebibliography}{10}

% 1.
\bibitem{BF03}
B. Farid, \emph{Phil. Mag.} \textbf{83}, 2829 (2003). \href{http://arxiv.org/abs/cond-mat/0211244}{arXiv:0211244}

% 2.
\bibitem{BF04a}
B. Farid, \emph{Phil. Mag.} \textbf{84}, 109 (2004). \href{http://arxiv.org/abs/cond-mat/0304350}{arXiv:0304350}

% 3.
\bibitem{BF04b}
B. Farid, \emph{Phil. Mag.} \textbf{84}, 909 (2004). \href{http://arxiv.org/abs/cond-mat/0308090}{arXiv:0308090}

% 4.
\bibitem{PWA59}
P.~W. Anderson, \emph{Phys. Rev.} \textbf{115}, 2 (1959).

% 5.
\bibitem{ThWR62}
Th. W. Ruijgrok, \emph{Physica} \textbf{28}, 877 (1962).

% 6.
\bibitem{JH63}
J. Hubbard, \emph{Proc. Roy. Soc.} (London), A\hspace{0.6pt}\textbf{276}, 238 (1963).

% 7.
\bibitem{LW60}
J.~M. Luttinger, and J.~C. Ward, \emph{Phys. Rev.} \textbf{118}, 1417 (1960).

% 8.
\bibitem{JML60}
J.~M. Luttinger, \emph{Phys. Rev.} \textbf{119}, 1153 (1960).

% 9.
\bibitem{ID03}
I. Dzyaloshinskii, \emph{Phys. Rev.} B~\textbf{68}, 085113 (2003).

% 10.
\bibitem{BF07a}
B. Farid, \textsl{On the Luttinger theorem concerning number of particles in the ground states of systems of interacting fermions}, \href{http://arxiv.org/abs/0711.0952}{arXiv:0711.0952v1}.

% 11.
\bibitem{BF07-12}
\emph{a}. B. Farid, \textsl{Reply to ``Comment `On the Luttinger theorem \dots fermions', \textrm{arXiv:0711.0952v1}, by B. Farid'', by A. Rosch}, \href{http://arxiv.org/abs/0711.3195}{arXiv:0711.3195}. \\
\emph{b}. B. Farid, and A.~M. Tsvelik, \textsl{Comment on ``Breakdown of the Luttinger sum rule within the Mott-Hubbard insulator'', by J. Kokalj and P. Prelov\v{s}ek, \emph{Phys. Rev.} B~\textbf{78}, 153103 (2008) [arXiv:0803.4468]}, \href{http://arxiv.org/abs/0909.2886}{arXiv:0909.2886}.\\
\emph{c}. B. Farid, \textsl{Comment on ``Violation of the Luttinger sum rule within the Hubbard model on a triangular lattice'', by J. Kokalj and P. Prelov\v{s}ek, \emph{Eur. Phys. J.} B~\textbf{63}, 431 (2008) [arXiv:arXiv:0709.0263]}, \href{http://arxiv.org/abs/0909.2887}{arXiv:0909.2887}.\\
\emph{d}. B. Farid, \textsl{Comment on ``Absence of Luttinger's Theorem'', by Kiaran B. Dave, Philip W. Phillips and Charles L. Kane, arXiv:1207.4201}, \href{http://arxiv.org/abs/1211.5612}{arXiv:1211.5612}.

% 12.
\bibitem{EMH89b}
E. M\"{u}ller-Hartmann, \emph{Z. Phys.} B~\textbf{76}, 211 (1989).

% 13.
\bibitem{MV89}
W. Metzner, and D. Vollhardt, \emph{Phys. Rev. Lett.} \textbf{62}, 324 (1989).

% 14.
\bibitem{EMH89a}
E. M\"{u}ller-Hartmann, \emph{Z. Phys.} B~\textbf{74}, 507 (1989).

% 15.
\bibitem{GKKR96}
A. Georges, G. Kotliar, W. Krauth, and M.~J. Rozenberg, \emph{Rev. Mod. Phys.} \textbf{68}, 13 (1996).

% 16.
\bibitem{PF03}
P. Fazekas, \emph{Lecture Notes on Electron Correlation and Magnetism} (World Scientific, Singapore, 2003).

% 17.
\bibitem{J06}
J.~L.~W.~V. Jensen, \emph{Acta Math.} \textbf{30}, 175 (1906).

% 18.
\bibitem{WT90}
W. Thirring, \emph{Found. Phys.} \textbf{20}, 1103 (1990).

% 19.
\bibitem{FW03}
A.~L. Fetter, and J.~D. Walecka, \emph{Quantum Theory of Many-Particle Systems} (Dover, New York, 2003).

% 20.
\bibitem{Note1}
Following the conditions under which $\{\vert\Phi_{N_{\sigma}\mp 1,N_{\b\sigma};\bm{k}}\rangle\}$ are defined (Eqs.~(\protect\ref{e29}) and (\protect\ref{e210})), in many (\emph{not} all) instances in this paper the statement `$\forall\bm{k}$' is to be understood as excluding those $\bm{k}$ at which either $\mathsf{n}_{\sigma}(\bm{k}) = 0$ or $\mathsf{n}_{\sigma}(\bm{k}) = 1$. With reference to the expressions in Eqs.~(\protect\ref{e249}) and (\protect\ref{e250}), we note however that for \textsl{interacting} systems both $\mathsf{n}_{\sigma}(\bm{k}) = 0$ and $\mathsf{n}_{\sigma}(\bm{k}) = 1$ correspond to rather pathological cases, this on account of $A_{\sigma}(\bm{k};\varepsilon) \ge 0$, $\forall \bm{k},\varepsilon$.

% 21.
\bibitem{ZSS95}
V. Zlati\'{c}, K.~D. Schotte, and G. Schliecker, \emph{Phys. Rev.} B~\textbf{52}, 3639 (1995).

% 22.
\bibitem{BF02}
B.~Farid, \emph{Phil. Mag.} B~\textbf{82}, 1413 (2002). \href{http://arxiv.org/abs/cond-mat/0110481}{arXiv:0110481}

% 23.
\bibitem{BF99}
B. Farid, \emph{Phil. Mag.} B~\textbf{79}, 1097 (1999). \href{http://arxiv.org/abs/cond-mat/0004476}{arXiv:0004476}

% 24.
\bibitem{ABM57}
A.~B. Migdal, \emph{Soviet Phys. JETP}, \textbf{5}, 333 (1957).

% 25.
\bibitem{ML65}
D.~C. Mattis, and E.~H. Lieb, \emph{J. Math. Phys.} \textbf{6}, 304 (1965).

% 26.
\bibitem{JV94}
J. Voit, \emph{Rep. Prog. Phys.} \textbf{57}, 977 (1994).

% 27.
\bibitem{Note2}
See p.~989 in Ref.~\protect\cite{JV94}, where, in discussing the `$g$-ology', the momentum independence of $g_4$ is indicated as the condition for which $\Sigma_{\sigma}^{\textsc{hf}}(k) \equiv 0$ applies.

% 28.
\bibitem{BB97}
K.~B. Blagoev, and K.~S. Bedell, \emph{Phys. Rev. Lett.} \textbf{79}, 1106 (1997).

% 29.
\bibitem{YDS05}
S. Yunoki, E. Dagotto, and S. Sorella, \emph{Phys. Rev. Lett.} \textbf{94}, 037001 (2005).

% 30.
\bibitem{GM58}
V.~M. Galitskii and A.~B. Migdal, \emph{Soviet Phys. JETP}, \textbf{7}, 96 (1958).

% 31.
\bibitem{HS91}
N.~B. Haaser, and J.~A. Sullivan, \emph{Real Analysis} (Dover, New York, 1991).

% 32.
\bibitem{SG88}
K. Sch\"onhammer, and O. Gunnarsson, \emph{Phys. Rev.} B~\textbf{37}, 3128 (1988).

% 33.
\bibitem{HM97}
C.~J. Halboth, and W. Metzner, \emph{Z. Phys.} B~\textbf{102}, 501 (1997).

% 34.
\bibitem{YY99}
S. Yoda, and K. Yamada, \emph{Phys. Rev.} B~\textbf{60}, 7886 (1999).

% 35.
\bibitem{ZEG96}
V. Zlati\'{c}, P. Entel, and S. Grabowski, \emph{Europhys. Lett.} \textbf{34}, 693 (1996).

% 36.
\bibitem{PN64}
P. Nozi\`eres, \emph{Theory of Interacting Fermi Systems} (W.~A. Benjamin, New York, 1964).

% 37.
\bibitem{JFC84}
J.~F. Cornwell, \emph{Group Theory in Physics}, Vol. I (Academic Press, London, 1984).

% 38.
\bibitem{AGD75}
A.~A. Abrikosov, L.~P. Gorkov, and I.~E. Dzyaloshinski, \emph{Methods of Quantum Field Theory in Statistical Physics} (Dover, New York, 1975).

% 39.
\bibitem{WW62}
E.~T. Whittaker, and G.~N. Watson, \emph{A Course of Modern Analysis}, 4th edition (Cambridge University Press, 1962).

% 40.
\bibitem{JML61}
J.~M. Luttinger, \emph{Phys. Rev.} \textbf{121}, 942 (1961).

% 41.
\bibitem{KL60}
W. Kohn, and J.~M. Luttinger, \emph{Phys. Rev.} \textbf{118}, 41 (1960).

% 42.
\bibitem{NO98}
J.~W. Negele, and H. Orland, \emph{Quantum Many-Particle Systems} (Westview Press, Boulder, Colorado, 1998).

% 43.
\bibitem{Note3}
In Ref.~\protect\cite{BF07a} we have rigorously demonstrated that at least in the case of a lattice model, such as the single-band Hubbard Hamiltonian, $\sum_{\nu = 1}^{\infty} \t{\Sigma}_{\sigma}^{(\nu)}(\bm{k};z;[\{G_{\sigma'}\}])$ is \textsl{uniformly} convergent for almost all $\bm{k}$ and $z$. In other words, $\t{\Sigma}_{\sigma}(\bm{k};z)$ as defined in terms of an infinite sum of contributions arising from skeleton self-energy diagrams evaluated in terms of the exact single-particle Green functions $\{G_{\sigma} \| \sigma\}$, is also a well-defined function.

% 44.
\bibitem{ETC65}
E.~T. Copson, \emph{Asymptotic Expansions} (Cambridge University Press, 1965).

% 45.
\bibitem{HAL74}
H.~A. Lauwerier, \emph{Asymptotic Analysis} (Mathematisch Centrum, Amsterdam, 1974).

% 46.
\bibitem{IZ80}
C. Itzykson, and J.-B. Zuber, \emph{Quantum Field Theory} (McGraw-Hill, New York, 1980).

% 47.
\bibitem{GST94}
G.-S. Tian, \emph{J. Phys.} A~\textbf{27}, 3635 (1994).

% 48.
\bibitem{Note4}
Note that for the $N$-particle paramagnetic uniform metallic GS under consideration one has $\mathcal{S}_{\textsc{f}}^{(0)} = \mathcal{S}_{\textsc{f};\sigma}^{\textsc{hf}}$, whereby, following the relationship in Eq.~(\protect\ref{e241}), $\mathcal{S}_{\textsc{f};\sigma} \subseteq \mathcal{S}_{\textsc{f}}^{(0)}$.

% 49.
\bibitem{OS90}
M. Ogata, and H. Shiba, \emph{Phys. Rev.} B~\textbf{41}, 2326 (1990).

% 50.
\bibitem{BvdL91}
D. Baeriswyl, and W. von der Linden, in \emph{The Hubbard Model -- Recent Results}, edited by Mario Rasetti (World Scientific, Singapore, 1991), pp. 135-150.

% 51.
\bibitem{Note5}
To be precise, while the authors of Ref.~\protect\cite{BvdL91} state that ``For small densities, where the Fermi surface [in $d=2$] corresponds essentially to a circle, one can show analytically that $n^{(2)}(\bm{k})$ [the second-order contribution to $\mathsf{n}_{\sigma}(\bm{k})$] saturates for $\bm{k} \to \bm{k}_{\textsc{f}}$ [$\bm{k} \to \bm{k}_{\textsc{f};\sigma}$]'', they do not state however whether or not the saturated value of $\mathsf{n}_{\sigma}^{(0)}(\bm{k}) + \mathsf{n}_{\sigma}^{(2)}(\bm{k})$ satisfies the strict inequalities $0\le \mathsf{n}_{\sigma}(\bm{k}) \le 1$, to be satisfied for all $\bm{k}$. From the data presented in Figs.~6 and 7 in Ref.~\protect\cite{BvdL91}, which correspond to quarter-filled bands, one can infer that this is \textsl{not} the case for a fixed value of $U$.

% 52.
\bibitem{Note6}
In Ref.~\protect\cite{BF04a} (see Fig.~3 herein) we have indicated that $\upalpha_{\sigma}^{-}$ and $\upalpha_{\sigma}^{+}$ approach zero for $\alpha\downarrow 0$, or $r_{\text{s}}\downarrow 0$, after taking the approximate value of $1.7$ (presented in E. Daniel, and S.~H. Vosko, \emph{Phys. Rev.} \textbf{120}, 2041 (1960)) over a relatively broad range of small values of $\alpha$ (here $\alpha$ denotes the coupling constant of the Coulomb interaction potential in $d=3$). On recalculating $\upalpha_{\sigma}^{\pm}$, we have established that the conclusion $\upalpha_{\sigma}^{\pm} \to 0$ for $\alpha \downarrow 0$ has been a consequence of performing the calculations reported in Ref.~\protect\cite{BF04a} with a small but \textsl{fixed} value of the \textsl{positive} quantity $\delta k$ in $k_{\textsc{f}}^{\pm} \doteq k_{\textsc{f}} \pm \delta k$. In other words, the values of $\upalpha_{\sigma}^{\pm}$ depend on the order in which the limits $\alpha\downarrow 0$ ($r_{\text{s}}\downarrow 0$) and $\delta k \downarrow 0$ are taken. Indeed, by calculating $\upalpha_{\sigma}^{\pm}$ for $\delta k/k_{\textsc{f}} \ll \alpha$, we have obtained $\upalpha_{\sigma}^{\pm} = 1.676 \pm 10^{-3}$ for $\alpha\downarrow 0$ (for clarity, the $\pm$ on the RHS of this equality has \textsl{no} bearing on the $\pm$ on the LHS), in conformity with the above-mentioned approximate value of $1.7$. Interestingly, the exact value of $\upalpha_{\sigma}^{\pm}$ specific to effecting $\alpha\downarrow 0$ ($r_{\text{s}}\downarrow 0$) \textsl{after} having effected the limit $\delta k \downarrow 0$, turns out to be equal to $G(0)/2$, where $G(x)$ is the so-called Kulik function (appendix A in P. Gori-Giori and P. Ziesche, \emph{Phys. Rev.} B~\textbf{66}, 235116 (2002)), for which one has $G(0) = 3.353\,337\,262\dots$, whereby $\upalpha_{\sigma}^{\pm} = 1.676\,668\,631\,444\,736\dots$~. Remarkably, the explicit expressions for $1-\mathsf{n}_{\sigma}(k_\textsc{f}-\delta k)$ and $\mathsf{n}_{\sigma}(k_\textsc{f}+\delta k)$ for $\delta k \downarrow 0$ (see the above-mentioned appendix A) reveal these as being functions of $k_{\textsc{f}}^{\pm}$, $r_{\text{s}}$ and the ratio $\delta k/\sqrt{r_{\text{s}}}$. This fact uncovers the mechanism by which, insofar as $\upalpha_{\sigma}^{\pm}$ are concerned, the processes of effecting the limits $r_{\text{s}}\downarrow 0$ and $\delta k \downarrow 0$ do not commute: effecting the limit $r_{\text{s}}\downarrow 0$ for a fixed positive value of $\delta k$, one probes the behaviour of the Kulik function $G(x)$ in the neighbourhood of $x=+\infty$ (to leading order, $G(x)$ decays like $A/x^2$ for $x\to\infty$, where $A = \frac{\pi}{6} (1-\ln 2)$), while effecting the limit $\delta k\downarrow 0$ for a fixed positive value of $r_{\text{s}}$, one probes the behaviour of the Kulik function $G(x)$ in the neighbourhood of $x=0$ (to leading order, $G(x) - G(0) \sim B x \ln(x)$, where $B = \pi (\frac{\pi}{4} + \sqrt{3})$).

% 53.
\bibitem{VT97}
Y.~M. Vilk, and A.-M.~S. Tremblay, \emph{J. Phys.} I (France) \textbf{7}, 1309 (1997).

% 54.
\bibitem{KAT01}
B. Kyung, S. Allen, and A.-M.~S. Tremblay, \emph{Phys. Rev.} B~\textbf{64}, 075116 (2001).

% 55.
\bibitem{VLBCJS99}
C.~N. Varney, C.-R. Lee, Z.~J. Bai, S. Chiesa, M. Jarrell, and R. T. Scalettar, \emph{Phys. Rev.} B~\textbf{80}, 075116 (2009).

% 56.
\bibitem{LW68-03}
E.~H. Lieb, and F.~Y. Wu, \emph{Phys. Rev. Lett.} \textbf{20}, 1445 (1968); \emph{Physica} A\hspace{0.6pt}\textbf{321}, 1 (2003).

% 57.
\bibitem{vdLMR90}
W. von der Linden, I. Morgenstern, and H. de Raedt, \emph{Phys. Rev.} B~\textbf{41}, 4669 (1990).

% 58.
\bibitem{Note7}
In Ref.~\protect\cite{YY99} one encounters a notational confusion, corresponding to the identification of $\omega =0$ with both the exact chemical potential $\mu$ and the Hartree-Fock chemical potential $\mu^{\textsc{hf}}$. In this connection, with reference to the expressions in Eqs.~(6) and (9) of Ref.~\protect\cite{YY99}, note that $\omega = 0$ coincides with what according to the notation of the present paper is denoted by $\varepsilon = \mu$. In spite of this, $\omega = 0$ in Eqs.~(10) and (11) of Ref.~\protect\cite{YY99} coincides with what we in the present paper denote by $\varepsilon = \mu^{\textsc{hf}}$. See Sec.~3.0.1 in Ref.~\protect\cite{BF07a}.

%. 59.
\bibitem{HN99}
H. Nojiri, \emph{J. Phys. Soc. Jpn.} \textbf{68}, 903 (1999).

% 60.
\bibitem{BSG12}
J. B\"{u}nemann, T. Schickling, and F. Gebhard, \emph{Europhys. Lett.} \textbf{98}, 27006 (2012).

% 61.
\bibitem{MSSWB90}
A. Moreo, D.~J. Scalapino, R.~L. Sugar, S.~R. White, and N.~E. Bickers, \emph{Phys. Rev.} B~\textbf{41}, 2313 (1990).

% 62.
\bibitem{PTVF01}
W.~H. Press, S.~A. Teukolsky, W.~T. Vetterling, B.~P. Flannery, \emph{Numerical Recipes in Fortran 77. The Art of Scientific Computing}, 2nd edition (Cambridge University Press, 2001).

% 63.
\bibitem{Note8}
The present choice for $\h{\bm{n}}(\bm{k}_{\textsc{f};\sigma})$ is in keeping with our earlier choice \protect\cite{BF03,BF04a}. This choice has further the advantage of ensuring the condition $a_{\sigma} \ge 0$ for all points on $\mathcal{S}_{\textsc{f};\sigma}$.

% 64.
\bibitem{KK04}
A.~A. Katanin, and A.~P. Kampf, \emph{Phys. Rev. Lett.} \textbf{93}, 106406 (2004).

% 65.
\bibitem{IED96}
I.~E. Dzyaloshinskii, \emph{J. Phys.} I (France) \textbf{6}, 119 (1996).

% 66.
\bibitem{GGV96-7}
J. Gonz\'alez, F. Guinea, and M.~A.~H. Vozmediano, \emph{Europhys. Lett.} \textbf{34}, 711 (1996); \emph{Nucl. Phys.} B~\textbf{485}, 694 (1997).

% 67.
\bibitem{BF04c}
B. Farid, \emph{Comment on ``Quasiparticle Anisotropy and Pseudogap Formation from the Weak-Coupling Renormalization Group Point of View'', by A.~A. Katanin, and A.~P. Kampf, Phys. Rev. Lett. \textbf{93}, 106406
(2004)}, \href{http://arxiv.org/abs/cond-mat/0410453}{arXiv:0410453}.

% 68.
\bibitem{LM62}
E. Lieb, and D. Mattis, \emph{Phys. Rev.} \textbf{125}, 164 (1962).

% 69.
\bibitem{EMH95}
E. M\"{u}ller-Hartmann, \emph{J. Low Temp. Phys.} \textbf{99}, 349 (1995).

% 70.
\bibitem{JEH85}
J.~E. Hirsch, \emph{Phys. Rev.} B~\textbf{31}, 4403 (1985).

% 71.
\bibitem{RM85}
S. Rudin, and D.~C. Mattis, \emph{Phys. Lett.} \textbf{110}\hspace{0.7pt}A, 273 (1985).

% 72.
\bibitem{DHS03}
A. Damascelli, Z. Hussain, and Z.-X. Shen, \emph{Rev. Mod. Phys.} \textbf{75}, 473 (2003).

% 73.
\bibitem{HL69}
L. Hedin, and S. Lundqvist, in \emph{Solid State Physics}, Vol.~\textbf{23}, edited by F. Seitz, D. Turnbull and H. Ehrenreich (Academic Press, New York, 1969).

% 74.
\bibitem{HL67}
A.~B. Harris, and R.~V. Lange, \emph{Phys. Rev.} \textbf{157}, 295 (1967).

% 75.
\bibitem{Note9}
See Sections B.3 -- B.6 in appendix B of Ref.~\protect\cite{BF07a}. For instance, for the \textsl{continuum model} of the uniform-electron system in $d=3$, with electrons interacting through both a short-range and the long-range Coulomb potential, $\int_{-E}^{\mu} \rd\varepsilon\, \varepsilon^2\hspace{0.6pt} A_{\sigma}(\bm{k};\varepsilon)$ and $\int_{\mu}^{E} \rd\varepsilon\, \varepsilon^2\hspace{0.6pt} A_{\sigma}(\bm{k};\varepsilon)$ both diverge like $\ln(E)$ for $E\to\infty$ \protect\cite[Eq.~(239)]{BF02}.

% 77.
\bibitem{XJZ03}
X.~J. Zhou \emph{et al.}, \emph{Nature}, \textbf{423}, 398 (2003).

% 77.
\bibitem{AL01}
A. Lanzara \emph{et al.}, \emph{Nature}, \textbf{412}, 510 (2001).

% 78.
\bibitem{Note10}
If the abrupt change in $\mathsf{n}_{\sigma}(\bm{k})$ at $\bm{k} = \bm{k}_{\star}$ is not a true discontinuity, the distances of $\bm{k}_{\star}^{\pm}$ to $\bm{k}_{\star}$ are to be of the order of the inverse of the magnitude of the radial component of $\bm{\nabla}\mathsf{n}_{\sigma}(\bm{k})$ at $\bm{k} = \bm{k}_{\star}$.

% 79.
\bibitem{HTG04}
J. Hwang, T.~Timusk, and G.~D. Gu, \emph{Nature}, \textbf{427}, 714 (2004).

% 80.
\bibitem{Note11}
Date of publication of Ref.~\protect\cite{HTG04}, i.e. 19 Feb. 2004 (announced on 25 Feb. 2004 on \href{http://arxiv.org/abs/cond-mat/0402612}{\emph{arXiv}}), is to be compared with 6 Aug. 2003, the date on which Ref.~\protect\cite{BF04b} was published on \href{http://arxiv.org/abs/cond-mat/0308090}{\emph{arXiv}}.

% 81.
\bibitem{KB07}
K. Byczuk, M. Kollar, K. Held, Y.-F. Yang, I.~A. Nekrasov, Th. Pruschke, and D. Vollhardt, \emph{Nature Physics}, \textbf{3}, 168 (2007).

% 82.
\bibitem{AM07}
A. Macridin, M. Jarrell, T. Maier, and D.~J. Scalapino, \emph{Phys. Rev. Lett.} \textbf{99}, 237001 (2007).

% 83.
\bibitem{Note12}
In Ref.~\protect\cite{BF04a} we express $\t{v}(\|\bm{q}\|)$ as $\mathsf{g}\, \t{w}(\|\bm{q}\|)$, where $\mathsf{g}$ is the dimensional coupling constant of interaction. With $\mathsf{g} = U$ in the case of the conventional Hubbard Hamiltonian, one can easily compare the more general results in Ref.~\protect\cite{BF04a} with their counterparts in Ref.~\protect\cite{BF03}.

% 84.
\bibitem{Note13}
The Hubbard model being a lattice model with four possible single-particle states per site, its $N$- and $(N\pm 1)$-particle Hilbert spaces are countable and their dimensions finite for any $N <\infty$. Hence the explicit reference to the limit $N=\infty$. For continuum models, in contrast, the energies $\mu_{N;\sigma}^{\mp}$ are branch points even for $N <\infty$.

% 85.
\bibitem{ECT52}
E.~C. Titchmarsh, \emph{The Theory of Functions}, 2nd edition (Cambridge University Press, 1952).

% 86.
\bibitem{AS72}
M. Abramowitz, I.~A. Stegun, Editors, \emph{Handbook of Mathematical Functions}, 9th printing (Dover, New York, 1972).

% 87.
\bibitem{BF97a}
B. Farid, \emph{Phil. Mag.} B~\textbf{76}, 145 (1997).

% 88.
\bibitem{BF97b}
B. Farid, \emph{Solid State Commun.} \textbf{104}, 227 (1997). \href{http://arxiv.org/abs/cond-mat/9604138}{arXiv:9604138}

% 89.
\bibitem{BF99b}
B. Farid, \emph{Phil. Mag. Lett.} \textbf{79}, 581 (1999).

% 90.
\bibitem{KS65}
W. Kohn, L.~J. Sham, \emph{Phys. Rev.} \textbf{140}, A~1133 (1965).

% 91.
\bibitem{SG95}
A. Schindlmayr, and R.~W. Godby, \emph{Phys. Rev.} B~\textbf{51}, 10427 (1995).

% 92.
\bibitem{SGN95}
K. Sch\"onhammer, O. Gunnarsson, and R.~M. Noack, \emph{Phys. Rev.} B~\textbf{52}, 2504 (1995).

\end{thebibliography}

%________________________

\end{document}